\documentclass[journal]{IEEEtran}

\usepackage{graphicx}
\usepackage{amsmath}
\usepackage{amssymb}
\usepackage{bm}

\usepackage{booktabs}
\usepackage{multirow}
\usepackage{array}

\usepackage{textcomp}      
\usepackage{siunitx}       
\usepackage{url}
\usepackage{xcolor}

\usepackage[hidelinks,breaklinks=true]{hyperref}

\usepackage{algorithm}
\usepackage{algpseudocode}

\usepackage{enumitem}

\usepackage[noadjust]{cite}

\title{From Full and Partial Intraoral Scans to Crown Proposal: A Classification-Guided Restoration Assistance Pipeline}

\author{Rabin~Kunwar$^{1,*}$, Dikshya~Parajuli$^{1}$, Rujal~Acharya$^{1}$, Romik~Gosai$^{1}$, Prince~Panta$^{2}$, Kundan~Siwakoti$^{1}$, Shuvangi~Adhikari$^{1}$, Saugat~Kafley$^{1}$, Louis~Digiorgio$^{4}$, Amit~Regmi$^{1,6,\dagger}$, Akio~Tanaka$^{5}$, Masahiko~Inada$^{4,\ddagger}$, Yuriko~Komagamine$^{3}$, Kennta~Kashiwazaki$^{3}$, and Manabu~Kanazawa$^{3}$%

\thanks{$^{1}$Accelerated Komputing Pvt. Ltd., Bhaktapur, Nepal.\quad $^{2}$University of Pittsburgh, Pittsburgh, PA, USA.\quad $^{3}$Institute of Science Tokyo, Tokyo, Japan.\quad $^{4}$Emium Co. Ltd., Tokyo, Japan.\quad $^{5}$GodelBlock Inc., Tokyo, Japan.\quad $^{6}$Carnegie Mellon University, Pittsburgh, PA, USA.\protect\\ $^{*}$Lead author: rabin@ank-world.com.\quad $^{\dagger}$Corresponding author: amit@ank-world.com.\quad $^{\ddagger}$Corresponding author: masahiko.inada@emium.co.jp.}}
\begin{document}
\maketitle

\begin{abstract}
Single-unit crown restoration is one of the most common procedures in clinical dentistry, and CAD/CAM workflows now perform most of the design directly from intraoral scans. For single-unit cases, partial scans are often preferred over full-arch scans because they have fewer stitching errors, yet most existing tooth segmentation networks are trained on full arches and perform poorly on partial scans, while current end-to-end generative crown methods often produce over-smoothed surfaces that lose fine occlusal detail. We propose an end-to-end pipeline for clinical restoration design that takes a raw intraoral scan and target FDI tooth number as input and produces an initial, patient-specific crown proposal that the clinician subsequently finalizes. The pipeline is organized into three main phases: Phase I performs data preparation and pose standardization, Phase II performs full or partial segmentation routed by scan type, and Phase III generates the initial crown proposal through context-aware retrieval and Blender-based fitting. We address the partial-scan segmentation challenge through a \textit{classify-then-align} preprocessing strategy: a DGCNN classifier first categorizes the input scan into one of five anatomical types, after which a coarse-to-fine RANSAC+ICP registration standardizes the coordinate frame of the jaw before segmentation, followed by a graph-cut optimization to refine tooth--gingival boundaries.

Developed using 1{,}958 partial intraoral scans, the pipeline achieves a macro-average DSC of 0.9249, Recall of 0.8919, and Precision of 0.9615 across all 17 semantic classes; a fine-tuned full-arch model reaches DSC 0.9347. Region of interest evaluation shows strong spatial accuracy: the prepared tooth and its immediate mesial and distal neighbors achieve DSC values of 0.9468--0.9569 with sub-millimeter Centroid Errors (0.2666--0.2774~mm). These segmented centroids anchor a context-aware crown retrieval module that uses DGCNN feature embeddings and cosine similarity search over neighboring and opposing teeth, after which the retrieved crown is positioned via spline-guided sequential alignment and refined in the Blender Python API for interproximal and occlusal fit. The complete pipeline produces a patient-specific preliminary crown shell in approximately 2.5--3.5 minutes, providing an anatomical starting point that the clinician refines through margin cutting, contact verification, and surface finishing, offering a practical alternative to end-to-end generative approaches.
\end{abstract}


\section{Introduction}
\label{sec:intro}
\IEEEPARstart{I}{ntraoral} scanners (IOSs) have transformed dentistry \cite{SAMARANAYAKE2025383,Dilated_DGCNN_Reference}, but a major gap remains between clinical reality and current Artificial Intelligence (AI) capabilities \cite{Jana2023DentalMeshSegmentation}. In clinical practice, studies suggest that partial arch scans often have higher trueness (closeness to the actual dimensions) than complete arch scans due to the reduction of stitching errors over long spans \cite{Ender2016Effect, Park2016ScanArea, Imburgia2017IOS, Mangano2019IOS}. Consequently, partial scans are the preferred modality for single-unit restorations.

However, most current AI models are trained exclusively on perfect, full dental arches. Prominent examples include MeshSegNet \cite{Lian2020MeshSegNet}, Dental Arch Prior-assisted 3D Tooth Instance Segmentation (DArch) \cite{Qiu2022DArch}, Dynamic Graph CNN (DGCNN) \cite{DGCNN_Reference}, PointNet++ \cite{PointNetPP_Reference}, generative adversarial networks \cite{Hwang2021GAN} and TFormer \cite{Xiong2022TFormer}. A majority of the available Deep Learning (DL) tooth segmentation techniques make an implicit, restrictive assumption of a full jaw model and subsequently report accuracy based only on these full jaw models \cite{Jana2023DentalMeshSegmentation}. When these models are fed a partial scan that lacks this global context, they drastically underperform \cite{Jana2023DentalMeshSegmentation}. Standard segmentation models frequently misclassify teeth when dealing with cases with varying numbers of teeth \cite{Cui2021TeethNet}. This lack of robustness when applied to partial intraoral scans is considered a serious problem for deploying these algorithms in medical settings \cite{Jana2023DentalMeshSegmentation,Alsheghri2024ArchSeg}.
This vulnerability is largely attributed to the lack of orientation in raw clinical data. Recent ablation studies explicitly highlight how critical geometric standardization is for model performance. For instance, the \textit{ArchSeg} framework \cite{Alsheghri2024ArchSeg} demonstrated that without prior registration to a standard coordinate system, state-of-the-art models like MeshSegNet failed to correctly segment even a single complete tooth in partial scans containing fewer than eight teeth. Their results identified registration as the ``most important factor'' for segmentation success \cite{Alsheghri2024ArchSeg}. 

While CAD/CAM systems have streamlined the manufacturing of patient-specific dental prostheses, the digital design phase remains a significant bottleneck due to high manual labor requirements \cite{Samra2016CADCAM}. Technicians must still manually select and rigorously adjust generic library templates to ensure precise marginal fit and functional contacts \cite{DCrownFormer2024, MeshCompletion2025, Hosseinimanesh2025}. Because these templates lack initial personalization, experienced human technicians currently continue to outperform knowledge-based AI in producing designs that balance anatomical fidelity with necessary fracture strength \cite{Ding2022Morphology}.

Recent deep learning approaches, particularly transformer-based networks, have significantly accelerated this process but frequently suffer from ``over-smoothing,'' failing to reproduce fine occlusal grooves and cusps despite utilizing specialized losses and large amounts of training data \cite{Hosseinimanesh2025, DCrownFormer2024}. Furthermore, fully automated commercial frameworks exhibit significant inconsistency across different tooth types; while simpler geometries like incisors (e.g., positions 11, 12, 22) are predicted with high accuracy, complex posterior teeth (e.g., premolars and molars at 15, 16, 36, 46) show significantly higher dissimilarity metrics, indicating a failure to capture intricate occlusal surfaces \cite{IntelliDent2025}. In contrast, Statistical Shape Models (SSMs) are fundamentally restricted by their reliance on rigid mean shapes that cannot adapt to the unique anatomy of the patient \cite{FrontiersAutomated}. Therefore, a truly effective clinical framework must solve two distinct but interconnected challenges: first, robustly segmenting geometrically unoriented partial scans, and second, utilizing that patient-specific context to generate morphologically accurate restorations. To address these gaps, this work proposes a unified pipeline for crown proposal from full and partial intraoral scans. We tackle the segmentation challenge through a classify-then-align preprocessing strategy that automates registration without manual input, and the crown-generation challenge through a retrieval-based module that preserves fine occlusal detail offering a practical alternative. The following section reviews related work and details our contributions.

\section{Related Works}
\label{sec:related}

\subsection{Literature Review}
\label{ssec:related_review}
\begin{enumerate}[label=\arabic*.]
    \item \textbf{Segmentation:} Accurate dental arch segmentation is important for reliable digital dentistry workflows. However, deep learning models that achieve high accuracy on ideal, fully captured scans often struggle to maintain robust segmentation when processing partial or incomplete intraoral scans commonly obtained in clinical settings\cite{Jana2023DentalMeshSegmentation}. Although the development of novel architectures, such as Transformer-based models, has shown promise in improving feature extraction on large-scale datasets totaling 16,000 IOS scans \cite{Xiong2022TFormer}, a key challenge still remains in ensuring reliable performance on geometrically variable, missing tooth, and partial or incomplete intraoral scans. Another line of research focuses on feature-based segmentation models, which aim to learn distinctive geometric and morphological features for individual teeth for segmentation\cite{pmlr-v227-chen24b}, \cite{pmlr-v172-he22b}. These models show promising results but still tend to struggle in abnormal cases such as tooth crowding, diastema, damaged teeth, or missing teeth, where the learned feature representations become inconsistent.\\
    
    Recent studies suggest that this problem is best addressed not just by improving the segmentation model itself, but by standardizing the input. The work on ArchSeg, for instance, provides compelling evidence that a registration-first approach with aligning the arch to a standard coordinate system before segmentation is the \textbf{most important factor} for achieving robust results on partial or incomplete scans\cite{Alsheghri2024ArchSeg}. This insight shifts the challenge from purely model-based feature learning to robust geometric pre-processing. However, the ArchSeg framework relies on the information provided by the user, such as the range of tooth labels in the arch, to assist in registration\cite{Alsheghri2024ArchSeg}. Our work builds on this \textbf{registration-first} approach but overcomes its dependence on initial tooth labels by introducing a fully automated global alignment using a RANSAC+ICP step during pre-processing, inspired by the registration method proposed in \cite{Xu2021FastPointCloud}. This step creates a standardized coordinate system without any manual intervention, ensuring that the segmentation network receives consistent and well-aligned inputs which is an essential factor for reliable segmentation performance on unseen clinical data.
\item \textbf{Crown Generation:} The literature shows steady progress in automated crown design, but most methods still need meaningful clinical adjustment before delivery\cite{DCrownFormer2024,Hosseinimanesh2025,MeshCompletion2025,IntelliDent2025}. Reported limitations include loss of fine occlusal detail, variability across tooth types, and reduced reliability in complex posterior cases. For this reason, many practical workflows continue to use AI as an assistant. Likewise, our work follows a retrieval-based strategy and produces an initial crown proposal through automated segmentation, retrieval, and crown template positioning, while the clinician performs the final margin cutting, contact verification, and surface finishing. 

\end{enumerate}

\subsection{Our Contribution}
\label{ssec:related_contribution}

In this paper, we propose an automated pipeline for dental segmentation and prosthesis generation that works robustly on both full and partial intraoral scans. The overall workflow of our proposed method is illustrated in Figure~\ref{fig:Pipeline_flowchart}. Our main contributions are summarized as follows:
\begin{figure*}[!t]
    \centering
    \includegraphics[width=\textwidth,height=0.8\textheight,keepaspectratio]{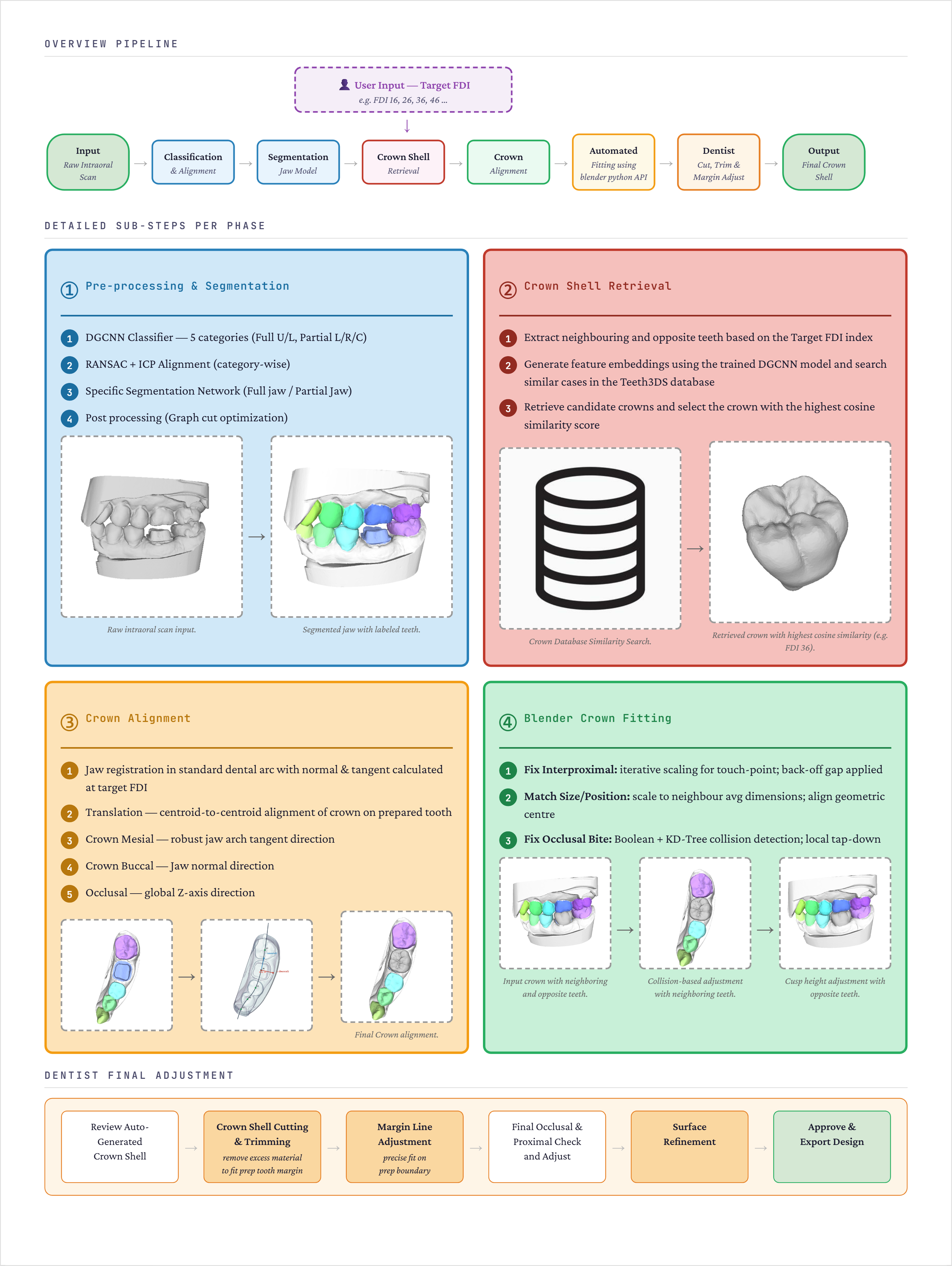}
    \caption{\textbf{Overview of the proposed crown generation pipeline.} The dentist first specifies the target \textbf{FDI tooth number} as user input. \textbf{(a)} Raw intraoral scans are classified and aligned using DGCNN + RANSAC/ICP. \textbf{(b)} A segmentation network isolates individual tooth surfaces. \textbf{(c)} A neighbour-context-aware retrieval module uses the FDI number to query the crown library and selects the most morphologically similar shell. \textbf{(d)} The shell is aligned sequentially (translation, mesial, buccal, occlusal) and fitted via Blender Python API (interproximal gap correction, size matching, occlusal bite adjustment). \textbf{(e)} The dentist physically cuts and trims the crown shell, adjusts the margin line to fit the prepared tooth boundary, and verifies occlusal and proximal contacts before export.}
    \label{fig:Pipeline_flowchart}
\end{figure*}

\begin{itemize}
    \item \textbf{Classification-guided Alignment:} We introduce a pre-processing step that uses a Dynamic Graph CNN (DGCNN) to classify the input scan into one of five specific anatomical categories: \textit{Full Upper Jaw}, \textit{Full Lower Jaw}, \textit{Partial Left}, \textit{Partial Right}, and \textit{Partial Center}. Based on this classification, we automatically align the scan to a fixed, standard orientation using RANSAC and ICP algorithms. Unlike the ArchSeg framework~\cite{Alsheghri2024ArchSeg}, which requires users to manually specify tooth label ranges to assist registration, our approach solves the alignment problem for partial data without any manual input.
    
    \item \textbf{Segmentation Models:} We replace the single general-purpose model approach used in previous studies with a targeted strategy. Using the initial classification results, we route the data to one of two dedicated networks: \textit{Full}, or \textit{Partial}. Furthermore, because our pre-processing step ensures that inputs are always aligned in a fixed, consistent direction, the models can generalize much better to unseen data. This combination of alignment and specialization results in significantly higher accuracy and fewer errors compared to a universal model that struggles with unoriented and partial scans as noted in~\cite{Jana2023DentalMeshSegmentation}, \cite{Alsheghri2024ArchSeg}.
    
    \item \textbf{Initial Crown Proposal:} We extend our pipeline to include a context-aware initial crown proposal. Initiated by a single, minimal user input identifying the target FDI number, the system automatically evaluates the geometric features of the neighboring teeth to retrieve the most morphologically suitable crown template from a library, and proper Buccal and Lingual alignment of the crown is done with respect to the jaw. This template is then automatically scaled and positioned using the Blender Python API. By executing the entire automated process from raw scan to positioned initial shell in roughly 2.5 to 3.5 minutes, this semi-automated workflow provides an advanced starting point that significantly reduces the clinician's manual design burden before final adjustments and export.

\end{itemize}
\section{Methodology}
\label{sec:methods}
Our proposed pipeline is divided into three primary phases. Phase I focuses on data preparation and pose standardization: (1) Dataset Preparation and Preprocessing, (2) Global Pose Standardization for Improved Segmentation.  Phase II focuses on segmentation: (3) Segmentation of Full and Partial Jaw Scans followed by graph-cut optimization. Phase III focuses on generating the Initial Crown Proposal: (4) Context-aware crown retrieval, (5) Crown-to-Preparation Alignment, and (6) Automated Crown Adjustment using the Blender API.

\subsection{Phase I: Data Preparation and Pose Standardization}
\label{ssec:phase1}

\subsubsection{Dataset Preparation}
\label{sssec:dataset_prep}
\begin{figure}[h!]
    \centering
    \includegraphics[width=\linewidth]{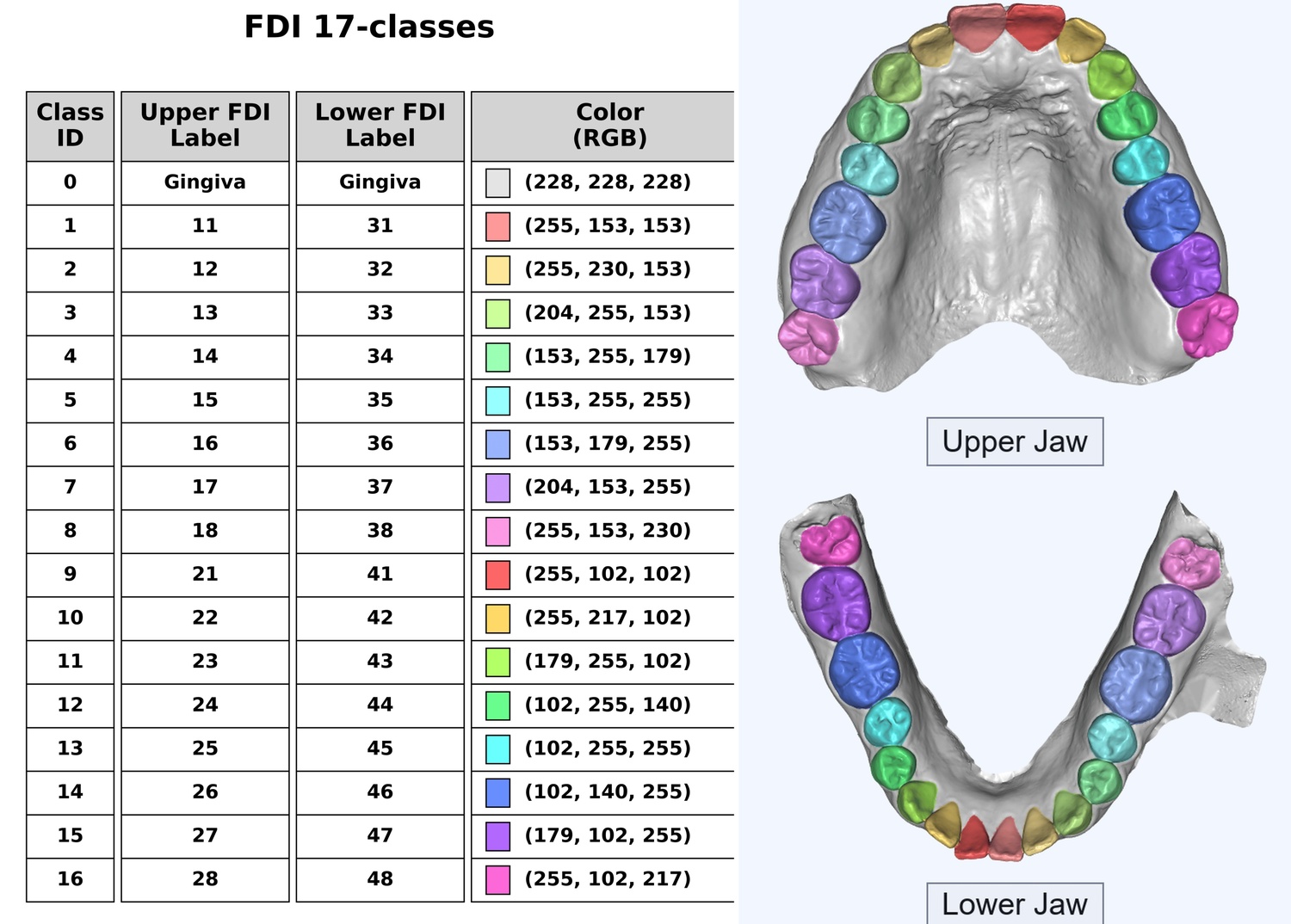}
    
    \caption{\textbf{Ground Truth Annotation Schema.} Visualization of the semantic labeling strategy based on the FDI notation system. 
    \textbf{(Top)} The Maxillary arch, containing indices 11--18 (upper right) and 21--28 (upper left). 
    \textbf{(Bottom)} The Mandibular arch, containing indices 31--38 (lower left) and 41--48 (lower right). 
    Each FDI label is associated with a unique color used during the semantic segmentation training process.}
    \label{fig:fdi_mapping}
\end{figure}

The data set was manually segmented by a team of annotators using internal annotation software. Ground truth labels were assigned according to the Fédération Dentaire Internationale (FDI) notation system. As illustrated in Fig.~\ref{fig:fdi_mapping}, each tooth type in the Maxillary (Upper) and Mandibular (Lower) arches is assigned a unique semantic class ID and a corresponding visualization color to facilitate multi-class segmentation training.

\subsubsection{Dataset Characteristics}
\label{sssec:dataset_chars}
Our dataset comprises both full-arch and partial-arch scans to reflect clinical variety.

\paragraph{Full Jaw Datasets}
The characteristics of the full jaw dataset are summarized in Table~\ref{tab:dataset_stats}. This subset consists of 301 total scans, with a nearly equal distribution between Full Upper (148) and Full Lower (153) arches.

\paragraph{Partial Jaw Datasets}
The characteristics of the partial jaw segments, categorized into Central, Left, and Right partials, are also summarized in Table~\ref{tab:dataset_stats}. This diversity ensures the pipeline remains robust even when the global arch context is missing.

\paragraph{Labeling Scheme and Class Distribution}
Following the approach used in DilatedToothSegNet~\cite{Dilated_DGCNN_Reference}, we do not distinguish between upper and lower jaws during training. Instead, we adopt a single 16-class system in which similar tooth types (e.g., an upper molar and a lower molar) share the same class ID. This shared labeling effectively doubles the training data per tooth type and is particularly helpful for wisdom teeth (Class IDs 8 and 16), which are rare in clinical scans. For the full-arch model, we address the limited dataset ($N=301$) by fine-tuning the best-performing partial-arch weights rather than training from scratch, while keeping the same unified labeling system.

\begin{figure*}[t]
    \centering
    \begin{minipage}[t]{0.495\textwidth}
        \centering
        \includegraphics[width=\linewidth]{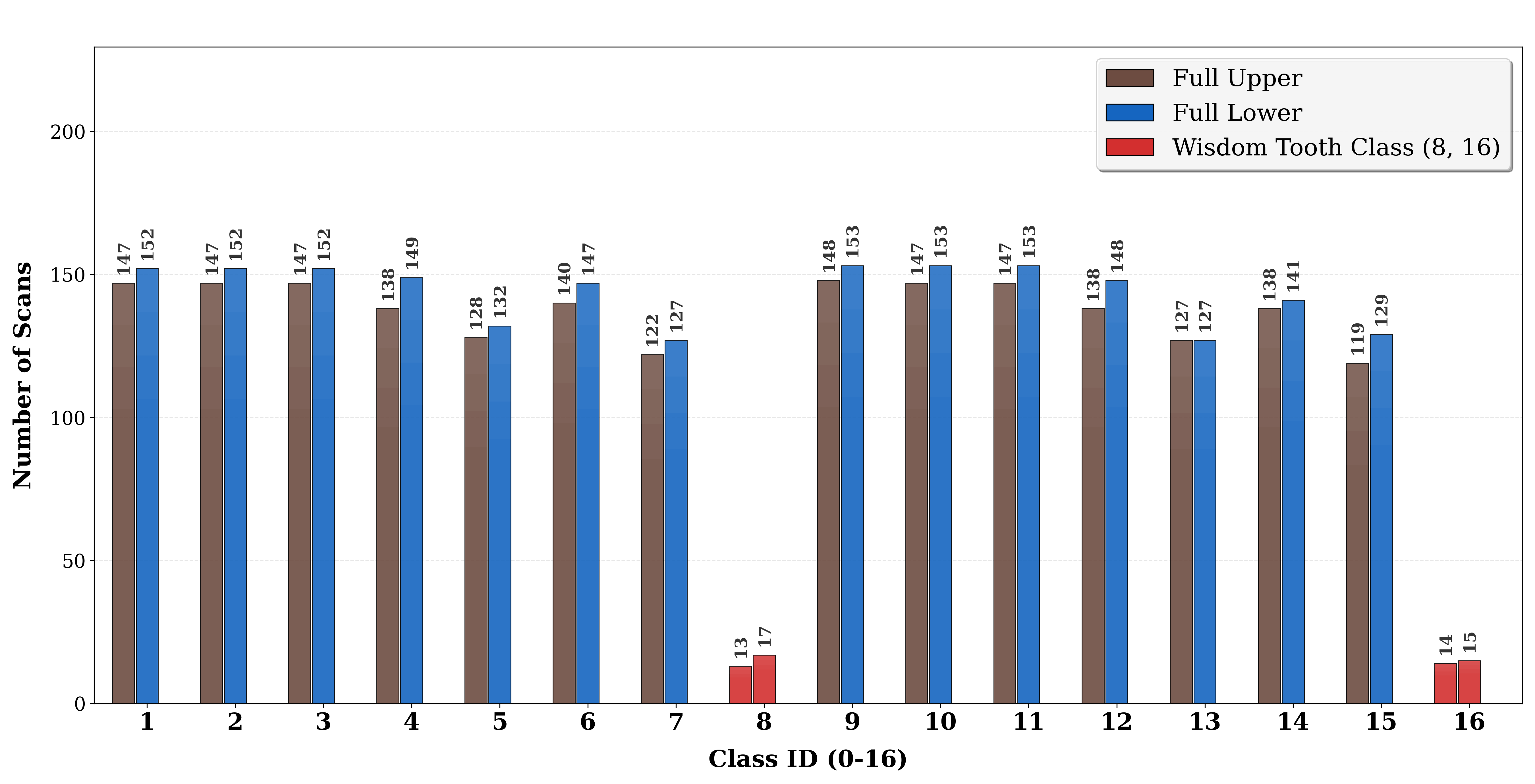}
        \caption{Distribution of full-jaw tooth classes highlighting wisdom teeth. Red bars indicate Class IDs 8 and 16, illustrating the data imbalance inherent in clinical intraoral full scans.}
        \label{fig:full_class_dist}
    \end{minipage}\hfill
    \begin{minipage}[t]{0.495\textwidth}
        \centering
        \includegraphics[width=\linewidth]{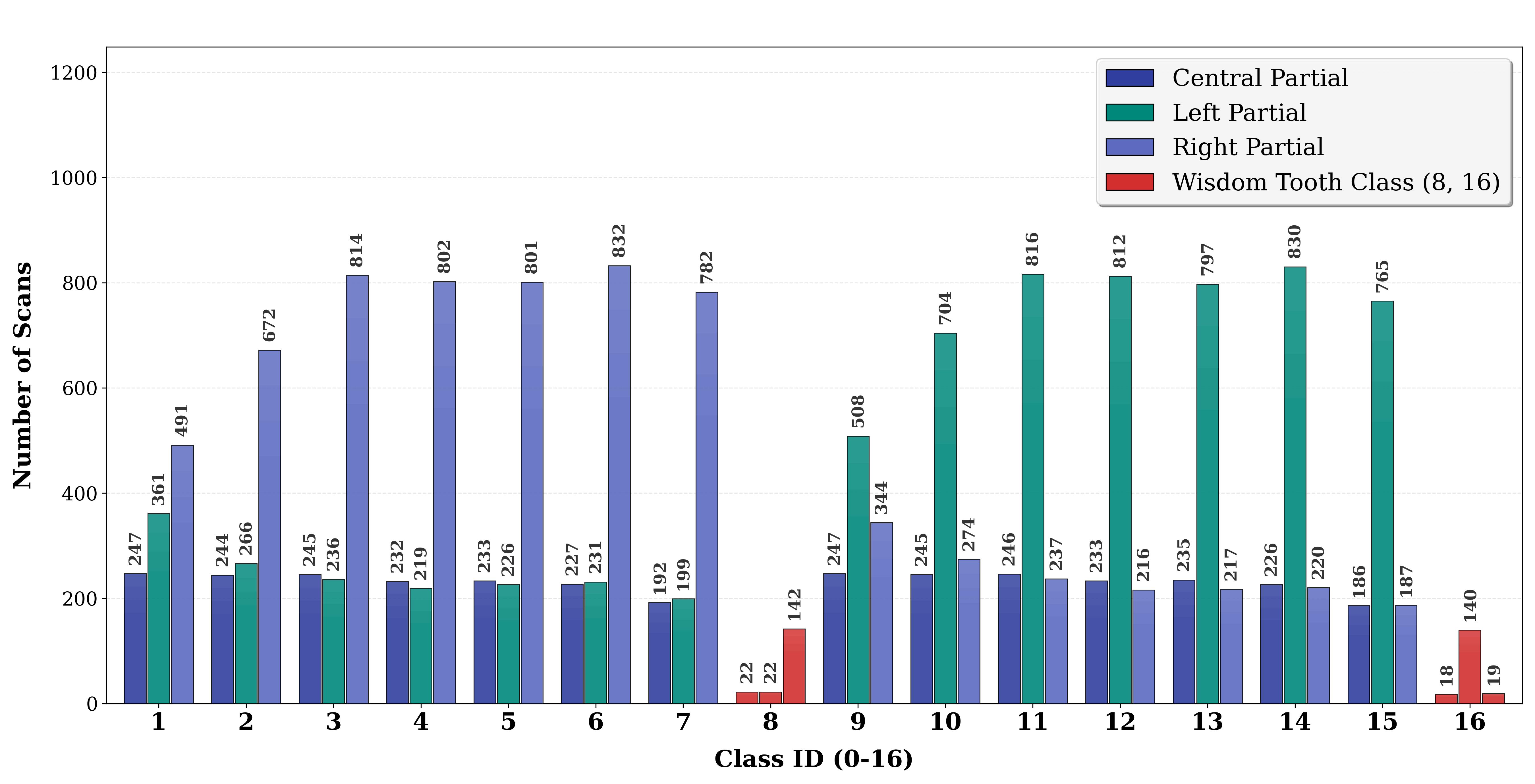}
        \caption{Distribution of partial-jaw tooth classes highlighting wisdom teeth. Red bars indicate Class IDs 8 and 16, illustrating the data imbalance inherent in clinical intraoral partial scans.}
        \label{fig:class_dist}
    \end{minipage}
\end{figure*}

As shown in Fig.~\ref{fig:full_class_dist} and Fig.~\ref{fig:class_dist}, there is a clear class imbalance across the 16 dental classes in both modalities. First molars and incisors are well represented, while wisdom teeth (Class IDs 8 and 16, highlighted in red) appear at drastically lower frequency. This reflects clinical reality, since third molars are commonly extracted or never erupt; the corresponding mitigation strategy (targeted augmentation on the training set) is described in Section~\ref{sec:experiments}.

\begin{table}[!t]
\renewcommand{\arraystretch}{1.3}
\caption{Dataset characteristics across full-jaw and partial-jaw scan categories. Counts include scan totals, prepared teeth, missing teeth, and outlier teeth.}
\label{tab:dataset_stats}
\centering
\footnotesize
\begin{tabular}{lrrrr}
\toprule
\textbf{Scan Category} & \textbf{Scans} & \textbf{Prepared} & \textbf{Missing} & \textbf{Outliers} \\
\midrule
\multicolumn{5}{l}{\textit{Full Jaw}} \\
\midrule
Full Upper & 148 & 64 & 74 & 80 \\
Full Lower & 153 & 91 & 68 & 49 \\
\textbf{Subtotal} & \textbf{301} & \textbf{155} & \textbf{142} & \textbf{129} \\
\midrule
\multicolumn{5}{l}{\textit{Partial Jaw}} \\
\midrule
Central Partial & 247 & 115 & 55 & 124 \\
Left Partial & 859 & 314 & 251 & 613 \\
Right Partial & 852 & 484 & 144 & 633 \\
\textbf{Subtotal} & \textbf{1{,}958} & \textbf{913} & \textbf{450} & \textbf{1{,}370} \\
\midrule
\textbf{Total} & \textbf{2{,}259} & \textbf{1{,}068} & \textbf{592} & \textbf{1{,}499} \\
\bottomrule
\end{tabular}
\vspace{2pt}\\
\raggedright
\footnotesize \textit{Note:} Outlier teeth come from extreme-case scans where the captured mesh is not watertight, e.g., teeth with holes, missing surface regions, or incomplete geometry.
\end{table}

\subsubsection{Global Pose Standardization for Improved Segmentation}
\label{sssec:pose_std}
Deep learning models for segmentation, such as DGCNN \cite{DGCNN_Reference} and its variants, often exhibit sensitivity to the input data's orientation. Intraoral scans are captured in arbitrary coordinate systems, which introduces significant variance that can degrade segmentation performance. Furthermore, aligning a partial scan to a generic full arch template often results in geometric mismatches. To mitigate this and ensure robust segmentation across diverse scans, we implement a \textit{``Classify-then-Align''} pre-processing strategy. This approach utilizes a hierarchical registration framework, inspired by principles discussed in \cite{vanNistelrooij2024ToothWear}, to standardize the orientation of the jaw scan before feeding it to the segmentation network.

The standardization process is composed of two sequential phases:

\paragraph{Preprocessing (i): Jaws Classification}
Prior to registration, it is essential to determine the anatomical category of the input scan. This classification step serves two critical functions: it selects the appropriate canonical reference template for the alignment phase, and it routes the data to the specific downstream segmentation model trained for that anatomy (either partial or full jaw model).

\begin{figure}[t]
    \centering
    \includegraphics[width=\columnwidth]{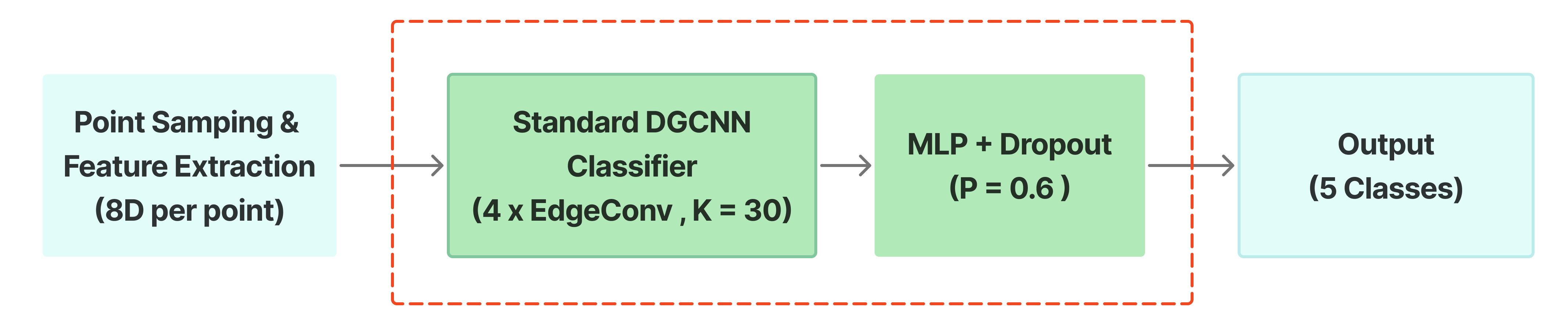}
    \caption{\textbf{Preprocessing stage (i):} Jaws classification of intraoral scans.}
    \label{fig:class_pipeline}
\end{figure}

To achieve this, we employ a deep classification network based on the Dynamic Graph CNN (DGCNN) architecture \cite{DGCNN_Reference}, as illustrated in Fig.~\ref{fig:class_pipeline}. This model categorizes the raw input point cloud into one of five topological classes: \textit{Full Upper}, \textit{Full Lower}, \textit{Partial Left}, \textit{Partial Right}, and \textit{Partial Center}.

\noindent\textbf{Input Representation.} The network processes the scan as a point cloud with $N$ nodes. To capture both spatial structure and local surface curvature, each point is represented by an 8-dimensional feature vector consisting of its Cartesian coordinates, estimated surface normals, and polar coordinates:
\begin{equation}
    \mathbf{x}_i = [x_i, y_i, z_i, n_{x_i}, n_{y_i}, n_{z_i}, r_i, \phi_i],
\end{equation}
where $(x_i, y_i, z_i)$ denote the Cartesian coordinates of the $i$-th point, $(n_{x_i}, n_{y_i}, n_{z_i})$ are the components of the corresponding surface normal vector, $r_i$ is the polar radius, and $\phi_i$ is the azimuth angle.

The model ($k=30$, embedding dims $= 1024$) was trained for 300 epochs using the AdamW optimizer ($lr=5 \times 10^{-4}$, weight decay $= 0.001$) and a Cosine Annealing scheduler. To address the inherent class imbalance where lateral partial scans (n=1,711) significantly outnumber full arch and central segments (n=547), we employed Focal Loss ($\gamma=1.5$) with label smoothing (0.2) and a Mixup augmentation strategy ($\alpha=1.0$). These techniques ensure the network learns robust geometric boundaries even for underrepresented anatomical types.
\\
\noindent\textbf{Classifier Performance.}
The classification module achieved a test accuracy of 94.2\% ($n=240$) (compared to 95.7\% during validation), providing a stable foundation for automated registration. As shown in Table~\ref{tab:classification_metrics}, the model demonstrated exceptional reliability for lateral partial scans, with F1-scores of 0.944 (Left) and 0.957 (Right). High precision in these categories is vital because partial scans lack the global arch context required for standard alignment.
\begin{table}[h]
\centering
\caption{Test Set Classification Metrics (8-Feature Best Accuracy Model)}
\label{tab:classification_metrics}
\renewcommand{\arraystretch}{1.2}
\setlength{\tabcolsep}{6pt}
\footnotesize
\begin{tabular}{lccc}
\toprule
\textbf{Scan Category} & \textbf{Precision} & \textbf{Recall} & \textbf{F1-Score} \\
\midrule
Left Partial    & 0.955 & 0.934 & 0.944 \\
Right Partial   & 0.957 & 0.957 & 0.957 \\
Central Partial & 0.833 & 0.909 & 0.870 \\
Full Lower      & 0.905 & 0.950 & 0.927 \\
Full Upper      & 0.920 & 0.920 & 0.920 \\
\midrule
\textbf{Macro Average} & \textbf{0.914} & \textbf{0.934} & \textbf{0.924} \\
\textbf{Accuracy}      &       &       & \textbf{94.2\%} \\
\bottomrule
\end{tabular}
\end{table}

While minor confusion occurred among the Full Upper, Full Lower, and Central Partial categories, these scans share a parabolic arch form, and their master templates share nearly identical canonical orientations. As a result, misclassifying an input within this group still leads to a closely oriented jaw scan during registration.

\paragraph{Preprocessing (ii): Classification-Guided Coarse-to-Fine Registration}

\noindent\textbf{Reference Template Selection.}
To ensure our registration targets were statistically representative of the general population, we utilized an internal dataset of $N=98$ high-quality full-arch scans (49 Maxillary and 49 Mandibular).

\noindent\textbf{Orientation Normalization.}
Raw intraoral scans often possess arbitrary coordinate systems. To standardize our private dataset, we applied the orientation normalization pipeline described in \cite{Teeth3DS_Reference}. This process aligns the occlusal plane to the global XY-plane and corrects anterior-posterior ambiguity using parabolic fitting, ensuring all 98 scans share a consistent anatomical pose prior to template selection.

\noindent\textbf{Reference Template Generation.}
With the dataset globally aligned, we performed a population-level analysis to select the Canonical Standard Templates.

\begin{figure}[t]
    \centering
    \includegraphics[width=\columnwidth,height=6cm,keepaspectratio]{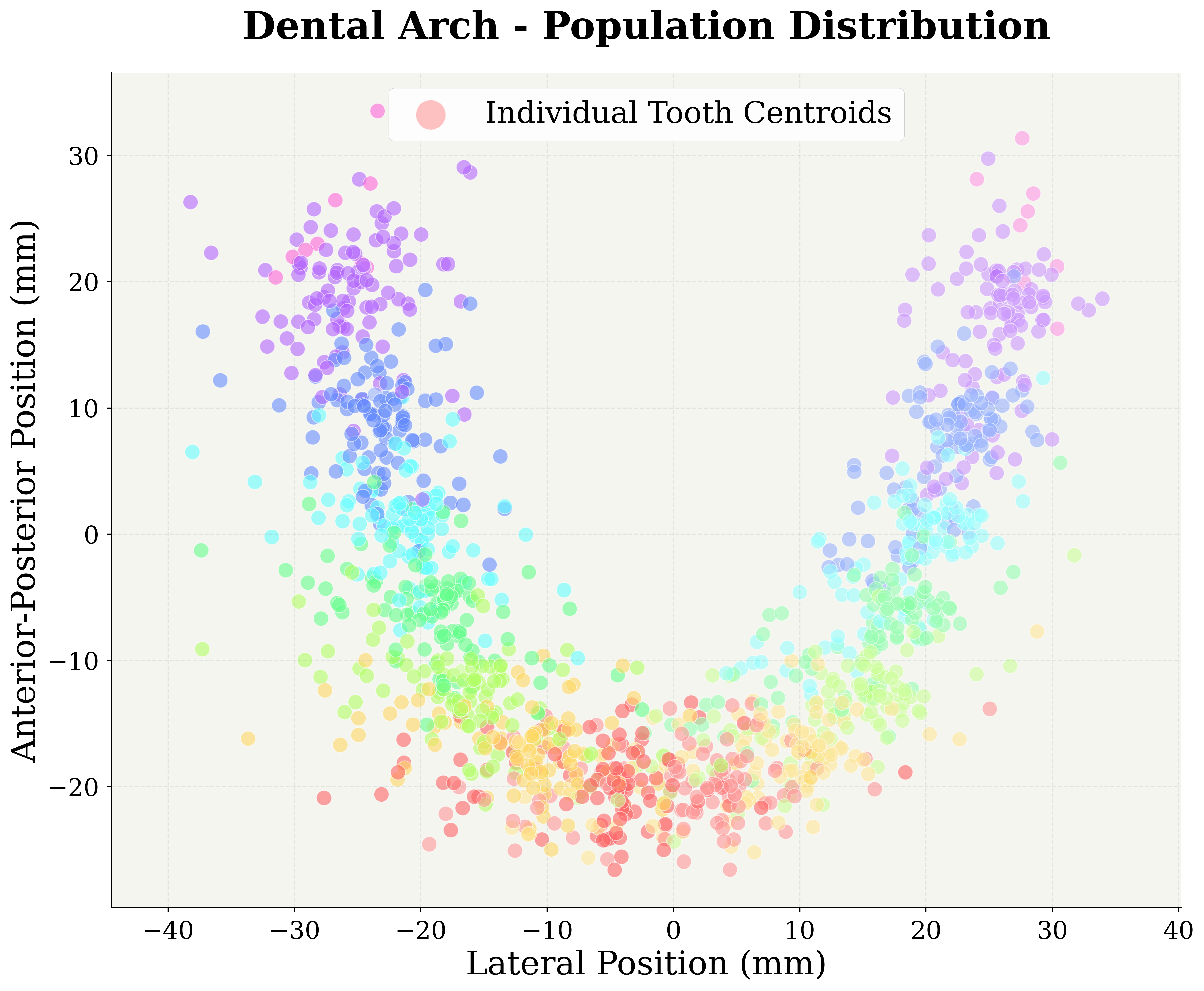}\\[6pt]
    \includegraphics[width=\columnwidth,height=6cm,keepaspectratio]{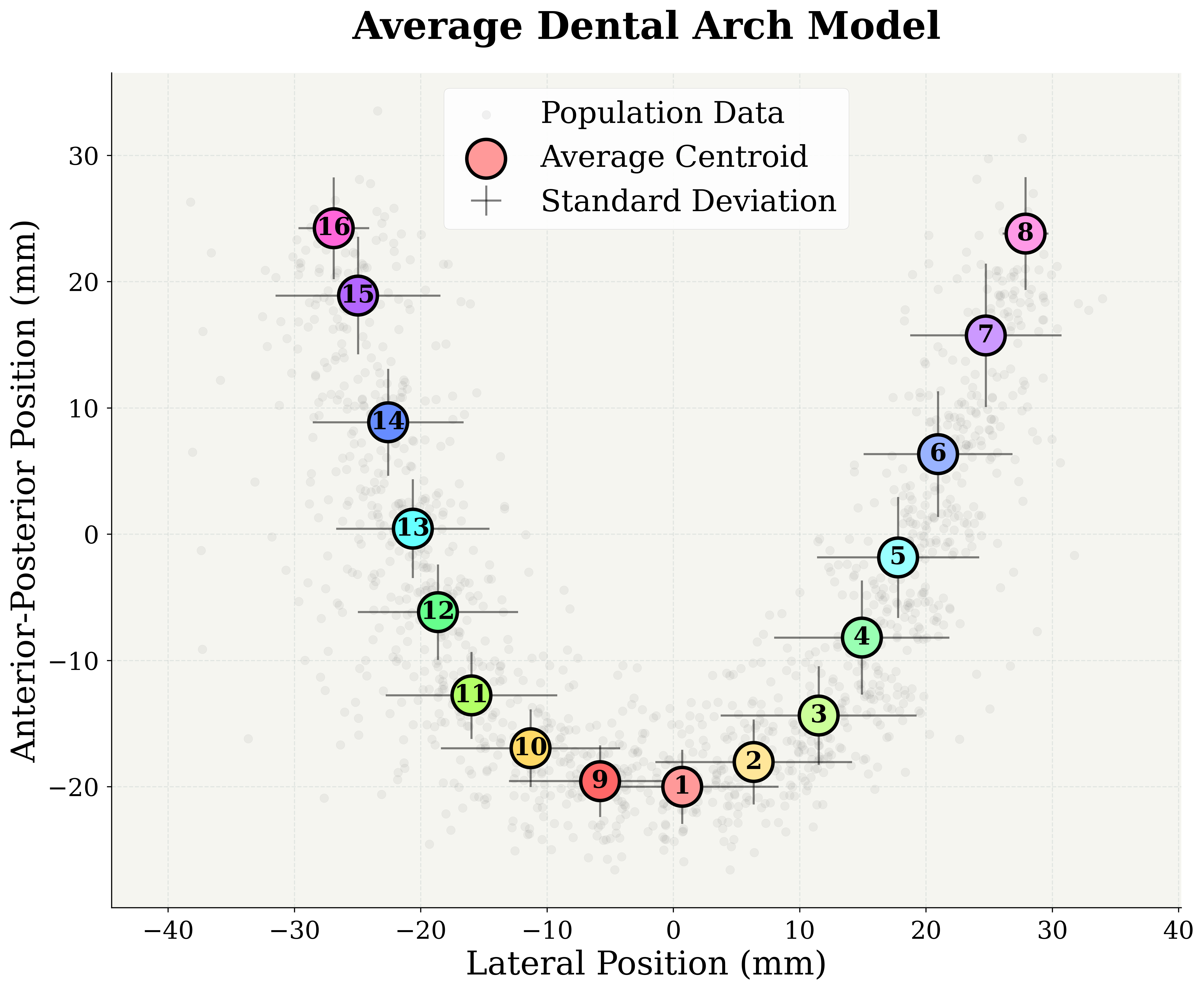}
    \caption{\textbf{Population Analysis.} (Top) Visualization of all individual tooth centroids after global alignment. (Bottom) The computed ``Global Average Centroid Curve'' representing the population mean.}
    \label{fig:pop_analysis}
\end{figure}

\begin{enumerate}
    \item \textbf{Centroid Extraction:} For every scan in the aligned dataset, we calculated the geometric center point (centroid) for each individual tooth (see Fig.~\ref{fig:pop_analysis}-a).
    
    \item \textbf{Finding the Average:} We averaged these points across the entire population ($N=98$). Since the Upper and Lower jaws in our standardized dataset share the same orientation, this allowed us to generate a \textbf{single Global Average Centroid Curve} that defines the standard arch shape for both the Maxilla and Mandible (see Fig.~\ref{fig:pop_analysis}-b).
\end{enumerate}

\begin{figure*}[t!]
    \centering
    \begin{minipage}[b]{0.32\textwidth}
        \centering
        \includegraphics[width=\linewidth]{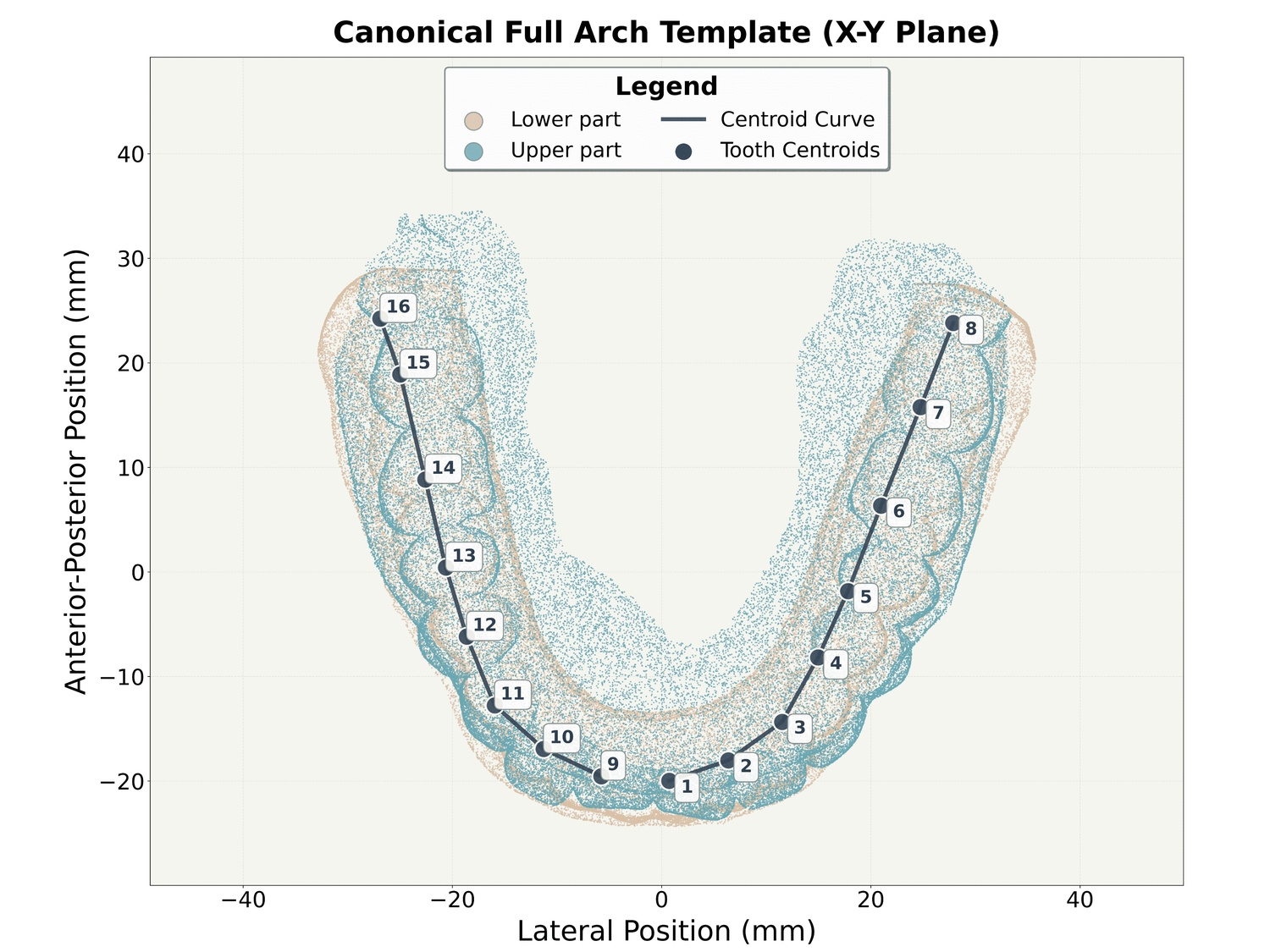}
    \end{minipage}
    \hfill
    \begin{minipage}[b]{0.32\textwidth}
        \centering
        \includegraphics[width=\linewidth]{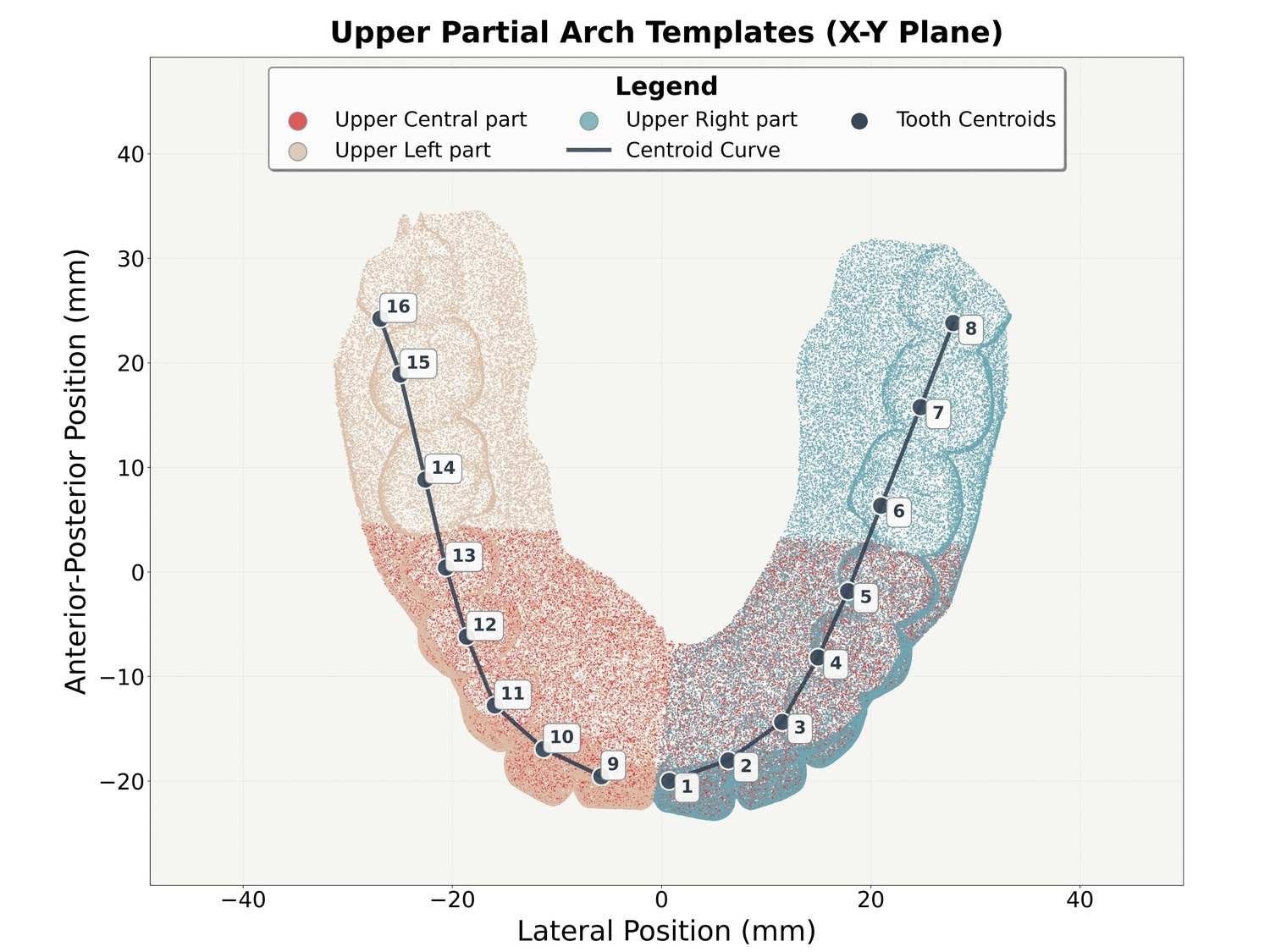}
    \end{minipage}
    \hfill
    \begin{minipage}[b]{0.32\textwidth}
        \centering
        \includegraphics[width=\linewidth]{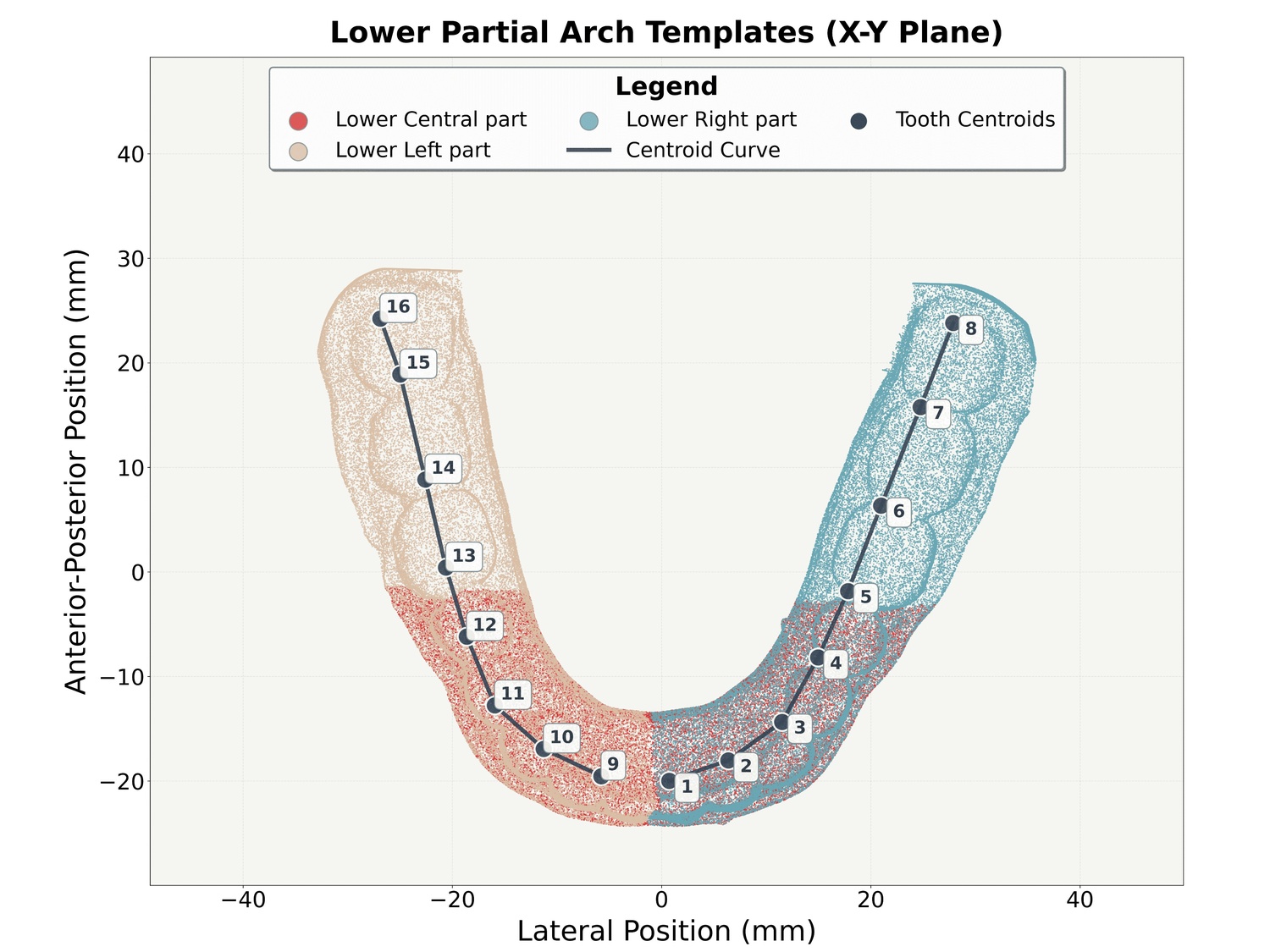}
    \end{minipage}
    
    \caption{\textbf{Canonical Reference Templates.} (a) The Canonical Full Upper and Lower jaws selected based on the best fit to the average curve. (b, c) The dictionary of Upper and Lower partial templates derived by cropping the Master templates, shown aligned to the global curve.}
    \label{fig:derived_templates}
\end{figure*}

\begin{enumerate}[resume] 
    \item \textbf{Selecting the Standard:} We searched for the specific Upper and Lower scans that best fit this single average curve. We selected the scans that minimized the total Euclidean distance to the population mean:
    \begin{equation}
        S_{\text{canonical}} = \mathop{\arg\min}\limits_{S} \sum_{t=1}^{16} \| C_{t,S} - \bar{C}_t \|_2
    \end{equation}
    The scans $S$ with the lowest error were chosen as our Canonical Master Templates (see Fig.~\ref{fig:derived_templates}-a).

    \item \textbf{Creating Partial Templates:} To ensure consistency in the standard templates, we derived distinct partial templates from the respective Standard full jaw templates:
    \begin{itemize}
        \item \textit{Upper Partials} (Left, Right, Center) were cut from the \textbf{Upper Master} (see Fig.~\ref{fig:derived_templates}-b).
        \item \textit{Lower Partials} (Left, Right, Center) were cut from the \textbf{Lower Master} (see Fig.~\ref{fig:derived_templates}-c).
    \end{itemize}
    This results in a library of 6 partial templates. During registration, the system checks the input against both the Upper and Lower options to automatically find the best anatomical match.
\end{enumerate}

\noindent\textbf{Alignment Pipeline.}
Following the semantic classification by the DGCNN, we perform a robust rigid registration to transform the input scan into a standard coordinate system. Unlike semi-automatic frameworks such as ArchSeg~\cite{Alsheghri2024ArchSeg}, which rely on user-defined tooth ranges to aid alignment, our approach automates the registration process entirely by leveraging the predicted class labels to select specific anatomical templates. The alignment pipeline proceeds in two stages:

\begin{enumerate}[label=\roman*.]
    \item \textbf{Coarse Alignment:} 
    Initial orientation is established using Global Registration with Fast Point Feature Histograms (FPFH)~\cite{Rusu2009}. A critical challenge in dental registration is \textit{feature aliasing}, where smooth, prepared abutment teeth lack distinct morphology and are geometrically conflated with anterior teeth (incisors) at standard resolutions. 
    
    To mitigate this, we calibrated the voxel downsampling size to $\nu = 0.8$~mm. While standard approaches often use coarser grids (1.5--2.0~mm), we found that the finer 0.8~mm resolution is necessary to capture the occlusal anatomy (cusps and fissures) of the healthy neighboring teeth. We compute FPFH features with a search radius of $r = 7\nu$ ($\approx 5.6$~mm), which enables the RANSAC estimator to distinguish unique posterior geometries from anterior curves, effectively preventing the algorithm from locking onto an incorrect arch.

    \item \textbf{Robust Fine Alignment:} 
    The coarse alignment is refined using a Point-to-Plane Iterative Closest Point (ICP) algorithm~\cite{Chen1991}. To handle clinical artifacts such as gingival noise or the geometric discrepancy between a prepared tooth in the input and a full tooth in the template we employ a \textbf{Tukey Biweight Loss} function ($k=0.5$). This robust estimator down weights outliers, ensuring the final alignment is driven strictly by the reliable, unchanged tooth surfaces rather than prepared stumps.
\end{enumerate}

\noindent\textbf{Implementation Details.}
The pipeline was implemented using the Open3D library~\cite{Zhou2018}. Our approach used DGCNN to accurately predict different categories of jaws, while the local curvature of maxillary and mandibular partial jaws can be geometrically similar. 

To resolve this, a scan predicted as \textit{Partial Left} is registered against both the canonical \textit{Upper Left partial} and \textit{Lower Left partial} templates. The alignment yielding the highest fitness score determines the final anatomical assignment.

Furthermore, to prevent misalignment errors along the dental arch, we tightened the RANSAC geometric constraints by setting the \textit{Edge Length Similarity} threshold to $0.95$. This enforces strict spatial consistency between the source and template triangles, ensuring that prepared molars are not wrongly aligned with geometrically smoother anterior regions.

\begin{figure*}[t]
    \centering
    \setlength{\tabcolsep}{2pt}
    \renewcommand{\arraystretch}{1.0}
    \newcommand{\regimg}[1]{\parbox[c][0.125\textwidth][c]{0.19\textwidth}{\centering\includegraphics[width=0.19\textwidth,height=0.125\textwidth,keepaspectratio]{#1}}}
    \begin{tabular}{c c c c c}
        \small \textbf{Upper Full Jaw} &
        \small \textbf{Lower Full Jaw} &
        \small \textbf{Left Quad} &
        \small \textbf{Right Quad} &
        \small \textbf{Anterior} \\[4pt]
        \regimg{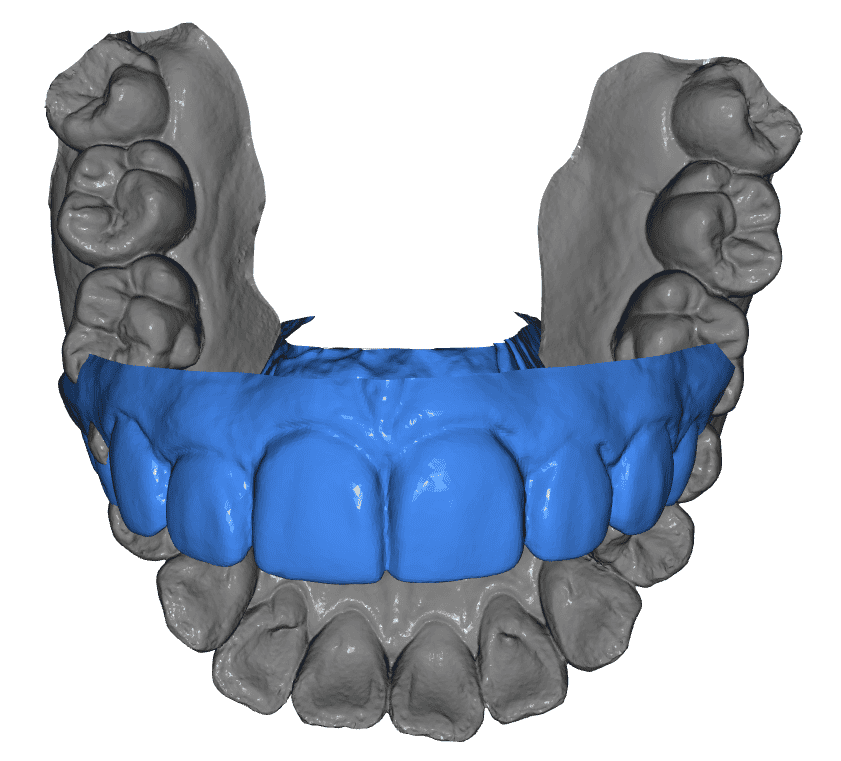} &
        \regimg{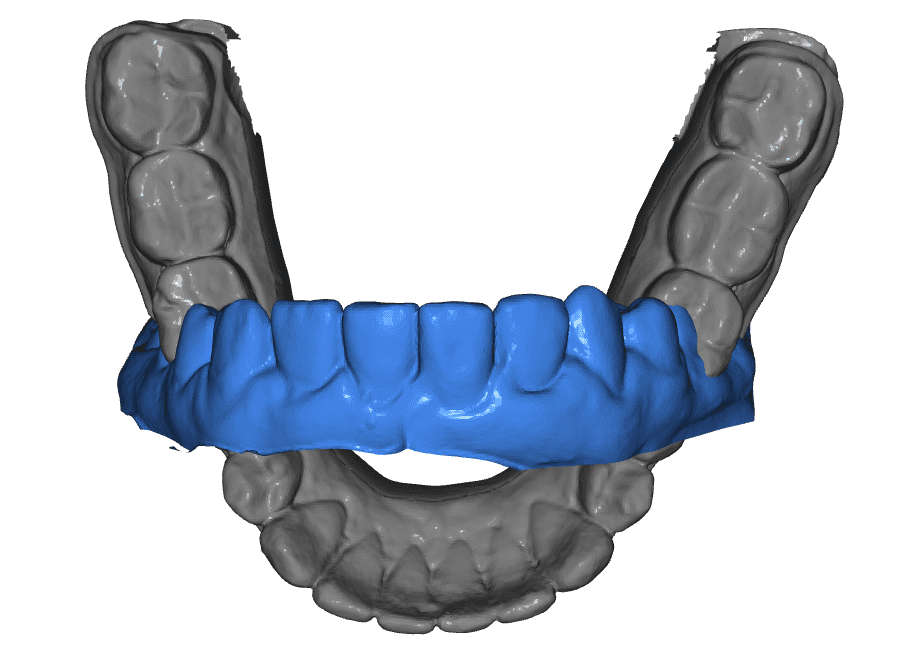} &
        \regimg{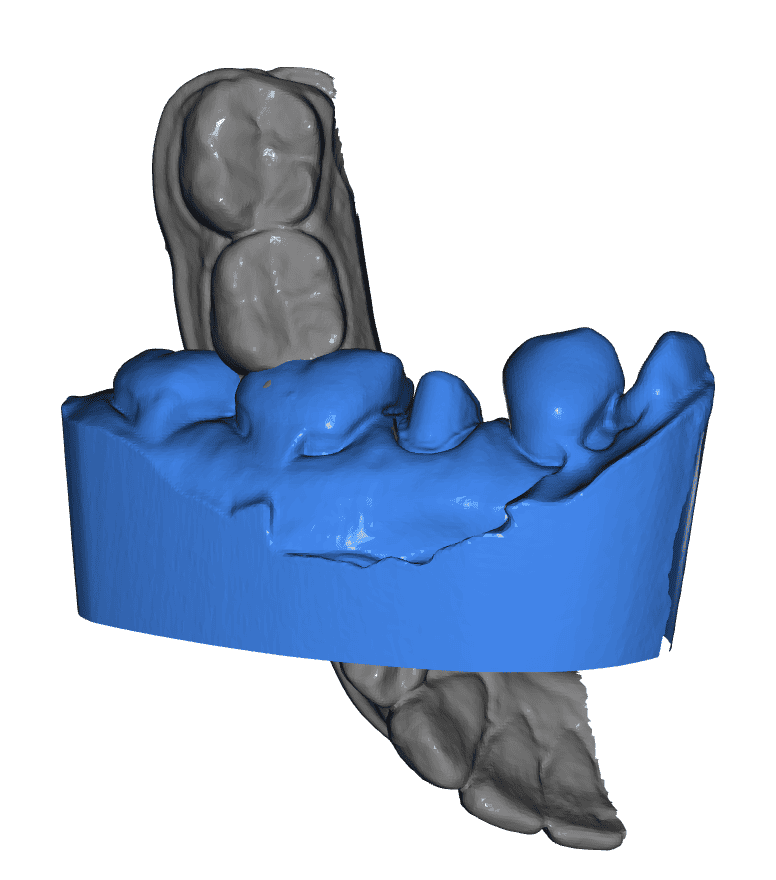} &
        \regimg{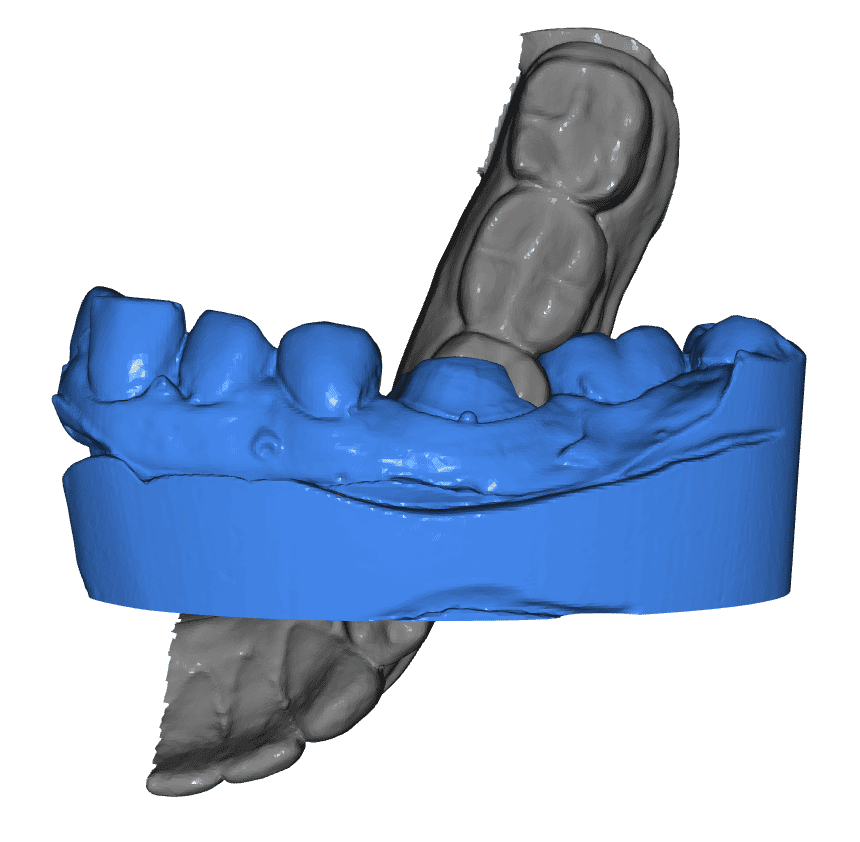} &
        \regimg{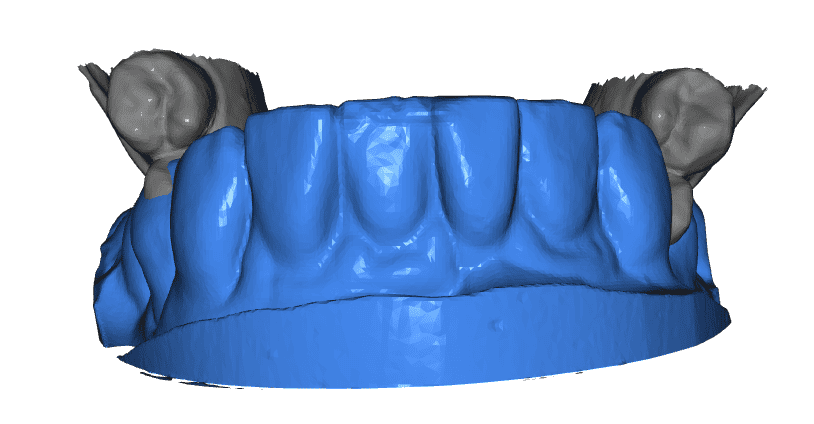} \\
        \scriptsize \textbf{Arbitrary Input} &
        \scriptsize \textbf{Arbitrary Input} &
        \scriptsize \textbf{Arbitrary Input} &
        \scriptsize \textbf{Arbitrary Input} &
        \scriptsize \textbf{Arbitrary Input} \\[6pt]
        \regimg{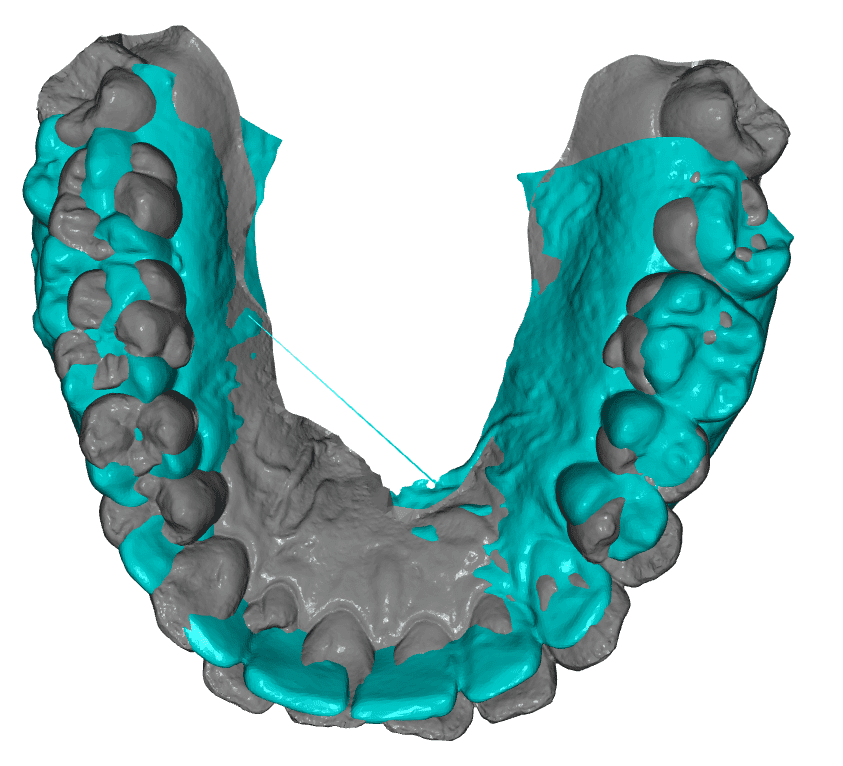} &
        \regimg{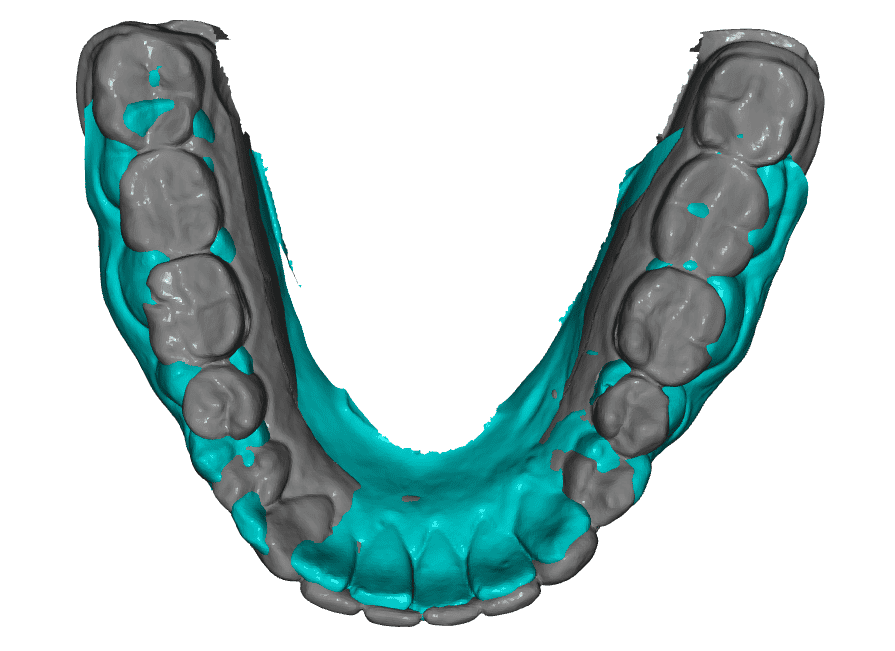} &
        \regimg{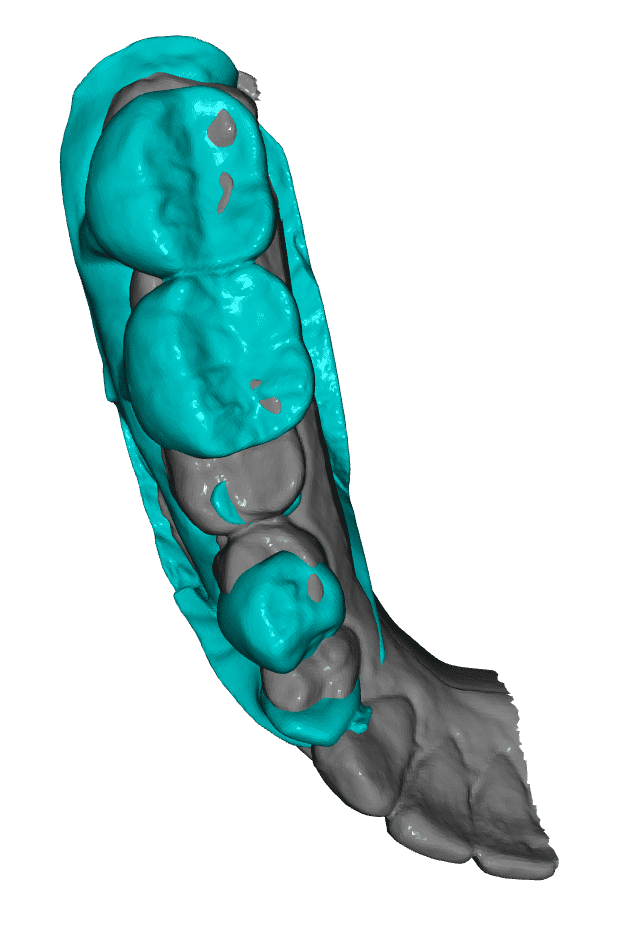} &
        \regimg{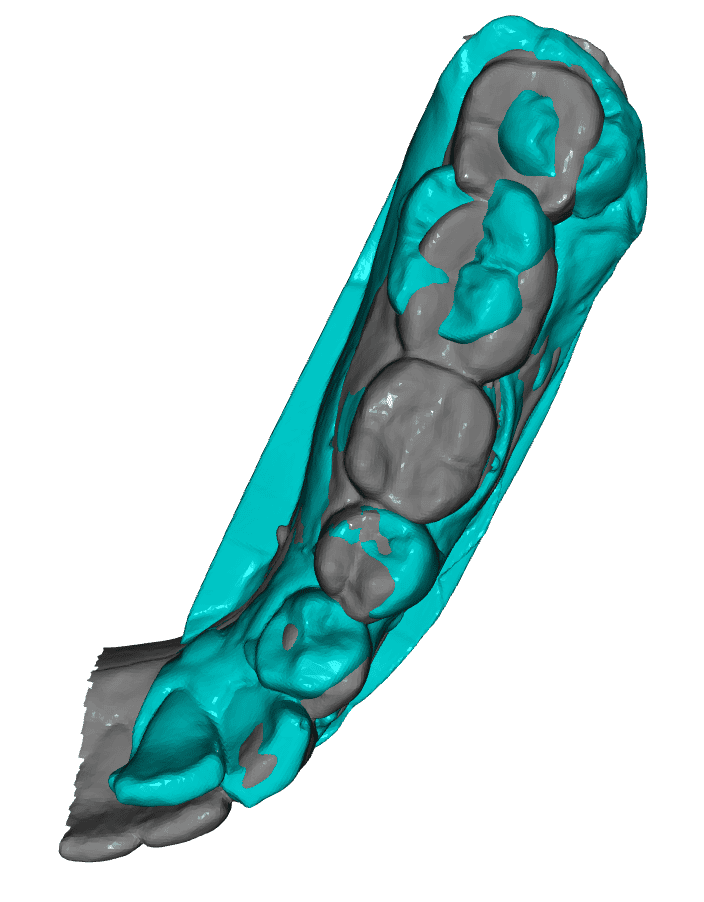} &
        \regimg{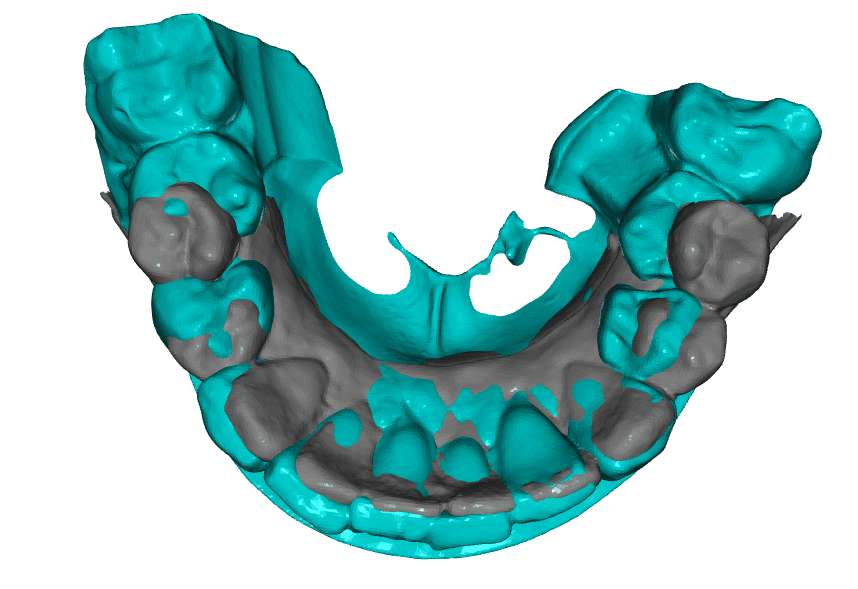} \\
        \scriptsize \textbf{Canonical Output} &
        \scriptsize \textbf{Canonical Output} &
        \scriptsize \textbf{Canonical Output} &
        \scriptsize \textbf{Canonical Output} &
        \scriptsize \textbf{Canonical Output}
    \end{tabular}
    \caption{\textbf{Qualitative registration validation on full and partial intraoral scans.}
    The top row shows arbitrarily oriented input scans overlaid with the corresponding reference template, while the bottom row shows the final canonical alignment produced by the proposed \textbf{Coarse-to-Fine} registration pipeline. Full-arch examples (upper and lower) and partial-arch examples (left, right, and anterior) are presented together in a single horizontal layout for direct comparison.}
    \label{fig:registration_examples}
\end{figure*}

In terms of computational performance, the alignment process averaged \textbf{5.01 seconds} for full arches, whereas partial arches required \textbf{7.49 seconds} due to the additional computational cost of registering against two competing templates. Qualitative examples of the registration results on both full and partial scans are shown in Figure~\ref{fig:registration_examples}.
\subsection{Phase II: Segmentation}
\label{ssec:phase2}
\subsubsection{Segmentation of Full and Partial Jaw Scans}
\label{sssec:methods_segmentation}
Following global pose standardization, the aligned scan is routed to one 
of two dedicated segmentation networks based on the classifier output: a 
\textbf{Full Arch Model} for complete upper and lower jaw scans, and a 
\textbf{Partial Arch Model} for left, right, and central partial scans. 
Both models are built on the DilatedToothSegNet architecture~\cite{Dilated_DGCNN_Reference}, 
which extends DGCNN~\cite{DGCNN_Reference} with dilated graph convolutions 
to capture both local and long-range geometric dependencies. Upper and lower 
tooth types sharing the same arch position are assigned a single class ID, 
effectively doubling the available training data per tooth type. Graph-cut 
boundary refinement is applied as post-processing to reduce label noise at 
tooth--gingival boundaries. Full training configuration, dataset splits, 
augmentation strategy, and quantitative results are reported in 
Section~\ref{sec:experiments}.
\subsection{Phase III: Initial Crown Proposal Generation}
\label{ssec:phase3}

\subsubsection{Context-Aware Crown Retrieval}
\label{sssec:methods_retrieval}
We employ a retrieval-based approach that selects an optimal crown morphology based on the patient's specific dental environment. This process utilizes a deep learning framework to match the patient's anatomy against crown templates.

\paragraph{Feature Embedding Construction}
We utilized a DGCNN~\cite{DGCNN_Reference} classifier trained on the open-source \textbf{Teeth3DS} dataset~\cite{Teeth3DS_Reference}. From this dataset, we processed over \textbf{24,000 individual teeth extracted from Teeth3DS}, training the network to classify them into 16 distinct tooth types. Once the model was trained, we removed the final classification head and utilized the network as a fixed feature extractor. We generated and stored a dense \textbf{256-dimensional feature embedding vector} for every individual tooth in the dataset, in which each tooth is represented by a unique numerical signature describing its shape and features.

\paragraph{Context Matching and Retrieval}
When a patient scan is segmented, we do not simply retrieve a generic template. Instead, we extract the patient's specific \textbf{context environment}, defined as the set of adjacent neighboring teeth and the opposing antagonist teeth. The retrieval pipeline proceeds as follows:
\begin{enumerate}
    \item \textbf{Context Embedding:} The neighbor teeth from the patient's private scan are passed through the pre-trained DGCNN~\cite{DGCNN_Reference} to generate their respective 256-dimensional feature vectors.

    \item \textbf{Macro-Average Similarity Search:} We query the open-source Teeth3DS dataset to identify a Teeth3DS jaw that possesses a dental environment nearly identical to the patient's private input. We calculate the similarity between the patient's clinical jaw and the candidate jaws from Teeth3DS by computing the \textbf{macro-average cosine similarity} of their corresponding context vectors. The Teeth3DS jaw with the highest aggregate score is selected as the most similar jaw.

    \item \textbf{Target Crown Selection:} From this best-matching jaw, we isolate the tooth corresponding to the patient's missing position. To obtain a clean, usable geometry for the final design, this extracted donor tooth is passed through the DGCNN~\cite{DGCNN_Reference} one final time to generate a 256-dimensional feature vector. This vector is compared against similar embedded feature vectors of the \textbf{Crown Library Database}, and the standard dental crown template with the highest cosine similarity is retrieved.
\end{enumerate}

\subsubsection{Crown-to-Preparation Alignment}
\label{sssec:methods_crown_align}
After segmentation identifies the prepared tooth and a suitable library crown is retrieved, the crown is placed using a multi-stage alignment strategy. This process, illustrated in Algorithm~\ref{alg:crown_alignment}, is divided into three steps: (1) Spline-Based Reference Guide, (2) Robust Target Extraction, and (3) Sequential Crown Alignment.
\begin{algorithm}[t]
\small
\caption{Patient-Specific Crown Alignment}
\label{alg:crown_alignment}
\begin{algorithmic}[1]
\Require Registered Jaw ($J$), Prep ID ($i_{prep}$), Retrieved Crown ($C_{ret}$)
\Require Crown Axes $\{\bm{n}_{c\text{-}mesial}, \bm{n}_{c\text{-}buccal}, \bm{n}_{c\text{-}occlusal}\}$
\Ensure Aligned crown ($C_{align}$)

\Statex \textbf{Step 1: Spline-Based Reference Guide}
\State $\mathcal{C}_{all} \gets \text{GetCentroids}(J)$
\State $\mathcal{A}(t) \gets \text{FitSpline}(\mathcal{C}_{all})$ \Comment{Fit dental arch curve}
\State $\bm{C}_{prep} \gets \text{GetCentroid}(J, i_{prep})$
\State $\bm{v}_{m\text{-}ref} \gets \text{Tangent}(\mathcal{A}, \bm{C}_{prep})$ \Comment{Reference mesial direction}
\State $\bm{v}_{b\text{-}ref} \gets \text{Normal2D}(\mathcal{A}, \bm{C}_{prep})$ \Comment{Reference buccal direction}

\Statex \textbf{Step 2: Robust Target Extraction}
\State $\mathcal{N} \gets \text{GetAllNormals}(J)$ \Comment{Full set of jaw vertex normals}
\State $\mathcal{N}_{prep} \gets \{\bm{n}_j \in \mathcal{N} \mid \text{label}(j) = i_{prep}\}$ \Comment{Prepared tooth normals only}
\State $\bm{v}_{m\text{-}robust} \gets \text{Mean}(\{\bm{n}_j \in \mathcal{N}_{prep} \mid \bm{n}_j \cdot \bm{v}_{m\text{-}ref} > \tau\})$
\State $\bm{v}_{b\text{-}robust} \gets \text{Mean}(\{\bm{n}_j \in \mathcal{N}_{prep} \mid \bm{n}_j \cdot \bm{v}_{b\text{-}ref} > \tau\})$

\Statex \textbf{Step 3: Sequential Crown Alignment}
\State $\bm{v}_{global\text{-}Z} \gets [0, 0, 1]$ \Comment{Standardized occlusal axis}
\State $C_{align} \gets \text{Translate}(C_{ret}, \bm{C}_{prep})$ \Comment{Match geometric centers}
\State $C_{align} \gets \text{Rotate}(C_{align}, \bm{n}_{c\text{-}mesial}, \bm{v}_{m\text{-}robust})$ \Comment{Align mesial-distal axis}
\State $C_{align} \gets \text{RotateAbout}(C_{align}, \bm{n}_{c\text{-}mesial}, \bm{n}_{c\text{-}buccal}, \bm{v}_{b\text{-}robust})$ \Comment{Align buccal axis (mesial-distal constrained)}
\State $C_{align} \gets \text{Rotate}(C_{align}, \bm{n}_{c\text{-}occlusal}, \bm{v}_{global\text{-}Z})$ \Comment{Align occlusal to global Z}
\State \Return $C_{align}$
\end{algorithmic}
\end{algorithm}

\begin{figure}[t]
    \centering
    \includegraphics[width=\columnwidth]{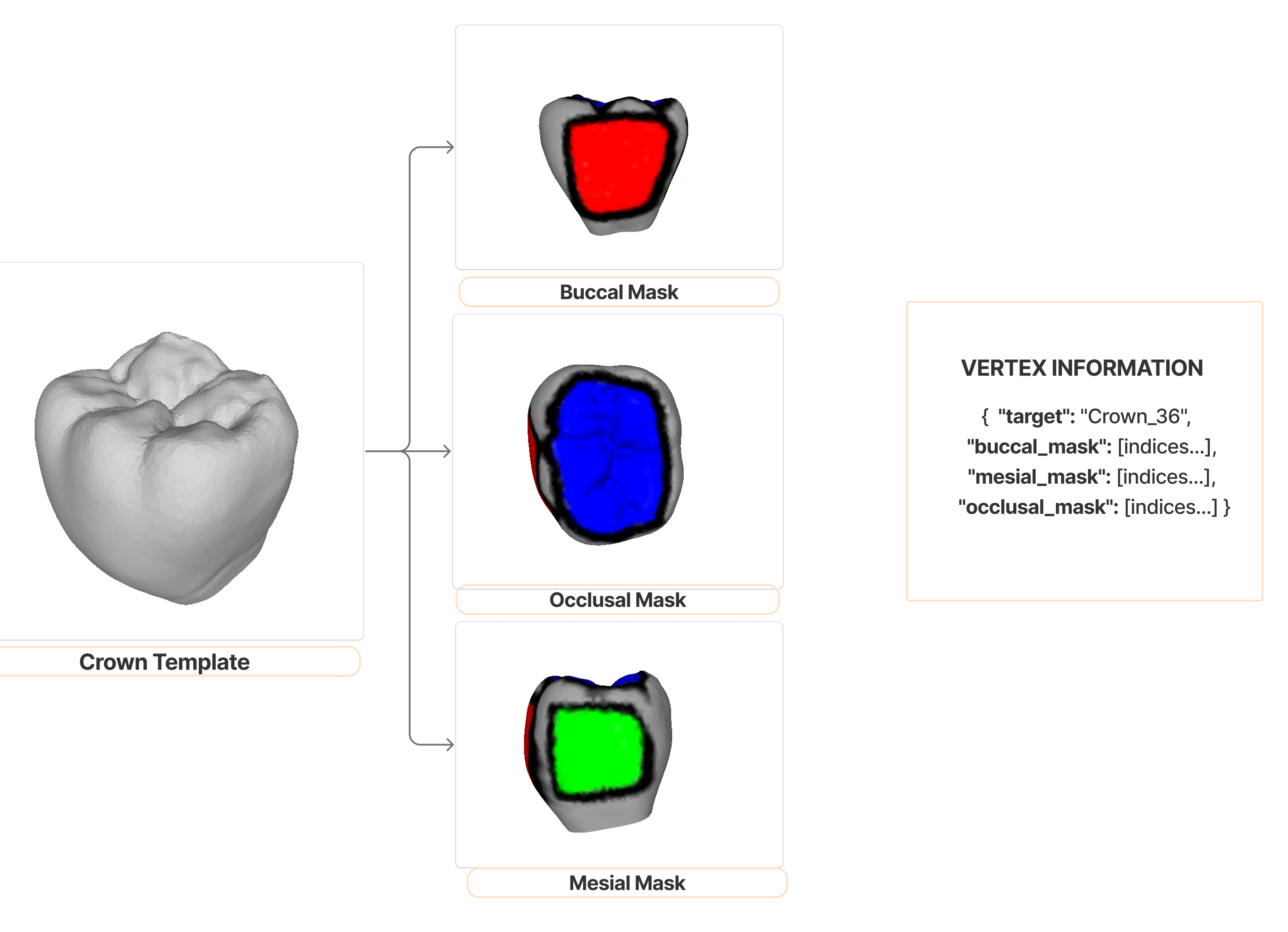}
    \caption{\textbf{Library Crown Pre-annotation.} Color-coded regions define the mesial (green), buccal (red), and occlusal (blue) faces.}
    \label{fig:crown_faces}
\end{figure}

\noindent\textbf{Library Crown Face Annotation.}
Before alignment can be applied to any new patient scan, each crown template in our library is pre-processed once to define its local orientation. As shown in Figure~\ref{fig:crown_faces}, we manually annotate three semantic regions on the mesh: the \textbf{mesial}, \textbf{buccal}, and \textbf{occlusal} faces. The mean surface normals of these annotated regions become the source vectors $\bm{n}_{c-mesial}, \bm{n}_{c-buccal}, \bm{n}_{c-occlusal}$ used in Step 3 of the algorithm.

\paragraph{Step 1: Spline-Based Reference Guide}
To align the crown within the standardized jaw coordinate system, we must first understand the unique layout of the patient's dental arch, and the following sequential steps were taken.
\begin{enumerate}[label=\arabic*.]
    \item We calculate the 3D center point (centroid), $\bm{C}_i$, for every tooth $i$ segmented in the standardized scan.
    \item A smooth 2D cubic spline, $\mathcal{A}$, is fitted through the $(x, y)$ coordinates of these centroids.
    \item From this dental curve, we focus on the location of the preparation site, $\bm{C}_{prep}$. At this point on the curve, we calculate two initial direction vectors within the $XY$ plane:
    \begin{itemize}
        \item \textbf{Reference Mesial Vector ($\bm{v}_{m-ref}$):} The normalized tangent to the curve, pointing ``forward'' (distal-to-mesial) along the arch.
        \item \textbf{Reference Buccal Vector ($\bm{v}_{b-ref}$):} The normalized 2D normal to the curve, pointing ``outward'' toward the cheek.
    \end{itemize}
\end{enumerate}

\paragraph{Step 2: Robust Target Extraction}
This spline is only an approximation. To more closely match the true 3D 
geometry of the prepared tooth surface, we use these 2D vectors as a guide 
to find the robust 3D orientation of the preparation surface itself:
\begin{enumerate}[label=\arabic*.]
    \item Let $\mathcal{N}$ denote the full set of vertex normals of the 
    standardized scan. We extract the subset corresponding to the 
    \textbf{prepared tooth only}: 
    $\mathcal{N}_{prep} = \{\bm{n}_j \in \mathcal{N} \mid \text{label}(j) = i_{prep}\}$.

    \item A \textbf{Robust Mesial Target ($\bm{v}_{m\text{-}robust}$)} is 
    found by averaging the normals in $\mathcal{N}_{prep}$ whose dot product 
    with $\bm{v}_{m\text{-}ref}$ exceeds $\tau$.

    \item A \textbf{Robust Buccal Target ($\bm{v}_{b\text{-}robust}$)} is 
    found similarly, using $\bm{v}_{b\text{-}ref}$ as the reference direction.
\end{enumerate}
The threshold $\tau = 0.6$, corresponding to an angle of 
$\approx \SI{53.1}{\degree}$, was determined empirically to provide a 
reliable subset of normals, effectively filtering out perpendicular or 
adjacent surfaces. This process yields more robust 3D vectors representing 
the actual orientation of the preparation, creating a stable target for 
alignment.

\begin{figure*}[t]
    \centering
    \includegraphics[width=\textwidth]{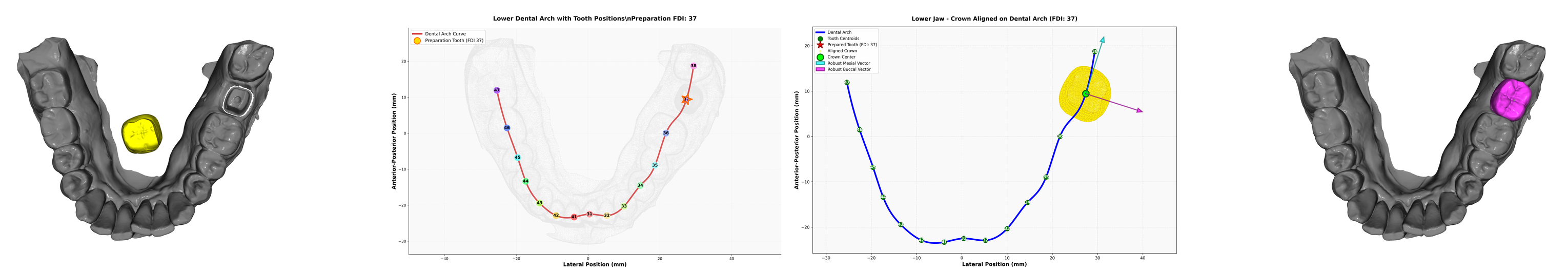}
    \caption{\textbf{Visual overview of the Patient-Specific Crown Alignment pipeline.}
    The figure illustrates the process in three stages:
    \textbf{(a) Initial Misalignment:} The retrieved library crown (yellow) is initially positioned far from the patient scan (Grey).
    \textbf{(b) Dental Arc:} a cubic spline is fitted to the dental arch tooth centroids. At the preparation site, robust \textit{Mesial} and \textit{Buccal} target vectors are extracted.
    \textbf{(c) Final Registration:} The crown has been translated and sequentially rotated to align its normal axes with these target vectors, resulting in an aligned crown (Pink).}
    \label{fig:crown_alignment_process}
\end{figure*}

\paragraph{Step 3: Sequential Crown Alignment}
With the robust target vectors established, the \textbf{Crown Library Database} (which has pre-defined mesial, buccal, and occlusal surfaces) is aligned in a specific, hierarchical order:
\begin{enumerate}[label=\arabic*.]
    \item \textbf{Translation:} The crown's geometric centroid is aligned with the preparation's centroid, $\bm{C}_{prep}$.
    \item \textbf{Mesial Alignment:} A rotation is applied to align the crown's mean mesial normal, $\bm{n}_{c-mesial}$, with the robust mesial target, $\bm{v}_{m-robust}$. This locks in the crown's front-to-back rotation.
    \item \textbf{Buccal Alignment:} A constrained rotation is performed only around the new mesial axis ($\bm{v}_{m-robust}$) to align the crown's mean buccal normal, $\bm{n}_{c-buccal}$, with the robust buccal target, $\bm{v}_{b-robust}$. This orients the crown's side-to-side rotation without disturbing the mesial alignment. Figure~\ref{fig:robust_alignment} illustrates the extracted 
mesial and buccal vertices before and after alignment, along with the 
resulting robust target vectors.
\begin{figure}[t]
    \centering
    \setlength{\tabcolsep}{2pt}
    \renewcommand{\arraystretch}{1.0}
    \begin{tabular}{cc}
        \includegraphics[width=0.46\columnwidth,height=3.2cm,keepaspectratio]{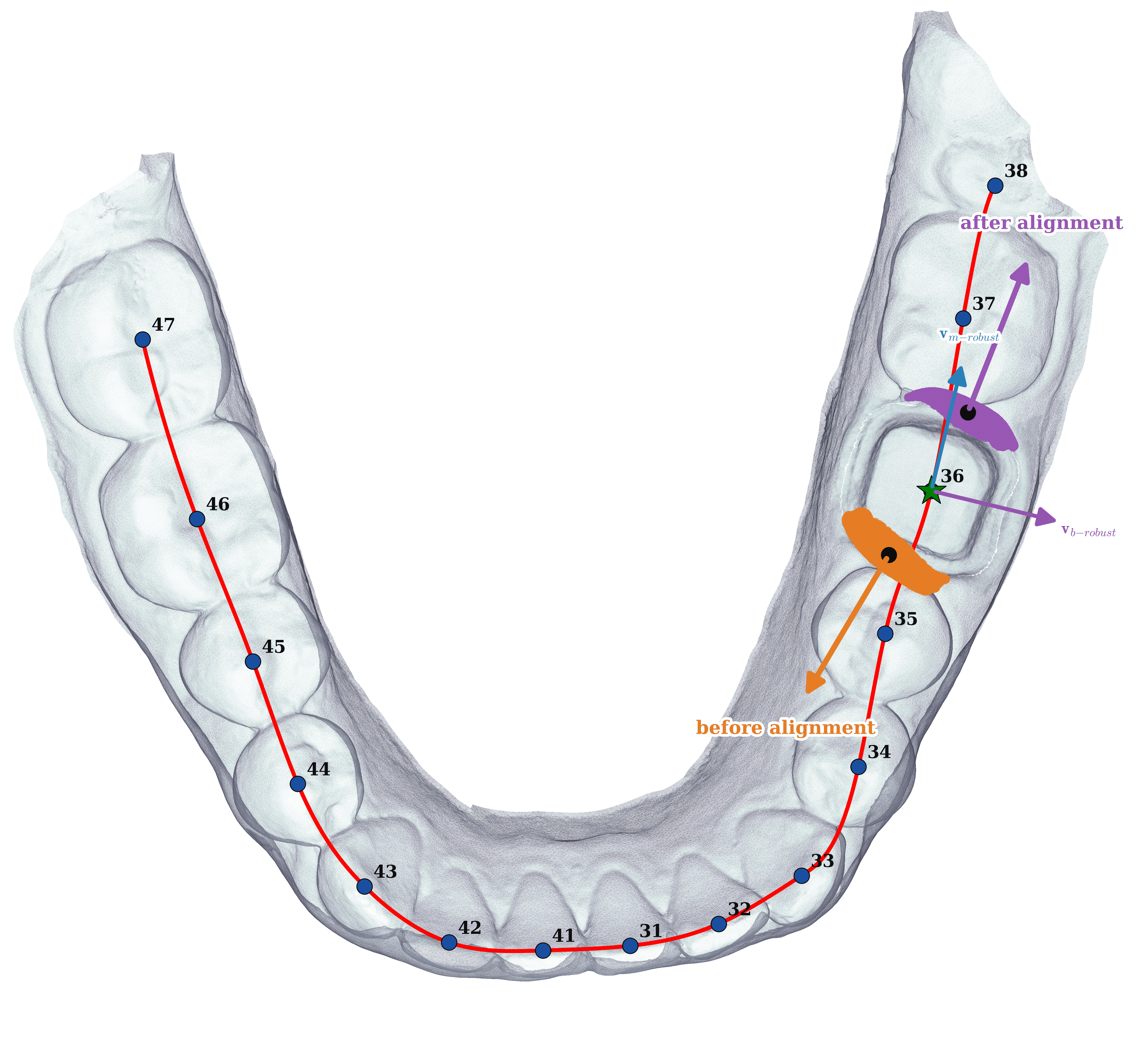} &
        \includegraphics[width=0.46\columnwidth,height=3.2cm,keepaspectratio]{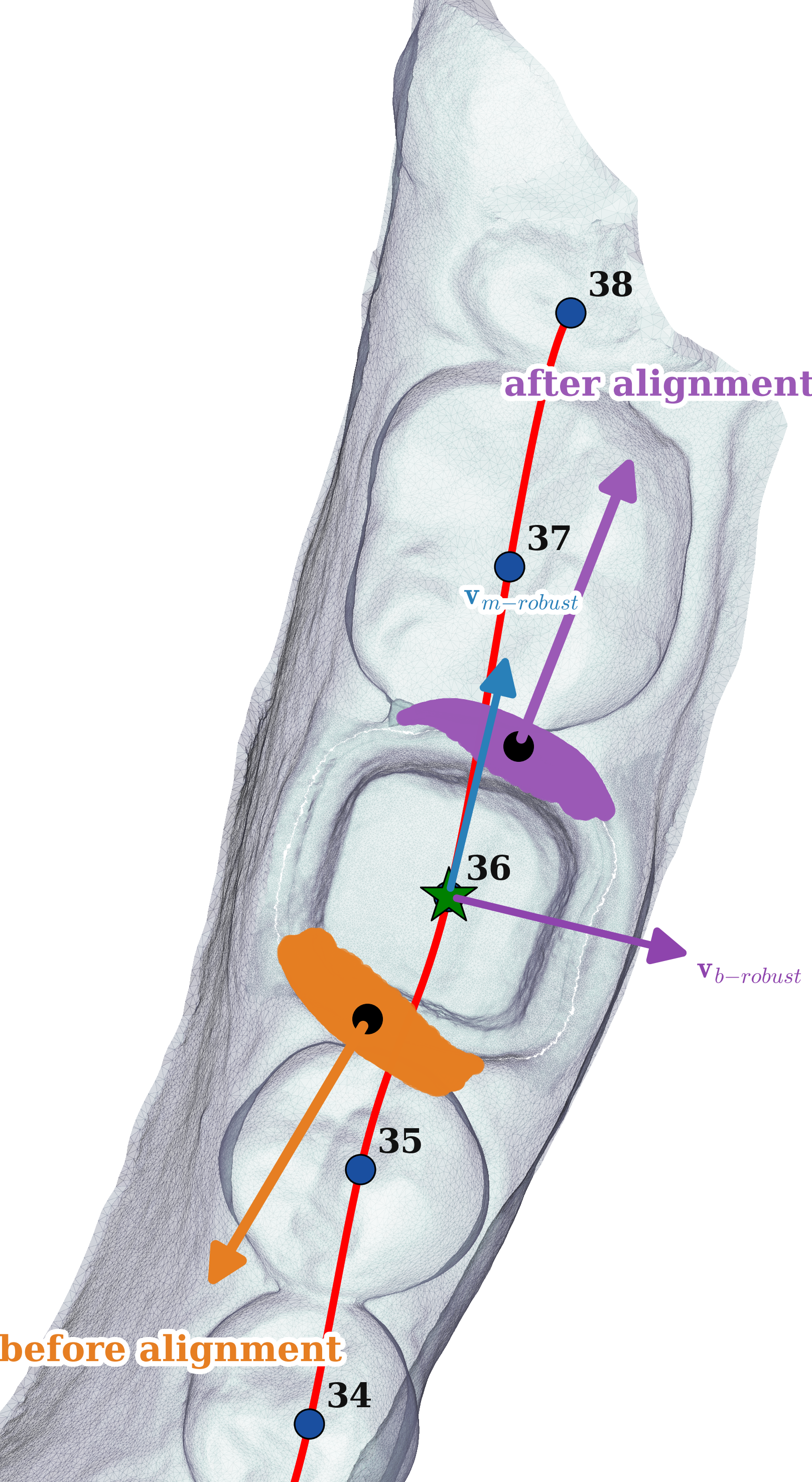} \\
        \multicolumn{2}{c}{\small \textbf{(a) Mesial alignment}} \\[4pt]
        \includegraphics[width=0.46\columnwidth,height=3.2cm,keepaspectratio]{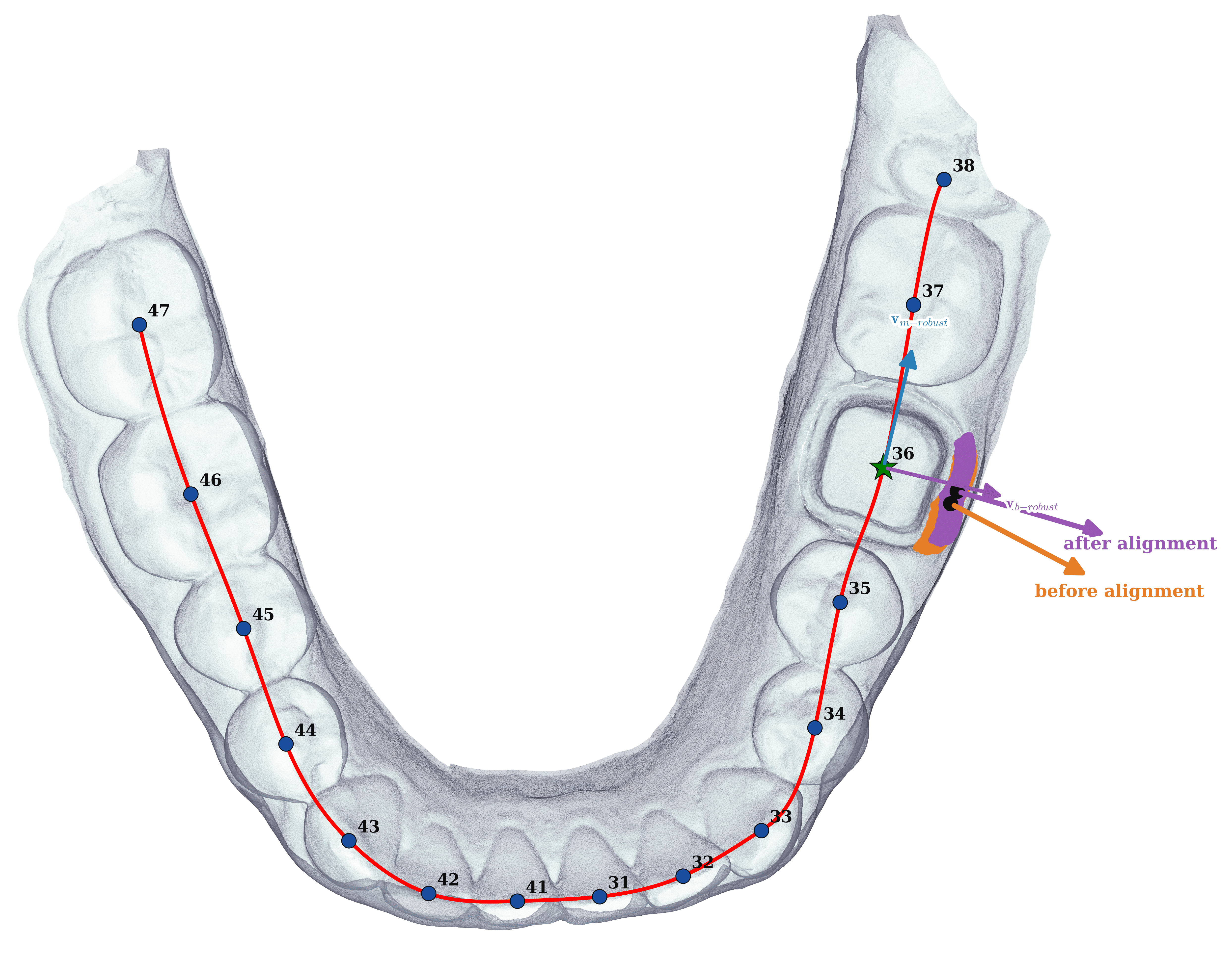} &
        \includegraphics[width=0.46\columnwidth,height=3.2cm,keepaspectratio]{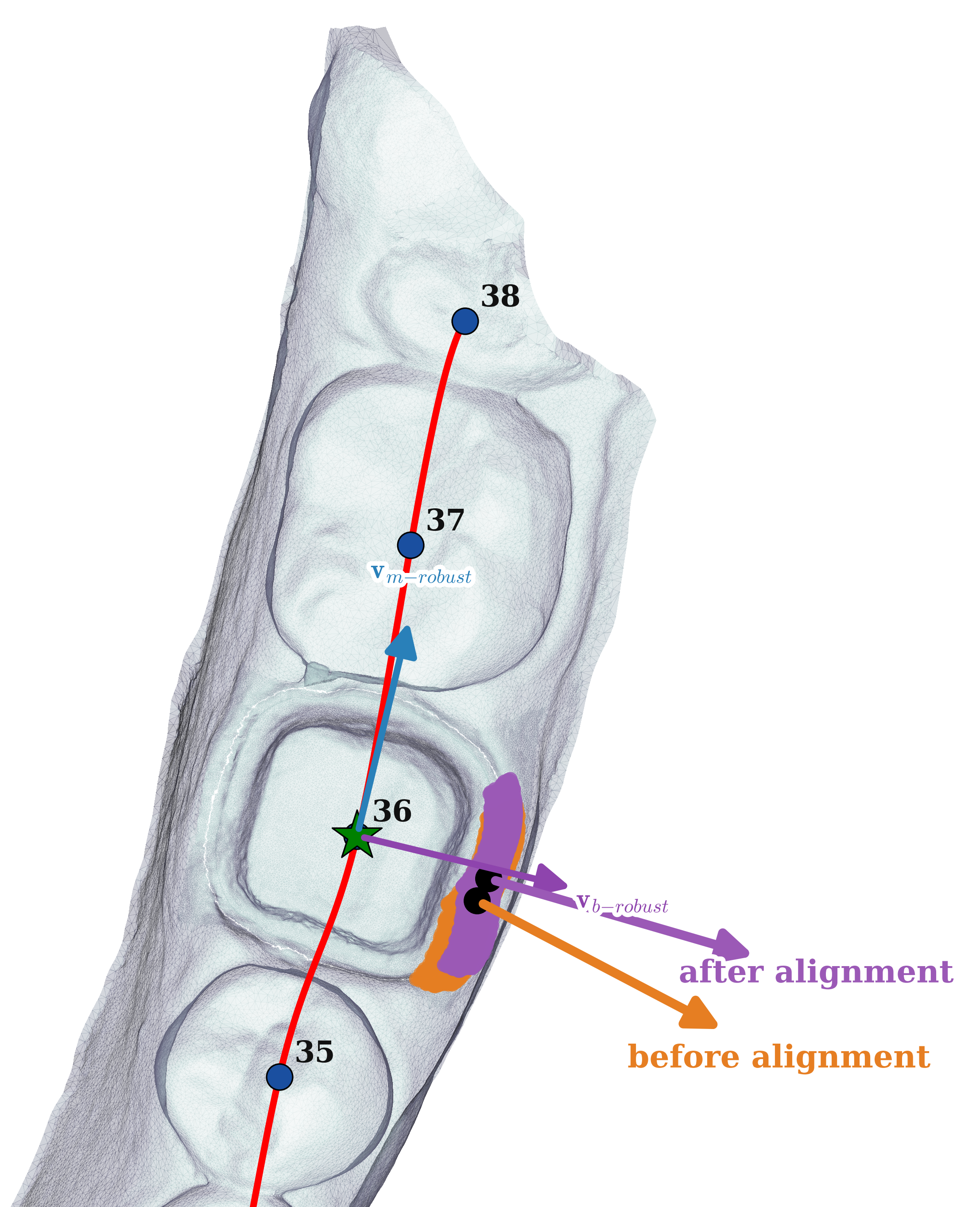} \\
        \multicolumn{2}{c}{\small \textbf{(b) Buccal alignment}} \\
    \end{tabular}
    \caption{\textbf{Robust target extraction at the preparation site.}
    \textbf{(a) Mesial alignment:} (left) full view showing orange and purple scatter of mesial vertices before and after alignment; (right) zoomed detail of the recovered target vector $\bm{v}_{m\text{-}robust}$ at the prepared tooth.
    \textbf{(b) Buccal alignment:} similarly showing the recovered buccal target $\bm{v}_{b\text{-}robust}$.}
    \label{fig:robust_alignment}
\end{figure}
    \item \textbf{Occlusal Alignment:} A final rotation aligns the crown's mean occlusal normal, $\bm{n}_{c-occlusal}$, with the global Z-axis $\bm{v}_{global-Z} = [0,0,1]^T$, ensuring it sits upright relative to the standardized occlusal plane.
\end{enumerate}
  Figure~\ref{fig:crown_alignment_process} provides a visual overview of this sequential alignment pipeline, showing the initial misalignment, the extracted dental arc with robust target vectors, and the final registered crown.

\subsubsection{Crown Adjustment and Scaling using the Blender API}
\label{sssec:methods_blender}
After the initial alignment, the crown must be adjusted to fit the patient's specific anatomy. We developed an automated workflow in the Blender Python API (bpy) for this final fitting, organized into three sequential steps as summarized in Algorithm~\ref{alg:crown_fitting}.

\begin{algorithm}[t]
\small
\caption{Crown Adjustment using the Blender Python API}
\label{alg:crown_fitting}
\begin{algorithmic}[1]
\Require Aligned crown ($C_{align}$), Neighbors ($N$), Antagonist ($J_{opp}$), FDI ($F_{id}$)
\Ensure Final crown shell for clinical review ($C_{final}$)

\State $C \gets C_{align}$ \Comment{Initialize from crown-alignment output}

\Statex \textbf{Step 1: Interproximal Adaptation}
\State $V_{int} \gets \text{GetIntersection}(C, N)$
\If{$V_{int} > 10^{-6}$} \Comment{Stage (i), Case A: initial collision}
    \While{$V_{int} > 10^{-6}$}
        \State $C \gets \text{Scale}(C, 0.99)$ \Comment{Shrink to contact}
        \State $V_{int} \gets \text{GetIntersection}(C, N)$
    \EndWhile
\Else \Comment{Stage (i), Case B: gap exists}
    \While{$V_{int} \le 10^{-6}$}
        \State $C \gets \text{Scale}(C, 1.01)$ \Comment{Grow to contact}
        \State $V_{int} \gets \text{GetIntersection}(C, N)$
    \EndWhile
\EndIf
\State $C \gets \text{Scale}(C, 0.99)$ \Comment{Stage (ii): apply functional gap}

\Statex \textbf{Step 2: Geometric Centering}
\State $P_{mid} \gets \text{Midpoint}(N)$
\State $C \gets \text{Align}(C, P_{mid})$ \Comment{Match crown center to neighbor midpoint}
\Statex \textbf{Step 3: Occlusal Correction}
\If{\text{IsPosterior}($F_{id}$)} \Comment{Mode A: molar / premolar}
    \State $S_{cusps} \gets \text{FindCusps}(C)$
    \State $K_{coll} \gets \text{BooleanIntersect}(S_{cusps}, J_{opp})$ \Comment{Direct collision check}
    \If{$K_{coll} = \emptyset$}
        \State $K_{coll} \gets \text{KDTreeQuery}(S_{cusps}, J_{opp})$ \Comment{Near-collision check}
    \EndIf
    \ForAll{$p \in K_{coll}$} \Comment{Each colliding cusp vertex}
        \State $p \gets \text{Deform}(p, -\delta)$ \Comment{Local tap-down}
    \EndFor
\Else \Comment{Mode B: anterior}
    \While{\text{HasAnyOcclusalInterference}($C, J_{opp}$)}
        \State $C \gets \text{Trans}(C, -\vec{z} \cdot \delta)$ \Comment{Global shift}
    \EndWhile
\EndIf
\State $C_{final} \gets C$ \Comment{Finalize geometry}
\State \Return $C_{final}$
\end{algorithmic}
\end{algorithm}
Let $C$ denote the working crown mesh, initialized from the aligned crown $C_{align}$. 
\paragraph{Step 1: Interproximal Fit Adaptation}
The first step ensures the crown fits precisely between its neighboring teeth $N$ using a two-stage iterative scaling procedure, guided by a volumetric intersection threshold of $V_{int}=10^{-6}$.

\noindent\textit{\textbf{Stage (i) --- Touch-Point Detection.}}
The algorithm first locates the scale at which the crown makes initial contact with its neighbors. If the initial intersection volume exceeds the threshold, the crown is iteratively scaled down by a factor of $0.99$ about its geometric center until contact disappears. If no initial collision exists, the crown is iteratively scaled up by $1.01$ until contact is detected. This locates the \emph{just-touching} scale.

\noindent\textit{\textbf{Stage (ii) --- Functional Gap Creation.}}
Once the touch-point scale is identified, a final $0.99\times$ shrink is applied about the geometric center, producing a minimal and consistent functional gap between the crown and its neighbors.

\paragraph{Step 2: Geometric Centering}
After interproximal adaptation, the crown is translated so that its geometric center aligns with the midpoint $P_{mid}$ of the two adjacent teeth in the occlusal direction.

\paragraph{Step 3: Occlusal Fit Correction}
The final step prevents interference with the opposing jaw $J_{opp}$ during occlusion. The algorithm applies one of two modes depending on the FDI tooth type $F_{id}$.

\begin{figure}[htbp]
    \centering
    \includegraphics[width=0.6\columnwidth]{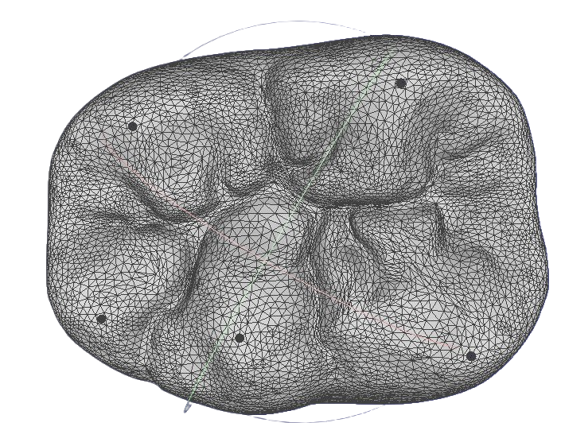}
    \caption{Identification of the top five cusps used for local ``tap-down'' adjustment to prevent occlusal interference.}
    \label{fig:cusp_detection}
\end{figure}

\noindent\textit{\textbf{Mode A --- Posterior teeth (molars and premolars).}}
For posterior teeth, the algorithm preserves the complex chewing surface by adjusting only the cusps rather than flattening the entire occlusal anatomy. As illustrated in Fig.~\ref{fig:cusp_detection}, cusps are identified by comparing each vertex on the tooth mesh to its connected neighbors: a vertex is classified as a cusp only if it lies higher than all of its surrounding vertices and its surface normal is oriented toward the opposing jaw. The highest cusps form the adjustment landmark set $S_{cusps}$. The algorithm then checks for direct collisions using a Boolean \texttt{INTERSECT} operation; if no direct collision is found, a KD-Tree query identifies cusps in close proximity to the opposing jaw, yielding the colliding or near-colliding subset $K_{coll}$. Only those cusps are ``tapped down'' locally by a small offset $\delta$, leaving the remainder of the occlusal anatomy intact.

\noindent\textit{\textbf{Mode B --- Anterior teeth (incisors and canines).}} 
For anterior teeth, which have a simpler chewing geometry, the algorithm bypasses cusp analysis and applies an iterative global shift along $-\vec{z}$ by offset $\delta$ until no collision with the opposing jaw remains.

This two-mode system ensures that all collisions are removed in a way that is appropriate for the tooth's specific shape and function. This adjusted mesh serves as a high-quality initial proposal, awaiting the clinician's final margin-line cutting and manual refinement. Executed on a single GPU, the complete computational journey from the raw intraoral scan input to this preliminary crown shell takes between 2.5 and 3.5 minutes, establishing an efficient CAD/CAM workflow.

\section{Experiments and Results}
\label{sec:experiments}

\subsection{Segmentation Experimental Setup}
\label{ssec:exp_setup}

\subsubsection{Training Configuration}
\label{sssec:exp_training}
All experiments were conducted on a single NVIDIA RTX 5080 GPU using PyTorch Lightning. The segmentation network is based on the DilatedToothSegNet~\cite{Dilated_DGCNN_Reference}. Each mesh face is represented by a 24-dimensional feature vector encoding the three vertex coordinates, face centroid, three vertex normals, and face normal; all normalized per scan. Graph-cut boundary refinement is applied as post-processing to reduce the label noise prediction, inspired by the refinement strategy applied in MeshSegNet~\cite{Lian2020MeshSegNet}. The model was optimized using AdamW with a learning rate of $1 \times 10^{-3}$ and weight decay of $5 \times 10^{-4}$. A Cosine Annealing learning rate scheduler was applied over 150 epochs with a minimum learning rate of $1 \times 10^{-6}$. A hybrid loss function combining Cross-Entropy and Dice loss with equal weighting (0.5 each) was employed to address class imbalance and improve boundary accuracy.

\subsubsection{Dataset Splits}
\label{sssec:exp_splits}
The complete dataset comprises 1,958 intraoral scans, and splitting was enforced at the patient level to prevent data leakage, ensuring all three partial views (central, left, and right) of each patient were assigned to a single split. This yielded 1,570 original training scans, 288 validation scans, and 100 test scans.

\subsubsection{Augmentation Strategy}
\label{sssec:exp_aug}
To address class imbalance, particularly the underrepresentation of wisdom teeth (FDI 38, FDI 48), a targeted augmentation strategy was employed exclusively on the training set. Scans containing underrepresented tooth classes received a higher number of augmented copies based on a class-frequency plan, resulting in 6,958 augmented samples and a total training set of 8,530 samples. Augmentation applied mild geometric perturbations (rotations of $\pm 5^\circ$ for X/Y and $\pm 15^\circ$ for Z, uniform scaling within $[0.9, 1.1]$, and translations of $\pm 5$mm X/Y, $\pm 2$mm Z) to preserve the standardized coordinate frame established during RANSAC+ICP registration. Validation and test sets contain only original, unaugmented scans.

\subsection{Segmentation Evaluation Metrics}
\label{ssec:exp_metrics}
Model performance was evaluated using standard segmentation and localization benchmarks. Given predicted labels $\hat{y}$ and ground truth labels $y$, for each class $c$, we define true positives $\text{TP}_c$, false positives $\text{FP}_c$, and false negatives $\text{FN}_c$.

\paragraph{Dice Similarity Coefficient (DSC)}
The primary metric for volumetric overlap:
\begin{equation}
    \mathrm{DSC}_c = \frac{2 \cdot \mathrm{TP}_c}{2 \cdot \mathrm{TP}_c + \mathrm{FP}_c + \mathrm{FN}_c}
\end{equation}

\paragraph{Precision and Recall}
Measure the correctness of positive predictions and sensitivity to true class membership, respectively:
\begin{equation}
\begin{aligned}
    \mathrm{Precision}_c &= \frac{\mathrm{TP}_c}{\mathrm{TP}_c + \mathrm{FP}_c}, \\
    \mathrm{Recall}_c    &= \frac{\mathrm{TP}_c}{\mathrm{TP}_c + \mathrm{FN}_c}.
\end{aligned}
\end{equation}

\paragraph{Centroid Error (CE)}
Used to validate the spatial reliability required for automated downstream workflows. CE is defined as the absolute 3D Euclidean distance (in millimeters) between the predicted and ground-truth centroids using world-scale coordinates:
\begin{equation}
    \text{CE}_c = \| \text{Centroid}(\hat{y}_c) - \text{Centroid}(y_c) \|_2
\end{equation}

All metrics are reported as macro-averages across all 17 semantic classes. Unless otherwise stated, all reported segmentation and localization results are computed after graph-cut optimization. Ground truth labels with fewer than 10 faces were treated as annotation artefacts and reassigned to the gingiva class prior to evaluation.

\paragraph{Statistical Reporting}
All per-scan metrics are summarized using four descriptive statistics. Let $\{m_i\}_{i=1}^{N}$ denote per-scan values of a given metric across $N$ test scans.
\begin{itemize}[leftmargin=*, itemsep=3pt, topsep=3pt]
    \item \textbf{Mean $\pm$ Std.}
    \begin{equation}
        \bar{m} = \frac{1}{N}\sum_{i=1}^{N} m_i, \qquad
        \sigma = \sqrt{\frac{1}{N-1}\sum_{i=1}^{N}(m_i-\bar{m})^2}
    \end{equation}

    \item \textbf{Median.}
    Reported as a robust central tendency measure, less sensitive to extreme penalty values than the mean.

    \item \textbf{95\% Bootstrap CI.}
    Per-scan DSC and CE distributions are non-Gaussian, bounded, skewed, and affected by extreme penalties; therefore, classical parametric intervals can be unreliable. We compute a non-parametric bootstrap confidence interval over $B = 10{,}000$ resamples with replacement:
    \begin{equation}
        \mathrm{CI}_{95} = \bigl[\hat{F}^{-1}(0.025),\;\hat{F}^{-1}(0.975)\bigr]
    \end{equation}
    where $\hat{F}$ is the empirical CDF of the bootstrap resample means $\{\bar{m}^{(b)}\}_{b=1}^{B}$, with $\bar{m}^{(b)}$ denoting the sample mean of the $b$-th resample drawn with replacement from $\{m_i\}_{i=1}^{N}$. A fixed random seed ensures reproducibility.

    \item \textbf{Miss Rate.}
    Applicable to CE only. If a tooth class is present in ground truth but predicted with zero faces in the context region, its centroid is undefined. In that case, CE is assigned the scan bounding-box diagonal distance (catastrophic localization failure penalty). Miss rate is the fraction of scans receiving this penalty.
\end{itemize}

\subsection{Segmentation Results}
\label{ssec:exp_results}

\subsubsection{Quantitative Results}
\label{sssec:exp_quant}
Using a DilatedToothSegNet architecture, the pipeline achieves an overall macro-average DSC of 0.9249, Precision of 0.9615, and Recall of 0.8919 across all 17 semantic classes.

To evaluate performance in realistic clinical scenarios, we focused on 41 test cases containing at least one prepared tooth. As reported in Table~\ref{tab:clinical_focused}, the model maintains consistently high segmentation quality in all three clinically relevant regions the prepared tooth and its immediate mesial and distal neighbors with a median DSC greater than 0.95 and a median centroid error below 0.25~mm in all cases. Crucially, zero complete segmentation failures were recorded across any context region for DilatedToothSegNet, confirming robust coverage even in geometrically challenging partial-arch scans.

The framework exhibits strong spatial localization, with median Centroid Error remaining sub-millimeter across all target regions (0.17--0.24~mm), ensuring that the local coordinate frame required for automated crown placement is anatomically grounded with minimal manual correction.

Table~\ref{tab:benchmark_comparison} compares DilatedToothSegNet against MeshSegNet under identical conditions: the same dataset, train/val/test split, augmentation strategy, and unseen test set. MeshSegNet was trained for 150 epochs following its original configuration, yet it falls considerably short across all metrics, with adjacent tooth centroid errors reaching several millimeters well beyond the sub-millimeter threshold required for reliable automated placement. This gap persists despite identical inputs and 1,958 original training scans, pointing to an architectural limitation in handling partial-arch geometry rather than a data or alignment issue.

\begin{table}[t]
\centering
\caption{Overall segmentation performance comparison.}
\label{tab:benchmark_comparison}
\renewcommand{\arraystretch}{1.25}
\setlength{\tabcolsep}{10pt}
\begin{tabular}{lccc}
\toprule
\textbf{Method} & \textbf{DSC} & \textbf{Precision} & \textbf{Recall} \\
\midrule
MeshSegNet~\cite{Lian2020MeshSegNet}       & 0.7932 & 0.7928 & 0.7944 \\
DilatedToothSegNet~\cite{Dilated_DGCNN_Reference} & \textbf{0.9249} & \textbf{0.9615} & \textbf{0.8919} \\
\bottomrule
\end{tabular}
\end{table}

\begin{table*}[t]
\centering
\caption{%
  Clinical focused metrics for 41 test cases containing one or more prepared teeth.
  Results reported as Mean\,$\pm$\,Std (Median) with 95\% bootstrap confidence
  intervals in brackets. CE\,=\,Centroid Error (mm). $^\dagger$Scans receiving
  bounding box diagonal penalty due to complete segmentation failure.
}
\label{tab:clinical_focused}
\renewcommand{\arraystretch}{1.12}
\setlength{\tabcolsep}{3pt}
\footnotesize
\begin{tabular*}{\textwidth}{@{\extracolsep{\fill}} l l c c c c}
\toprule
\textbf{Model} & \textbf{Region} & \textbf{DSC Mean\,$\pm$\,Std [95\% CI]} & \textbf{DSC Median} & \textbf{CE Mean\,$\pm$\,Std [95\% CI]} & \textbf{CE Median} \\
\midrule
\multirow{3}{*}{MeshSegNet}
& Prepared Tooth & $0.8255 \pm 0.2716\ [0.7338,\ 0.9002]$ & 0.9349 & $4.0471 \pm 14.0337^\dagger\ [0.7196,\ 8.8864]$ & 0.3692 \\
& Adjacent 1 -- Mesial & $0.9279 \pm 0.1594\ [0.8714,\ 0.9606]$ & 0.9638 & $0.6491 \pm 1.8943\ [0.2437,\ 1.3314]$ & 0.1566 \\
& Adjacent 2 -- Distal & $0.7197 \pm 0.3830\ [0.5603,\ 0.8569]$ & 0.9299 & $6.9001 \pm 17.7153^{\dagger\dagger}\ [1.3387,\ 14.6009]$ & 0.3599 \\
\midrule
\multirow{3}{*}{\textbf{DilatedToothSegNet}}
& Prepared Tooth & $\mathbf{0.9468 \pm 0.0318\ [0.9366,\ 0.9558]}$ & 0.9573 & $\mathbf{0.2666 \pm 0.2398\ [0.2015,\ 0.3464]}$ & 0.1732 \\
& Adjacent 1 -- Mesial & $\mathbf{0.9524 \pm 0.0205\ [0.9453,\ 0.9586]}$ & 0.9564 & $\mathbf{0.2702 \pm 0.1520\ [0.2239,\ 0.3213]}$ & 0.2438 \\
& Adjacent 2 -- Distal & $\mathbf{0.9569 \pm 0.0213\ [0.9481,\ 0.9644]}$ & 0.9645 & $\mathbf{0.2774 \pm 0.2098\ [0.2035,\ 0.3633]}$ & 0.1840 \\
\bottomrule
\multicolumn{6}{l}{\footnotesize $^\dagger$Miss rate: 2/41 scans (4.88\%). \quad $^{\dagger\dagger}$Miss rate: 2/25 scans (8.00\%) --- distal neighbour absent in 16 of 41 cases.}
\end{tabular*}
\end{table*}

\subsubsection{Qualitative Analysis: Challenging Cases}
\label{sssec:exp_qualitative}
\begin{figure*}[t]
    \centering
    \setlength{\tabcolsep}{3pt}
    \renewcommand{\arraystretch}{1.0}
    \begin{tabular}{r@{\hspace{6pt}}ccccc}
        & \small \textbf{Case 1}
        & \small \textbf{Case 2}
        & \small \textbf{Case 3}
        & \small \textbf{Case 4}
        & \small \textbf{Case 5} \\[4pt]
        \rotatebox{90}{\small \textbf{Ground Truth\hspace{6pt}}} &
        \includegraphics[width=0.17\textwidth]{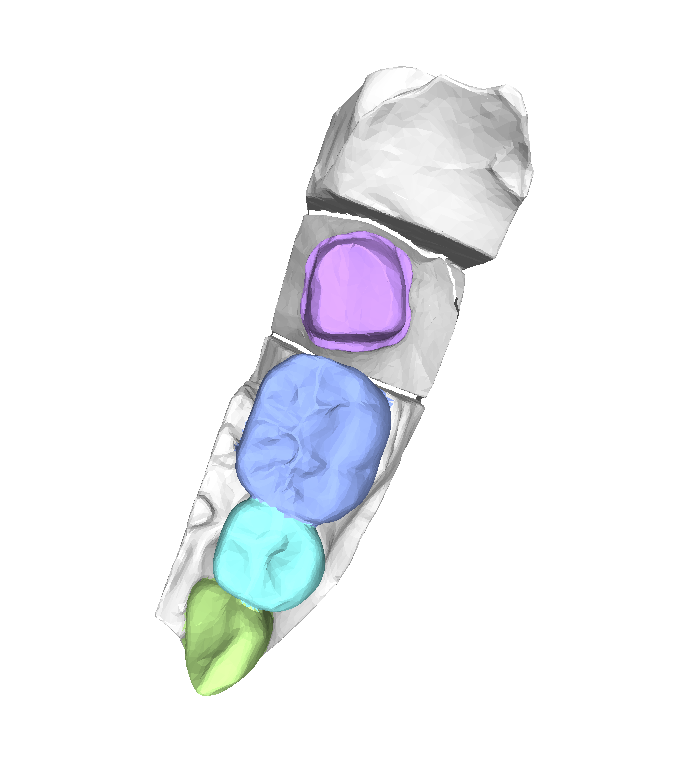} &
        \includegraphics[width=0.17\textwidth]{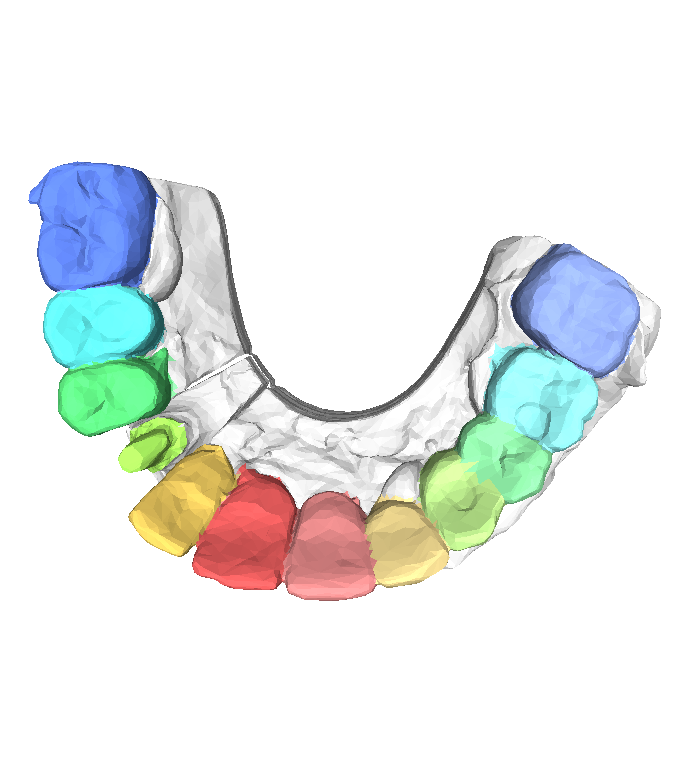} &
        \includegraphics[width=0.17\textwidth]{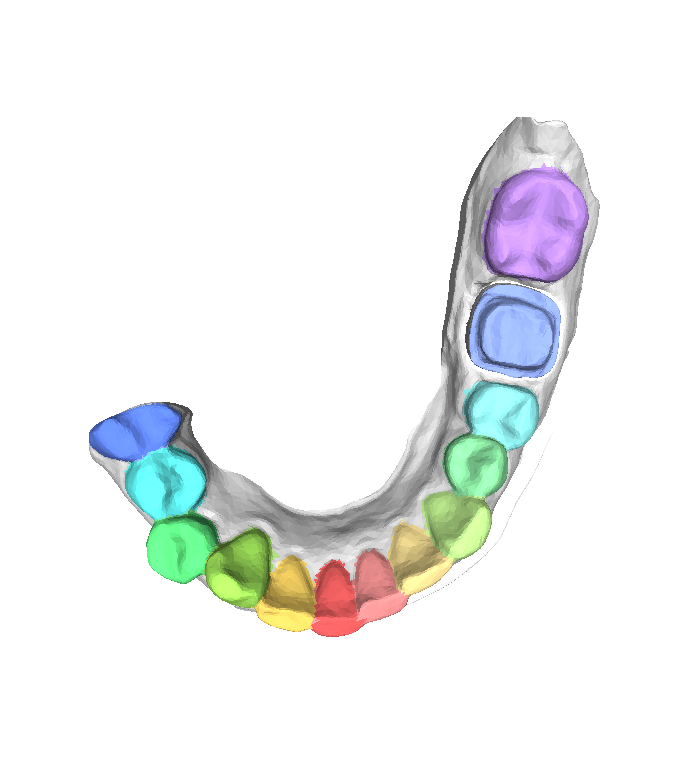} &
        \includegraphics[width=0.17\textwidth]{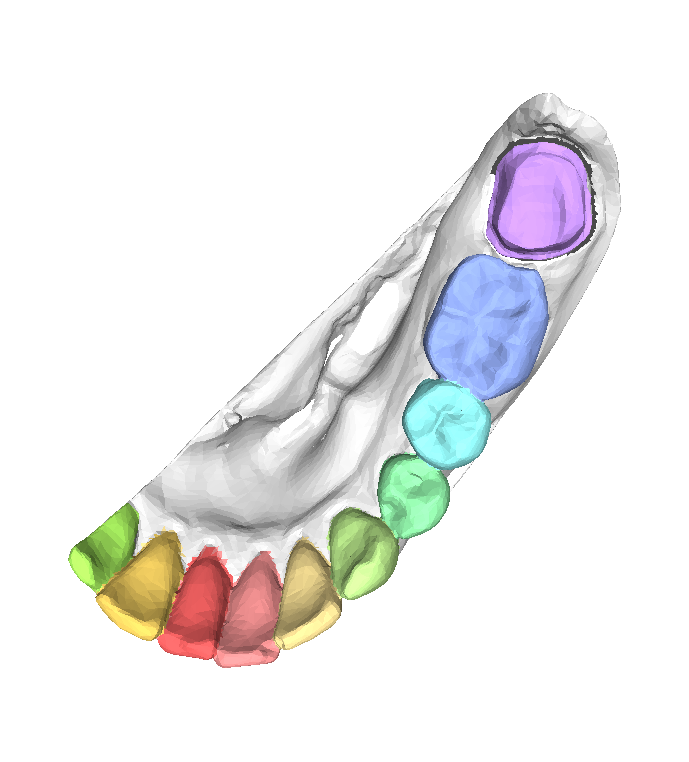} &
        \includegraphics[width=0.17\textwidth]{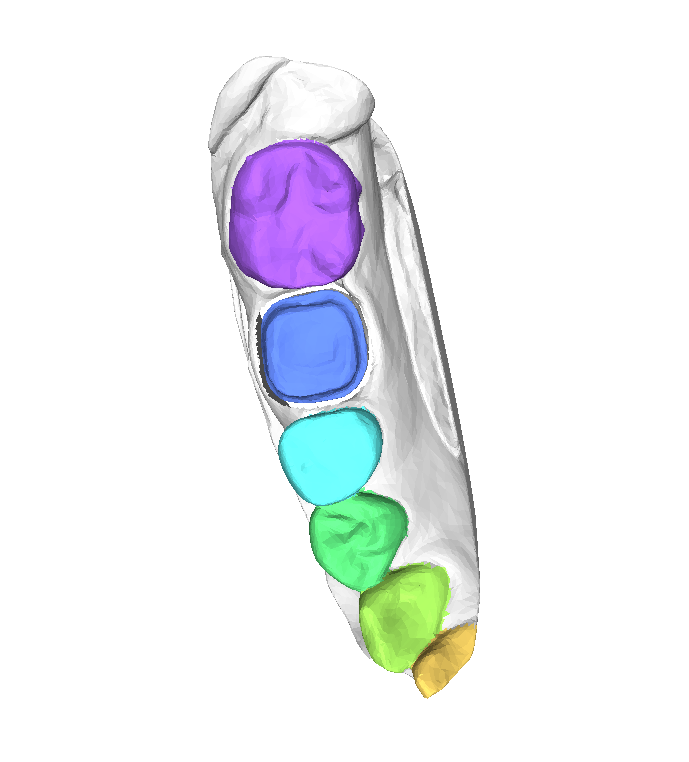} \\[6pt]
        \rotatebox{90}{\small \textbf{Prediction\hspace{6pt}}} &
        \includegraphics[width=0.17\textwidth]{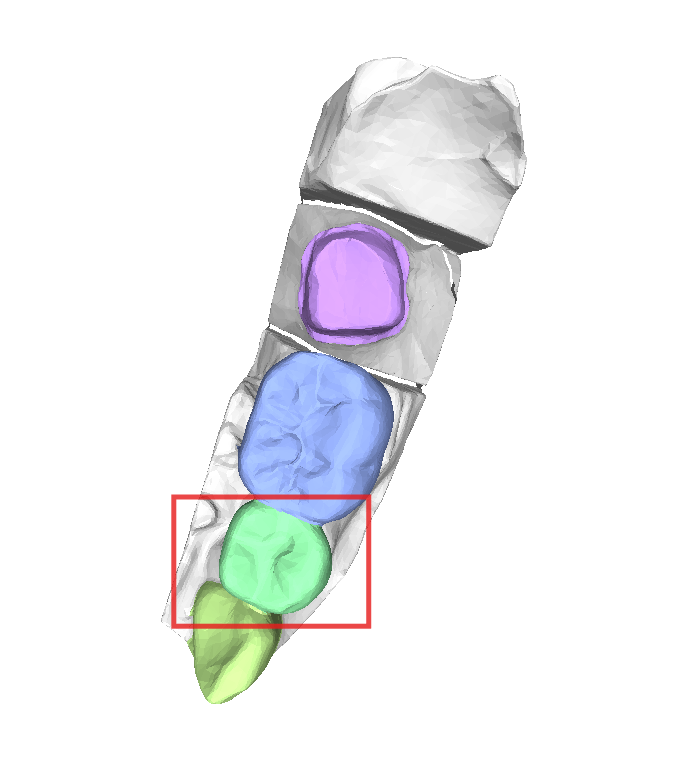} &
        \includegraphics[width=0.17\textwidth]{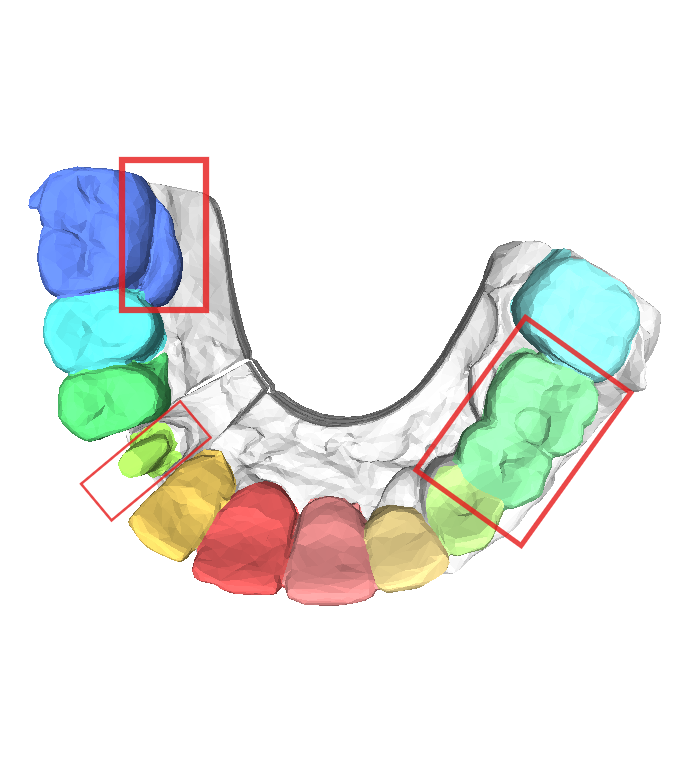} &
        \includegraphics[width=0.17\textwidth]{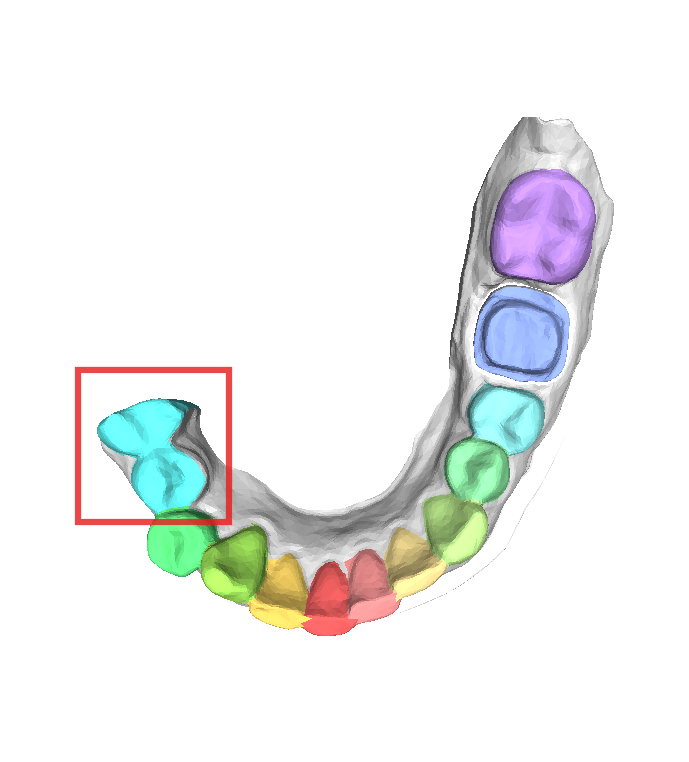} &
        \includegraphics[width=0.17\textwidth]{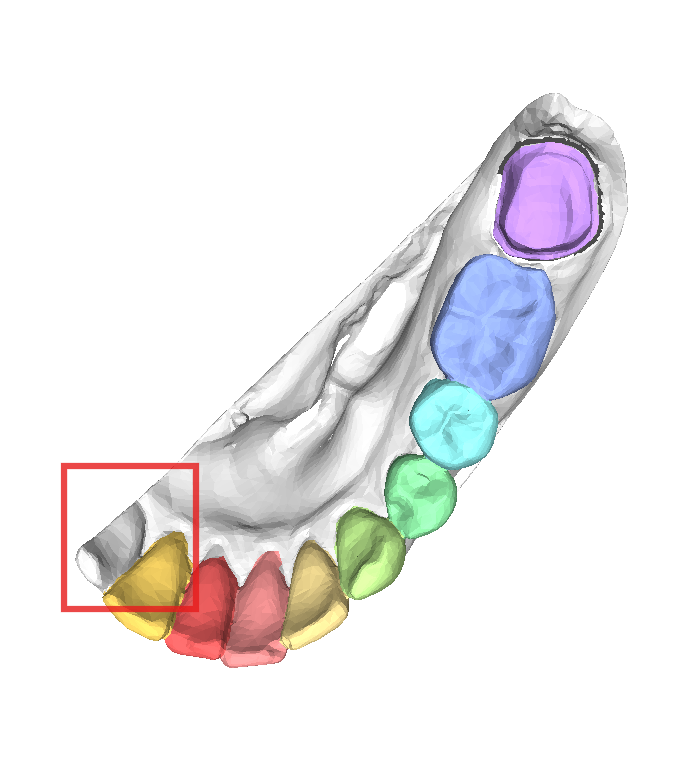} &
        \includegraphics[width=0.17\textwidth]{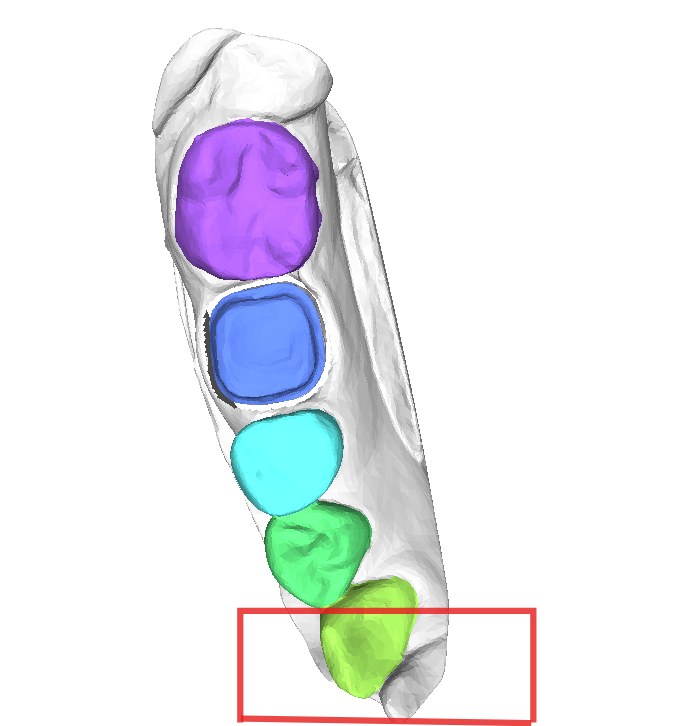} \\[6pt]
        & \scriptsize \begin{tabular}[t]{@{}l@{}}
            Jaw DSC: 0.6345 \\
            Ctx Dice: 0.9623 \\
            Ctx CE: 0.2024~mm
          \end{tabular} &
          \scriptsize \begin{tabular}[t]{@{}l@{}}
            Jaw DSC: 0.7435 \\
            Ctx Dice: 0.9401 \\
            Ctx CE: 0.2985~mm
          \end{tabular} &
          \scriptsize \begin{tabular}[t]{@{}l@{}}
            Jaw DSC: 0.8422 \\
            Ctx Dice: 0.9557 \\
            Ctx CE: 0.7081~mm
          \end{tabular} &
          \scriptsize \begin{tabular}[t]{@{}l@{}}
            Jaw DSC: 0.8586 \\
            Ctx Dice: 0.9563 \\
            Ctx CE: 0.1821~mm
          \end{tabular} &
          \scriptsize \begin{tabular}[t]{@{}l@{}}
            Jaw DSC: 0.8303 \\
            Ctx Dice: 0.9731 \\
            Ctx CE: 0.1211~mm
          \end{tabular} \\
    \end{tabular}
    \caption{\textbf{Qualitative analysis of the five lowest scoring scans.}
    \textbf{(Top row)} Ground truth annotations.
    \textbf{(Bottom row)} Model predictions.
    Per-case metrics are reported below each scan: \textit{Jaw DSC} denotes the overall macro-average arch DSC; \textit{Ctx Dice} and \textit{Ctx CE} denote the macro-average DSC and Centroid Error (mm) computed over the context area, defined as the prepared tooth and its immediate mesial (Adjacent~1) and distal (Adjacent~2) neighbors. Despite low overall arch DSC, the context area remains well-localized across all cases.}
    \label{fig:worst_cases}
\end{figure*}

Figure~\ref{fig:worst_cases} presents the five scans with the lowest overall arch DSC from the test set (partial jaw), comparing ground truth annotations against model predictions. Despite degraded global segmentation in these challenging cases, the context area --- comprising the prepared tooth and its immediate mesial and distal neighbors --- remains consistently well localized, confirming the clinical reliability of the pipeline at the restoration site.

\subsection{Full Jaw Segmentation via Fine-Tuning}
\label{ssec:exp_finetune}
While the primary focus of this work is partial jaw segmentation, a practical clinical pipeline must also handle full arch scans. Given the limited availability of full jaw data (301 scans total: 240 train, 31 val, 30 test), training from scratch was not feasible. Instead, the best-performing DilatedToothSegNet checkpoint pretrained on partial jaw scans was fine-tuned on the full jaw dataset using a two-phase strategy: the early encoder layers were frozen for the first 10 epochs to preserve learned geometric features, followed by full model fine-tuning for the remaining 90 epochs.

Across all 30 full jaw test scans, the fine-tuned model achieves an overall macro-average DSC of 0.9347, Recall of 0.9525, and Precision of 0.9211 after graph-cut refinement. Clinical performance was further evaluated on the 10 full jaw test cases containing at least one crown preparation site. As reported in Table~\ref{tab:finetune_clinical}, the fine-tuned model achieves median DSC above 0.94 and median Centroid Error below 0.41~mm across all three regions, with zero complete segmentation failures recorded.

\begin{table*}[t]
\centering
\caption{%
  Clinical focused metrics for 10 full jaw test cases after fine-tuning
  from partial jaw pretrained weights. Mean\,$\pm$\,Std with 95\%
  bootstrap CI in brackets, along with Median. CE\,=\,Centroid Error (mm).
  Miss rate\,=\,0\%.
}
\label{tab:finetune_clinical}
\renewcommand{\arraystretch}{1.12}
\setlength{\tabcolsep}{3pt}
\footnotesize
\begin{tabular*}{\textwidth}{@{\extracolsep{\fill}} l l c c c c}
\toprule
\textbf{Model} & \textbf{Region} & \textbf{DSC Mean\,$\pm$\,Std [95\% CI]} & \textbf{DSC Median} & \textbf{CE Mean\,$\pm$\,Std [95\% CI]} & \textbf{CE Median} \\
\midrule
\multirow{3}{*}{\shortstack{\textbf{DilatedToothSegNet}\\\textbf{(fine-tuned)}}}
& Prepared Tooth & $0.9281 \pm 0.0690\ [0.8832,\ 0.9631]$ & 0.9464 & $0.5097 \pm 0.4720\ [0.2813,\ 0.8207]$ & 0.4079 \\
& Adjacent 1 (Mesial) & $0.9559 \pm 0.0211\ [0.9432,\ 0.9676]$ & 0.9604 & $0.1955 \pm 0.1478\ [0.1170,\ 0.2872]$ & 0.1329 \\
& Adjacent 2 (Distal) & $0.9585 \pm 0.0340\ [0.9354,\ 0.9740]$ & 0.9679 & $0.2936 \pm 0.3387\ [0.1121,\ 0.5155]$ & 0.1214 \\
\bottomrule
\multicolumn{6}{l}{\scriptsize Miss rate = 0\%.}
\end{tabular*}
\end{table*}

\subsection{Crown Proposal Qualitative Analysis}
\label{ssec:exp_crown_qualitative}
We present a qualitative analysis using the challenging partial scan cases identified previously (Posterior Cases of Fig.~\ref{fig:worst_cases}). While the global arch segmentation metrics were lower in these specific cases, the local context area remained spatially accurate, which provides the geometric foundation for initial crown retrieval and placement.

\begin{figure*}[t]
    \centering
    \setlength{\tabcolsep}{0.8pt}
    \renewcommand{\arraystretch}{1.2}
    \newcommand{\crownpanel}[1]{%
        \includegraphics[width=0.238\textwidth]{#1}%
    }
    \begin{tabular}{m{0.05\textwidth} c c c c}
        & \scriptsize \textbf{Case 1 (FDI 36)} & \scriptsize \textbf{Case 2 (FDI 37)} & \scriptsize \textbf{Case 3 (FDI 46)} & \scriptsize \textbf{Case 4 (FDI 37)} \\
        \scriptsize \textbf{Input Scan} &
        \crownpanel{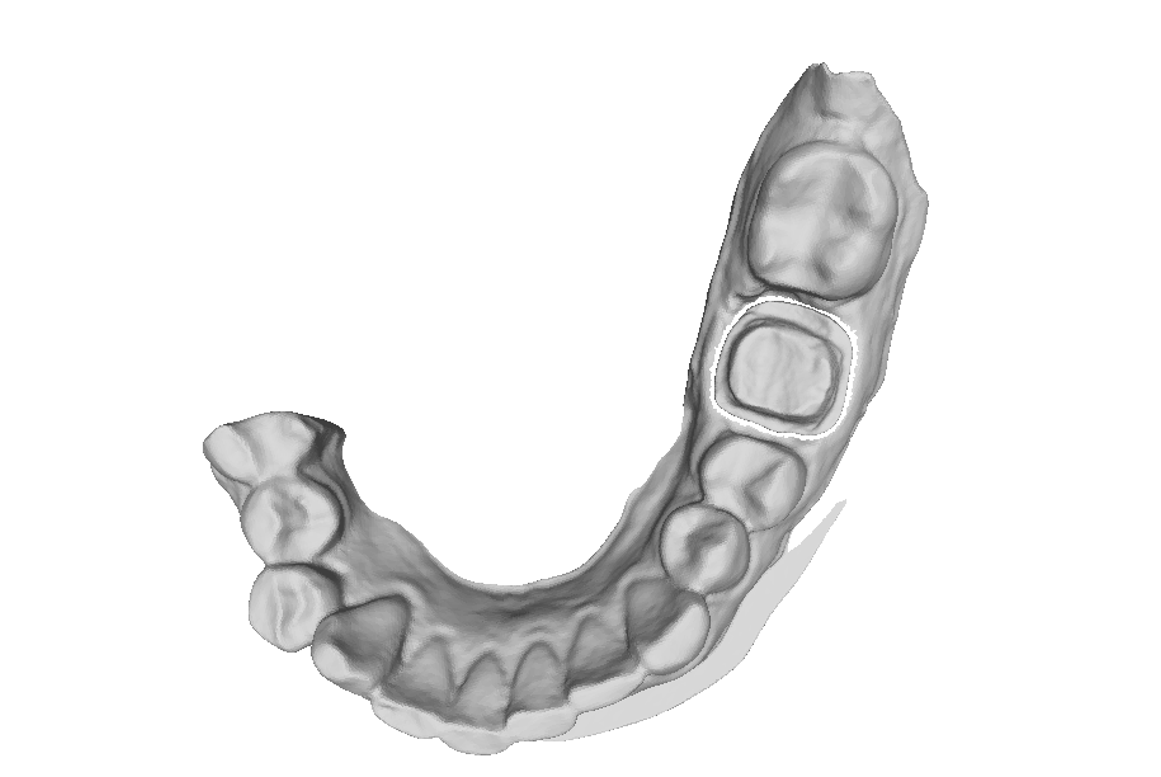} &
        \crownpanel{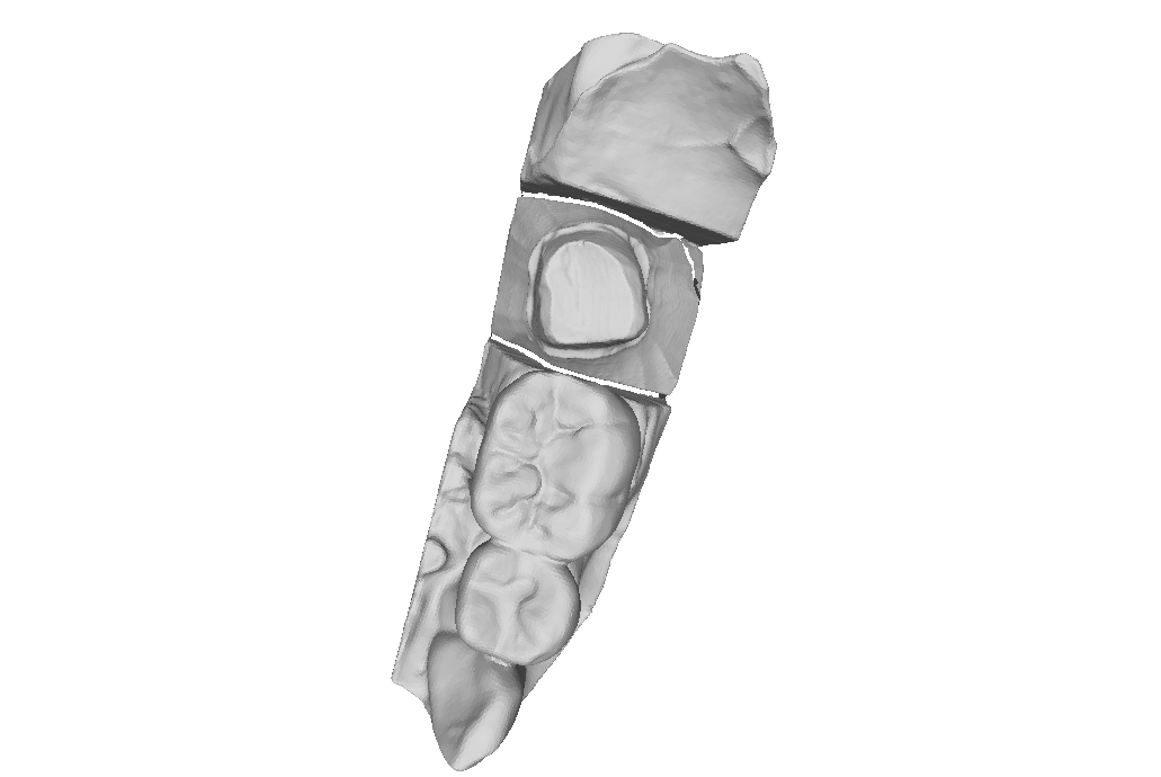} &
        \crownpanel{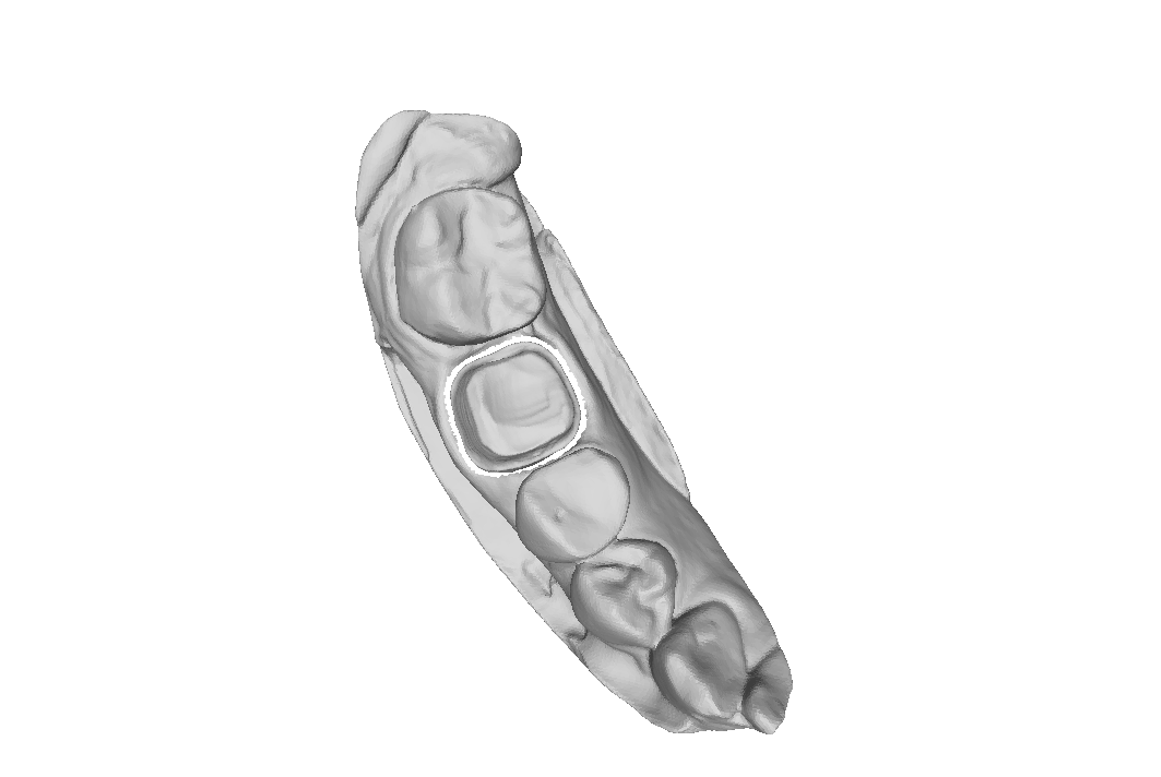} &
        \crownpanel{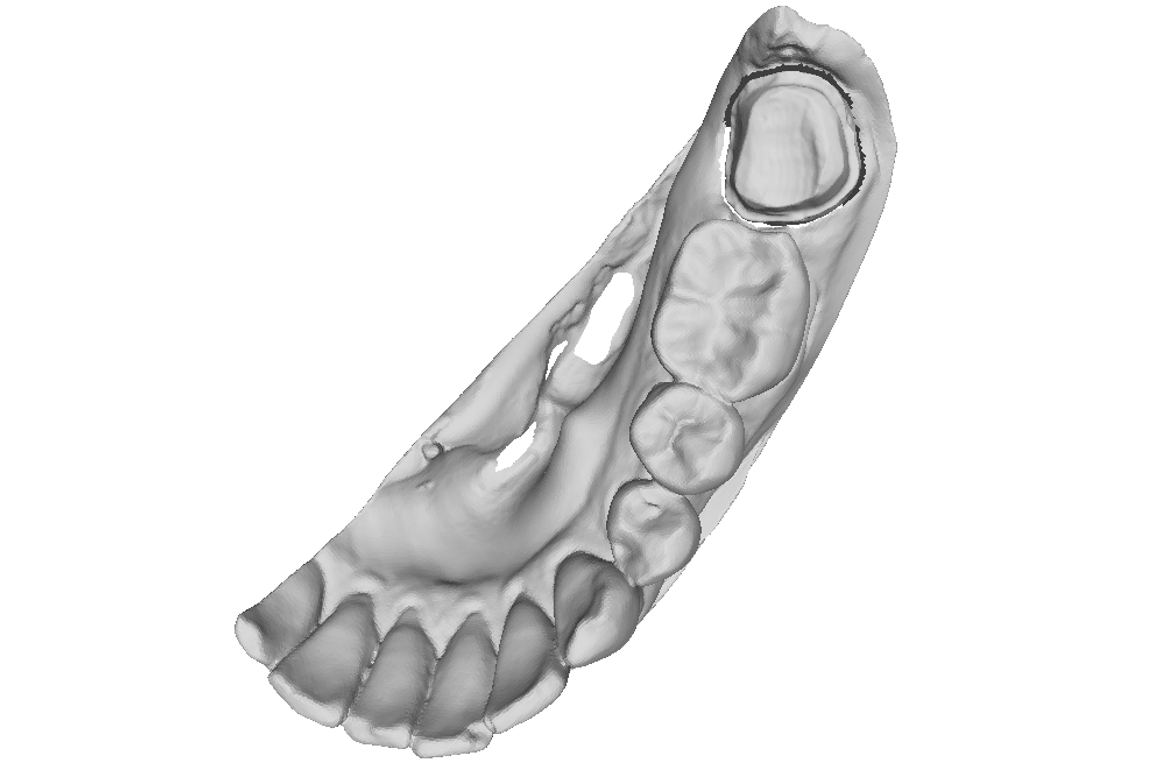} \\
        \scriptsize \textbf{Single-Jaw Crown Fit} &
        \crownpanel{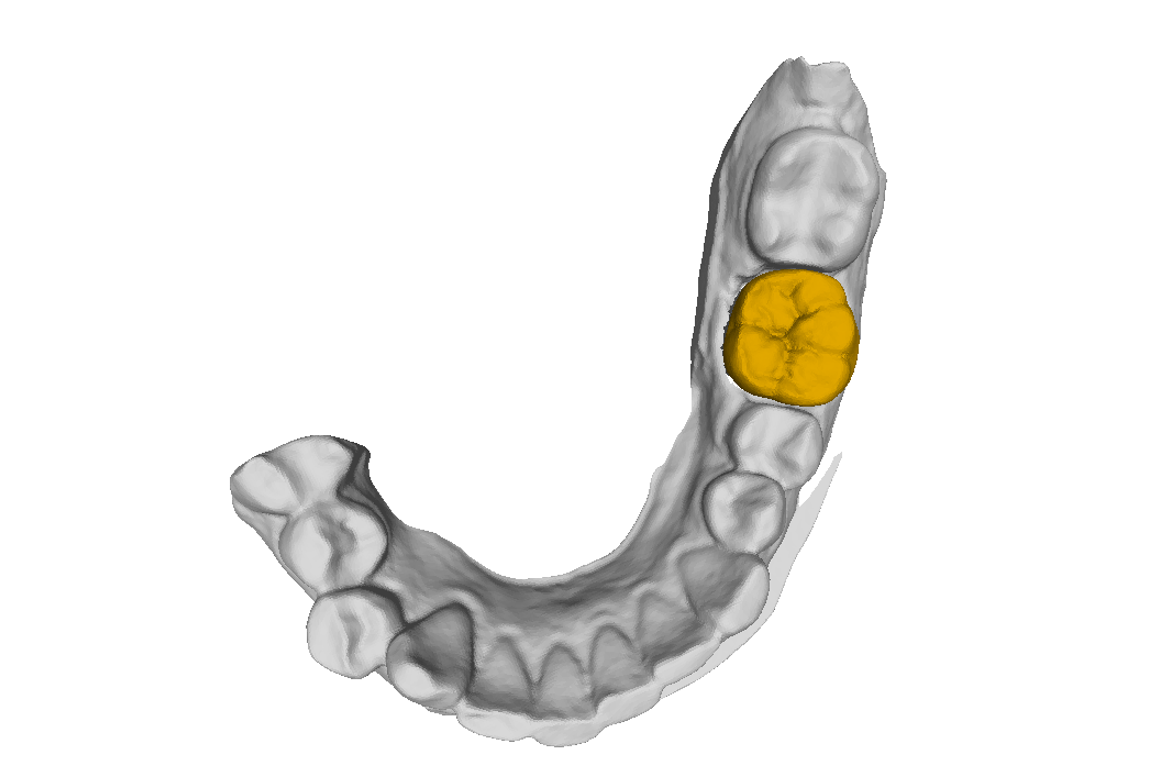} &
        \crownpanel{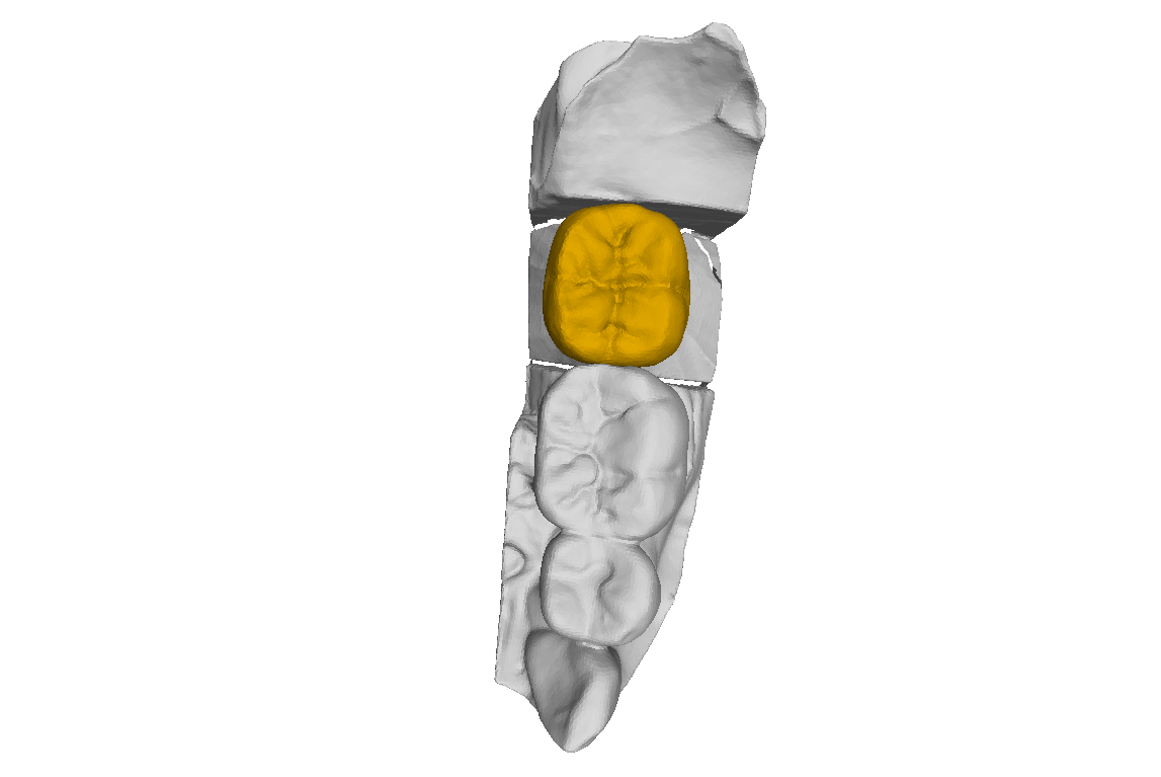} &
        \crownpanel{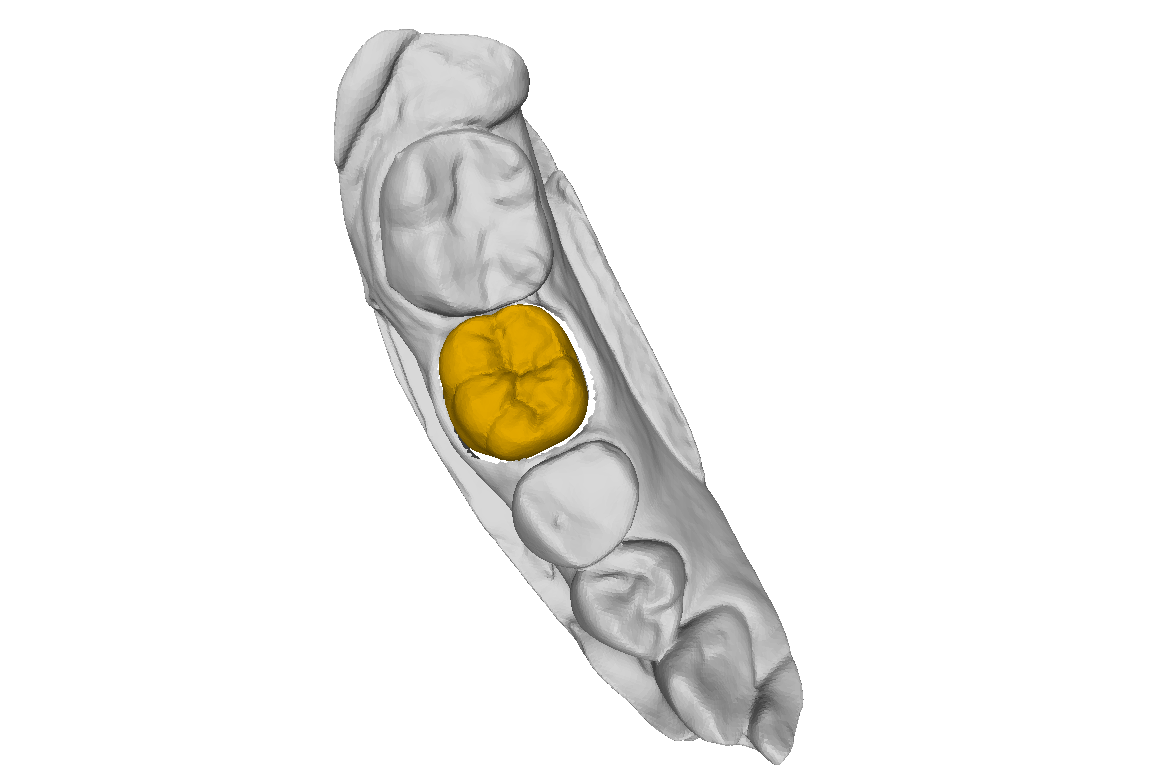} &
        \crownpanel{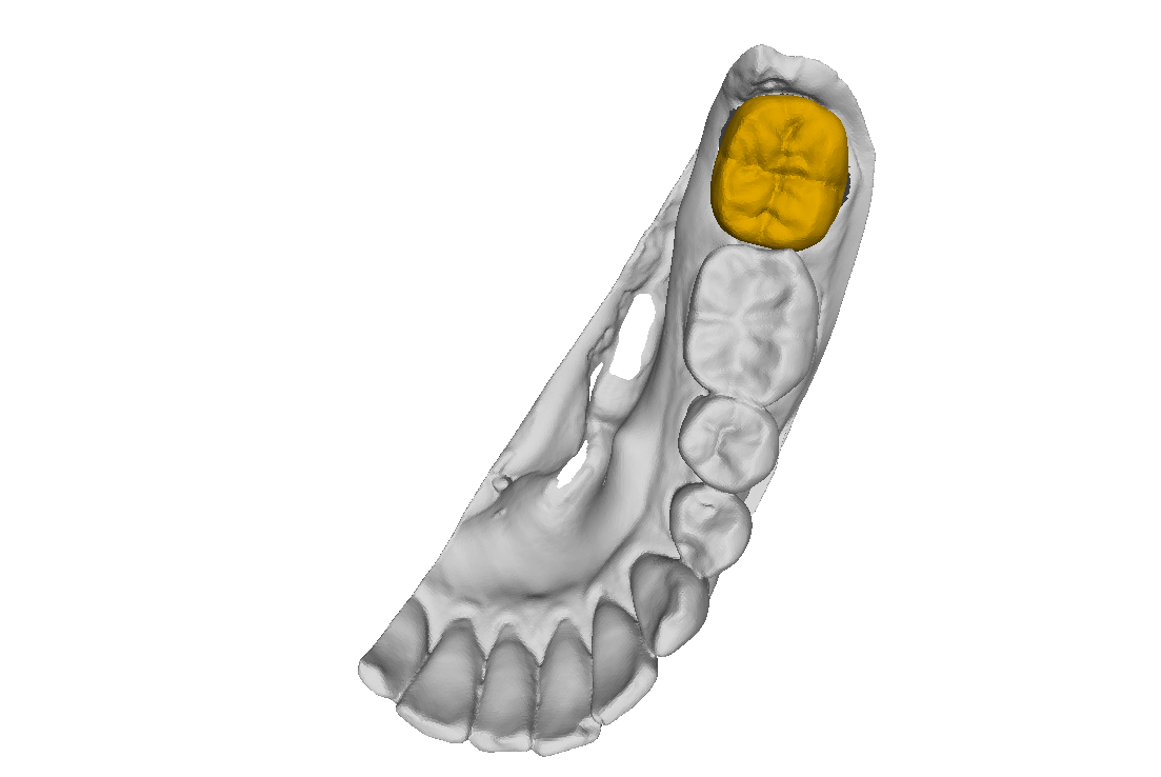} \\
        \scriptsize \textbf{Both-Jaws Occlusal Fit} &
        \crownpanel{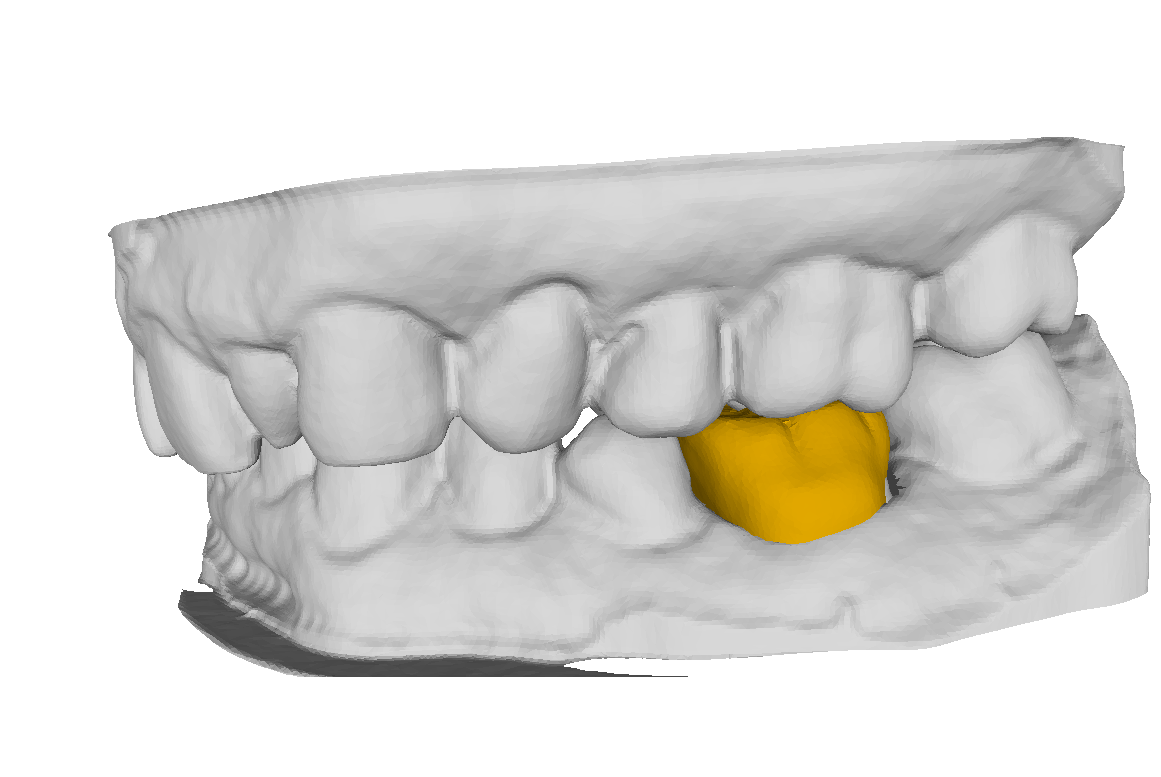} &
        \crownpanel{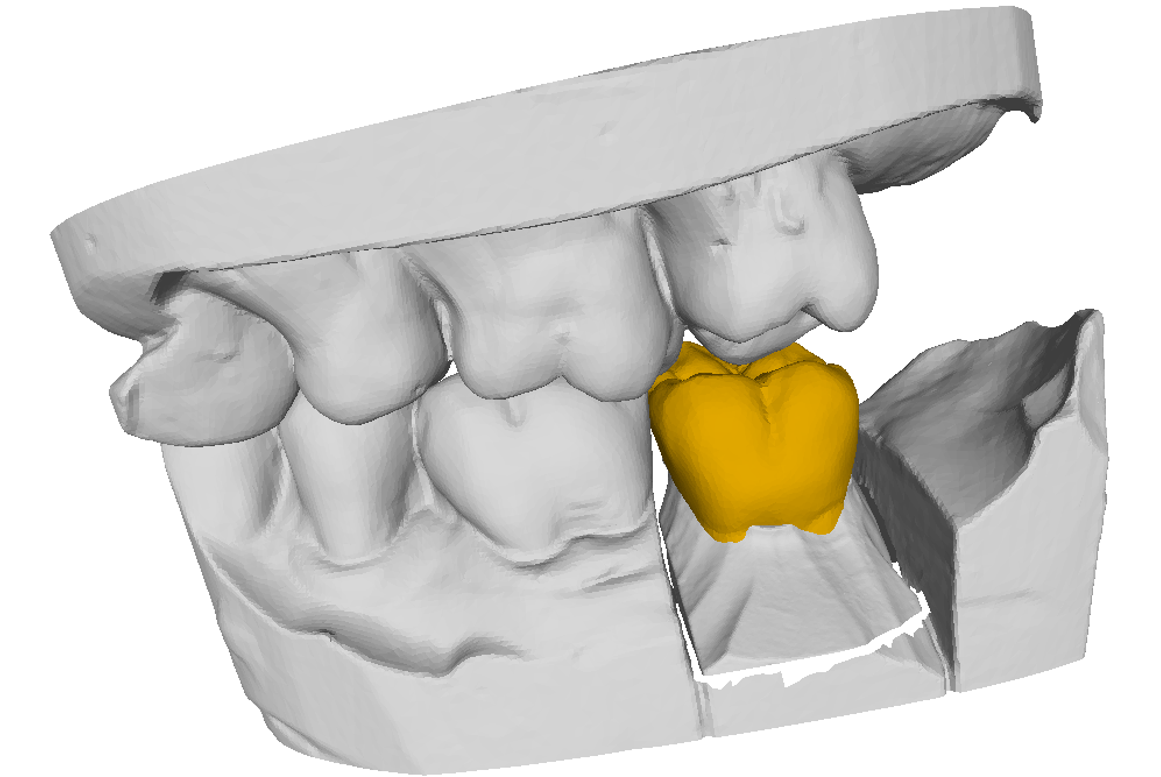} &
        \crownpanel{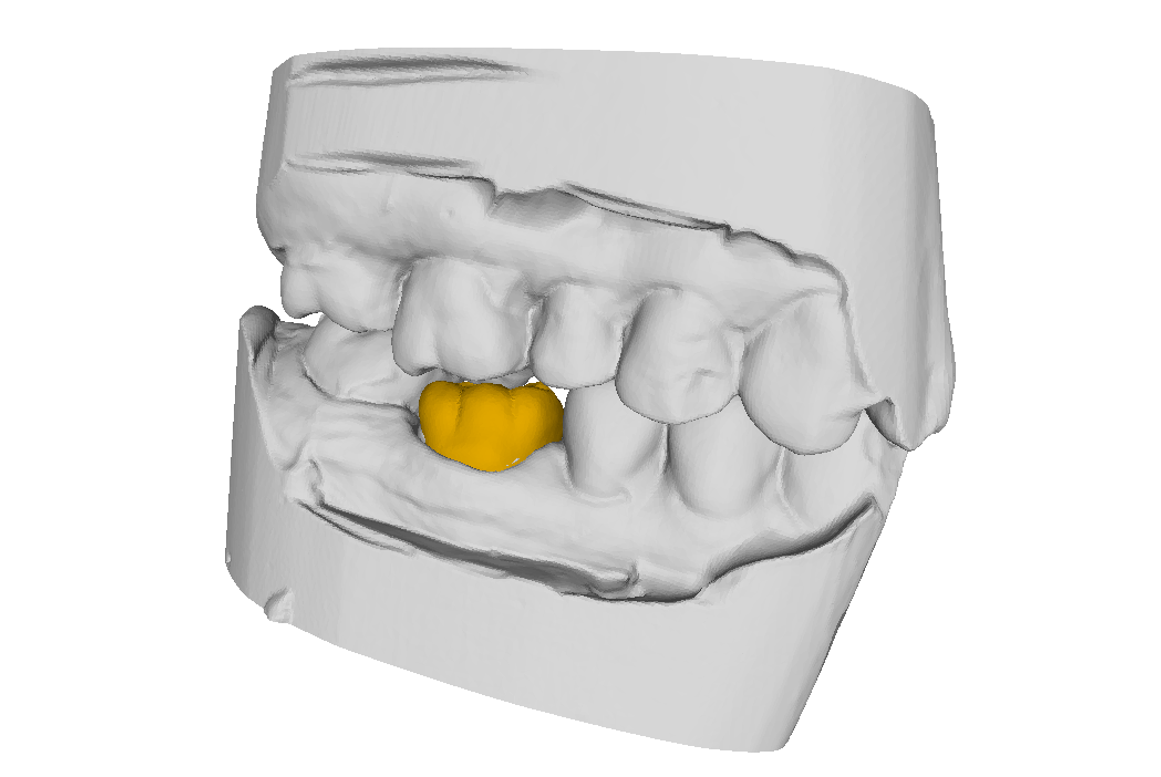} &
        \crownpanel{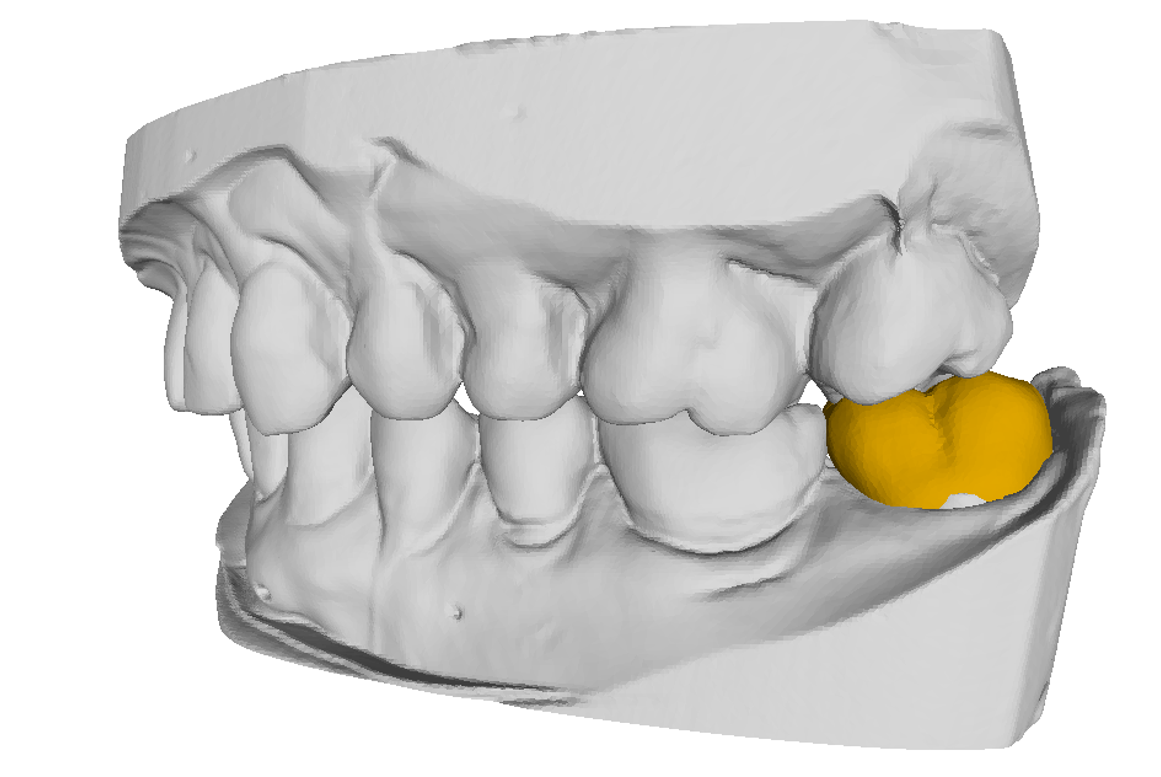} \\
    \end{tabular}
    \caption{\textbf{Qualitative evaluation} of the crown-generation workflow on challenging posterior partial scan cases identified previously in Fig.~\ref{fig:worst_cases}. For each case, the first row shows the initial partial scan with the preparation site, the second row shows the finalized crown obtained from our proposal in a single-jaw view, and the third row shows the occlusal relationship against the antagonist in a both-jaws view. The proposed pipeline provides a reliable anatomical baseline and streamlines the subsequent manual refinement stage.}
    \label{fig:qualitative_crown}
\end{figure*}

As illustrated in Fig.~\ref{fig:qualitative_crown}, the pipeline provides an initial anatomical starting point for crown design. In conventional CAD/CAM workflows, a single crown design can be time-intensive. Prior reports note that an experienced technician may require approximately 15 minutes for a single crown starting from template selection~\cite{madcrowner2026}, while broader estimates extend up to 1 hour depending on case complexity~\cite{madcrowner2026}. Controlled studies have reported active technician times of 74 to 92 minutes for complete posterior single-crown fabrication using CAD/CAM digital workflows~\cite{sailer2017}.

Our pipeline is intended to reduce these preparatory steps. The computational phase, including classification, registration, segmentation, crown retrieval, and initial placement, executes in approximately 2.5 to 3.5 minutes on a single GPU. Following this, a clinician performs necessary manual adjustments, such as margin line definition, contact verification, and occlusal refinement. In a single, informal preliminary observation of a molar case, this manual finishing step required approximately 3 minutes and 40 seconds.

We emphasize that this observation does not constitute a formal clinical time study. Finishing time will naturally vary with practitioner experience, tooth type, restoration complexity, and preparation morphology. Nevertheless, this preliminary evidence suggests that the proposed workflow can generate a reasonable geometric baseline, allowing the clinician to focus primarily on case-specific refinement rather than designing and initializing the crown entirely from scratch. A rigorous multi-operator time-motion study across different tooth categories remains necessary to quantify the true clinical efficiency gain.
\section{Ablation Studies}
\label{sec:ablation}

\subsection{Impact of Polar Features}
\label{ssec:exp_ablation}
We conducted an ablation study to evaluate the impact of the 8-dimensional feature vector used in the classification module. We compared the proposed 8D input against a 6D baseline for the exact same dataset that excludes polar coordinates ($r_i, \phi_i$). As shown in Table~\ref{tab:ablation_classification}, the inclusion of polar features resulted in a performance gain, increasing the test accuracy from 92.1\% to 94.2\% (on 240 same unseen datasets). This improvement suggests that the polar radius and azimuth provide the EdgeConv operators with critical global context regarding a point's relative position within the dental arch. This spatial information allows the network to more effectively distinguish between the global topological structures of full arches and localized partial quadrants used in our data cases.

\begin{table}[h]
\centering
\caption{Ablation study: impact of polar features on classification performance.}
\label{tab:ablation_classification}
\renewcommand{\arraystretch}{1.1}
\setlength{\tabcolsep}{3pt}
\scriptsize
\resizebox{\columnwidth}{!}{%
\begin{tabular}{lcccc}
\toprule
\textbf{Input Features} & \textbf{Accuracy} & \textbf{Precision} & \textbf{Recall} & \textbf{F1-Score} \\
\midrule
Cartesian + Normals (6D) & 92.1\% & 0.854 & 0.897 & 0.868 \\
\textbf{Cartesian + Normals + Polar (8D)} & \textbf{94.2\%} & \textbf{0.914} & \textbf{0.934} & \textbf{0.924} \\
\bottomrule
\end{tabular}
}
\end{table}

\subsection{Pre-Orientation Alignment}
\label{ssec:ablation}
A core design decision in this pipeline is the pre-orientation of partial jaw input scans into a canonical coordinate frame via RANSAC\,+\,ICP registration prior to training. To quantify its contribution, we trained an ablation model under identical architecture and evaluation conditions but with alignment removed. Since the model must now handle arbitrary scan orientations, augmentation was applied aggressively ($\pm 180^\circ$ all axes, $\pm 10$~mm translations all axes, scaling $[0.97, 1.03]$), expanding the partial jaw training set to 16,412 samples --- nearly double the 8,530 samples used in the aligned pipeline. The aligned pipeline, by contrast, uses only mild perturbations ($\pm 5^\circ$ X/Y, $\pm 15^\circ$ Z, $\pm 5$~mm X/Y, $\pm 2$~mm Z, scaling $[0.9, 1.1]$) that deliberately preserve the standardized coordinate frame. Training ran for 150 epochs at double the computational cost. Despite nearly doubling the computational cost and training data through aggressive augmentation, the unaligned model underperforms on partial jaw segmentation, highlighting the importance of canonical alignment, as shown in Table~\ref{tab:ablation}.

The most telling gap is in spatial localization. The aligned pipeline achieves a mean Prepared Tooth CE of 0.267~mm with zero complete failures, while the unaligned model reaches 0.697~mm mean CE. The effect is even more pronounced for Adjacent~2, where the unaligned model's mean CE degrades to 2.833~mm with a 4\% miss rate, indicating that aggressive augmentation cannot recover the spatial ambiguity introduced by inconsistent partial scan orientation. Overall macro-average DSC drops from 0.9249 to 0.8999, a gap that persists despite the unaligned model seeing nearly twice as many training samples. This confirms that canonical pre-orientation is not substitutable by data augmentation or additional compute.

\begin{table*}[t]
\centering
\caption{%
  Ablation study: effect of pre-orientation alignment on partial jaw segmentation.
  Both models use an identical DilatedToothSegNet architecture, dataset splits, and
  evaluation protocol. Results reported as Mean\,$\pm$\,Std with 95\% bootstrap CI in
  brackets, along with Median. CE\,=\,Centroid Error (mm).
  $^\dagger$Scans receiving bounding box diagonal penalty.
}
\label{tab:ablation}
\renewcommand{\arraystretch}{1.12}
\setlength{\tabcolsep}{3pt}
\footnotesize
\begin{tabular*}{\textwidth}{@{\extracolsep{\fill}} l l c c c c}
\toprule
\textbf{Model} & \textbf{Region} & \textbf{DSC Mean\,$\pm$\,Std [95\% CI]} & \textbf{DSC Median} & \textbf{CE Mean\,$\pm$\,Std [95\% CI]} & \textbf{CE Median} \\
\midrule
\multirow{3}{*}{W/O Alignment}
& Prepared Tooth & $0.8884 \pm 0.1737\ [0.8299,\ 0.9324]$ & 0.9554 & $0.6975 \pm 1.5522\ [0.3483,\ 1.2465]$ & 0.2218 \\
& Adjacent 1 -- Mesial & $0.9279 \pm 0.1136\ [0.8862,\ 0.9553]$ & 0.9602 & $0.4009 \pm 0.7667\ [0.2277,\ 0.6842]$ & 0.1843 \\
& Adjacent 2 -- Distal & $0.8975 \pm 0.2011\ [0.8072,\ 0.9546]$ & 0.9596 & $2.8330 \pm 8.1421^\dagger\ [0.7283,\ 5.4012]$ & 0.2208 \\
\midrule
\multirow{3}{*}{W/ Alignment}
& Prepared Tooth & $\mathbf{0.9468 \pm 0.0318\ [0.9366,\ 0.9558]}$ & 0.9573 & $\mathbf{0.2666 \pm 0.2398\ [0.2015,\ 0.3464]}$ & 0.1732 \\
& Adjacent 1 -- Mesial & $\mathbf{0.9524 \pm 0.0205\ [0.9453,\ 0.9586]}$ & 0.9564 & $\mathbf{0.2702 \pm 0.1520\ [0.2239,\ 0.3213]}$ & 0.2438 \\
& Adjacent 2 -- Distal & $\mathbf{0.9569 \pm 0.0213\ [0.9481,\ 0.9644]}$ & 0.9645 & $\mathbf{0.2774 \pm 0.2098\ [0.2035,\ 0.3633]}$ & 0.1840 \\
\bottomrule
\multicolumn{6}{l}{\footnotesize $^\dagger$Miss rate: 1/25 scans (4.00\%). Distal neighbour absent in 16/41 cases. Overall DSC: 0.8999 (w/o) vs 0.9249 (w/).}
\end{tabular*}
\end{table*}

\subsection{Graph-Cut Post-Processing}
\label{ssec:graphcut}
Following MeshSegNet~\cite{Lian2020MeshSegNet}, we also apply a graph-cut refinement step as post-processing to smooth spatially inconsistent predictions using mesh geometry. The smoothness parameter $\lambda_c$ was tuned on the validation set, and the selected value was applied unchanged to the test set to avoid data leakage.

Table~\ref{tab:graphcut} summarizes the effect on DilatedToothSegNet predictions. Consistent gains are observed across all clinically relevant regions, with no additional training cost.

\begin{table}[t]
\centering
\caption{%
  Effect of graph-cut post-processing on DilatedToothSegNet (partial jaw test set).
  Metrics are reported before (raw) and after graph-cut refinement.
  CE\,=\,Centroid Error (mm).
}
\label{tab:graphcut}
\renewcommand{\arraystretch}{1.10}
\setlength{\tabcolsep}{1pt}
\scriptsize
\resizebox{\columnwidth}{!}{%
\begin{tabular}{l c c c c}
\toprule
\textbf{Region} & \multicolumn{2}{c}{\textbf{Raw}} & \multicolumn{2}{c}{\textbf{Refined}} \\
\cmidrule(lr){2-3}\cmidrule(lr){4-5}
& \textbf{DSC} & \textbf{CE (mm)} & \textbf{DSC} & \textbf{CE (mm)} \\
\midrule
Overall (all teeth) & 0.9152 & --- & 0.9249 & --- \\
Prepared Tooth & 0.9358 & 0.2798 & 0.9468 & 0.2666 \\
Adjacent 1 (Mesial) & 0.9458 & 0.2928 & 0.9524 & 0.2702 \\
Adjacent 2 (Distal) & 0.9490 & 0.3016 & 0.9569 & 0.2774 \\
\bottomrule
\end{tabular}%
}
\end{table}

\section{Limitations and Future Work}
\label{sec:limitations}

\subsection{Segmentation}
\label{ssec:lim_segmentation}
\begin{figure*}[t]
    \centering
    \setlength{\tabcolsep}{2pt}
    \renewcommand{\arraystretch}{1.0}
    \begin{tabular}{ccccc}
        \includegraphics[width=0.196\textwidth]{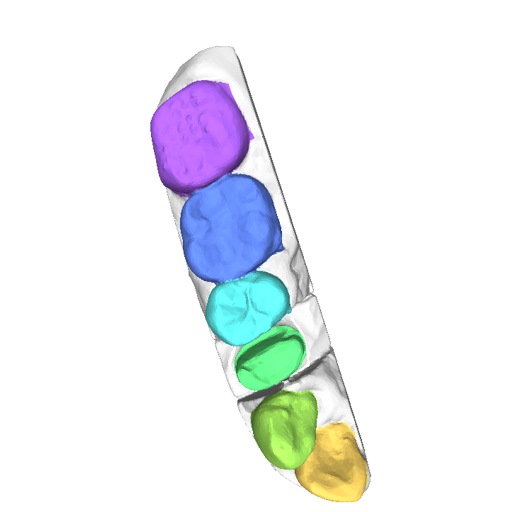} &
        \includegraphics[width=0.196\textwidth]{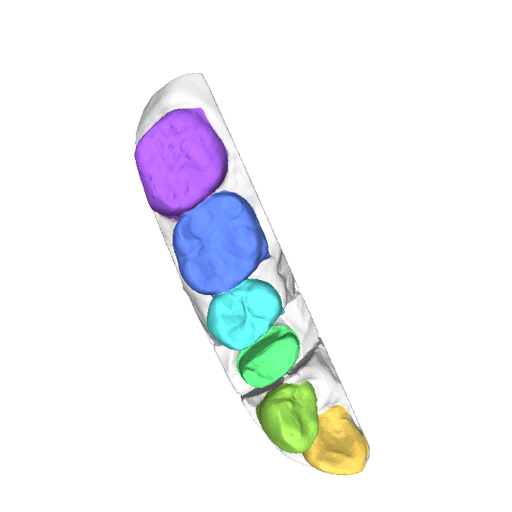} &
        \includegraphics[width=0.196\textwidth]{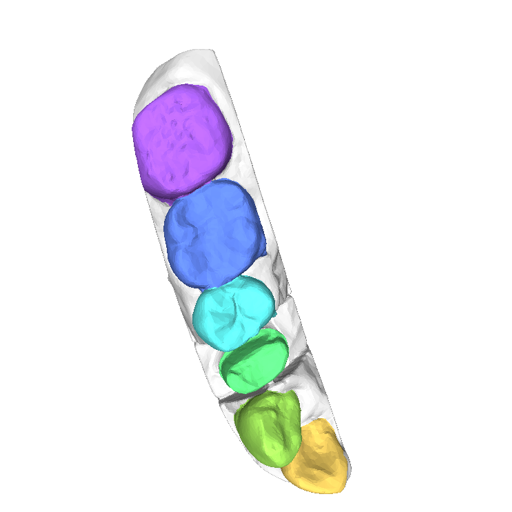} &
        \includegraphics[width=0.196\textwidth]{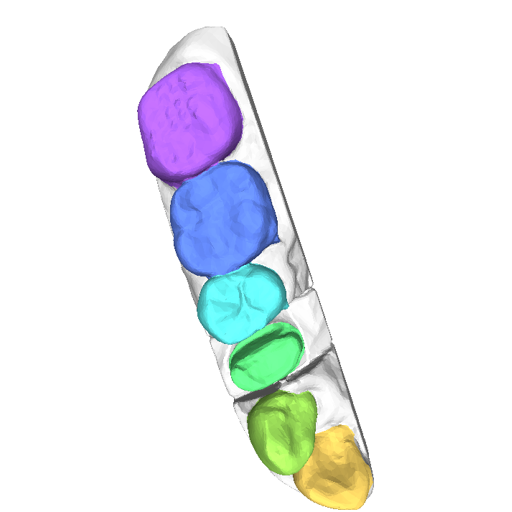} &
        \includegraphics[width=0.196\textwidth]{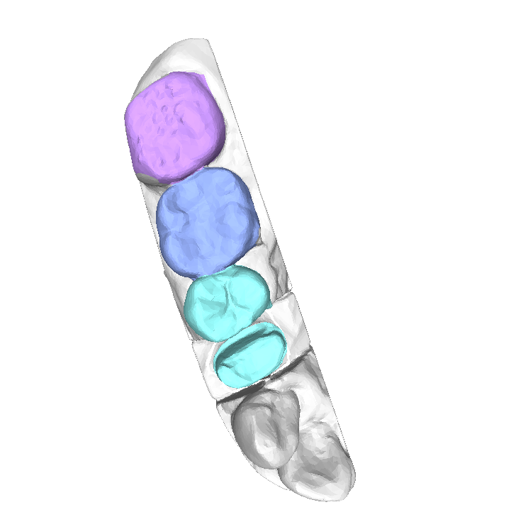} \\
        \scriptsize (a) GT &
        \scriptsize (b) Full: 0.9674 &
        \scriptsize (c) Central: 0.9664 &
        \scriptsize \begin{tabular}[t]{@{}c@{}}(d) \textbf{Left: 0.9686}\\(Aligned with correct template)\end{tabular} &
        \scriptsize \textcolor{red}{\textbf{(e) Right: 0.0838}}
    \end{tabular}
    \caption{\textbf{Misclassification impact on segmentation for a left-partial input.} The same input jaw is aligned with each candidate template (full, central-partial, partial-left, and partial-right), then passed through the same segmentation model. Compatible template alignments (b--d) preserve high DSC, while contralateral partial-right alignment (e) yields catastrophic failure.}
    \label{fig:misclassification_impact}
\end{figure*}

The proposed classify-then-align strategy introduces a dependency on the upstream classifier: if a scan is misclassified, registration is performed against an incorrect template before segmentation begins. As illustrated in Figure~\ref{fig:misclassification_impact}, most confusion modes produce negligible degradation (typically $\Delta$DSC $< 0.003$) because of shared global arch geometry across classes. In particular, the Central-partial template can still yield strong performance for some lateral partial scans when the input retains central/shared teeth that provide sufficient geometric anchors for registration. In contrast, direct Partial Left--Right confusion can cause catastrophic failure (DSC $= 0.0838$). Importantly, this high-impact error is also the least likely, consistent with strong lateral classification reliability in our result (F1: Left $= 0.944$, Right $= 0.957$). Although our dataset aggregates scans from multiple clinical sources, its scale remains constrained by the substantial effort required for manual FDI annotation. Wisdom teeth are still underrepresented despite targeted augmentation. In addition, registration may converge unreliably on severely edentulous or heavily restored arches, and the current pipeline does not have an automatic quality gate to detect such failures before segmentation proceeds. Furthermore, while DilatedToothSegNet was deliberately chosen based on its previously published superior benchmark performance, the lack of direct experimental comparison against more recent state-of-the-art architectures (such as Transformer-based networks) on our specific dataset remains a limitation.

Future work for segmentation should therefore include: (i) cost-sensitive classification that weights errors by downstream segmentation consequence, (ii) an automatic registration quality gate to flag unreliable alignments prior to inference, and (iii) confidence-aware template routing that explicitly uses central-tooth coverage to decide when Central-partial templates are safe substitutes for lateral cases. Public benchmarks such as Teeth3DS are composed of full-arch scans that are typically alignable with standard PCA-based initialization and thus do not capture the core challenge addressed here: robust partial-scan segmentation under missing-context conditions. A dedicated partial jaw benchmark spanning diverse scanner types and clinical conditions remains an important open need for the field.

\subsection{Crown Generation}
\label{ssec:lim_crown}
The main limitation of our crown retrieval stage is domain mismatch: embeddings learned from Teeth3DS full-arch data do not fully represent the morphological variability of real clinical cases, so the retrieved donor crown can be locally similar to neighboring teeth yet still remain globally inconsistent with the patient's overall arch form and jaw proportions. Alignment currently uses the global $Z$-axis as the occlusal target; this is effective for many posterior cases but less reliable for tilted or distally positioned preparations that deviate from vertical orientation. In addition, the crown-alignment stage depends on human annotation of the buccal, mesial, and occlusal regions on the crown templates. Any masking or annotation error in these manually defined regions can distort the extracted reference normals and consequently lead to incorrect crown orientation during alignment. Occlusal adjustment detects cusps using geometric peak cues only, without differentiating functional from non-functional cusps, which can lead to suboptimal corrections in complex molars. Most importantly, the crown retrieval, alignment, and fitting stages are qualitatively demonstrated in this study, and no quantitative comparison against expert-designed restorations has yet been reported.

Future work should replace the global $Z$-axis target with a patient-specific preparation normal for anatomy-driven seating, incorporate explicit cusp-type labeling for targeted occlusal correction, and fine-tune the embedding network on clinical data to reduce domain shift. A rigorous evaluation framework should then be established using objective metrics such as marginal gap distance, interproximal contact area, and occlusal clearance, together with practitioner studies measuring manual adjustment time and clinician acceptance across diverse tooth types and preparation styles. Such validation is essential to transition this pipeline from a promising research prototype toward clinically deployable decision support.

\section{Conclusion}
\label{sec:conclusion}
In conclusion, we developed a pipeline to assist dentists in designing single-unit crowns from both full and partial intraoral scans. Because AI models typically fail on partial scans due to a lack of standard orientation, we introduced a ``classify-then-align'' preprocessing step. This ensures that the segmentation model always receives the scan in the correct canonical position without requiring manual intervention.

Benefiting from this alignment, our dedicated segmentation models achieved a macro-average DSC of 0.9249 and maintained a sub-millimeter spatial localization (0.25 to 0.30~mm centroid error). Using this result, our system retrieves the best matching crown from a library and adjusts its interproximal and occlusal fit using the Blender Python API.

The entire computational process from the raw intraoral scan to a well-positioned preliminary crown shell takes approximately 2.5 to 3.5 minutes on a single GPU. Although the clinician must still perform the final clinical steps, such as cutting the margin line and verifying and finalizing contacts, our pipeline handles the most time-consuming digital setup. Specifically, it aligns the crown's mesial and buccal axes with those of the prepared tooth. This yields a fast, accurate starting point that significantly reduces manual effort and accelerates CAD/CAM workflows.

\bibliographystyle{IEEEtran}
\bibliography{references}

\end{document}